\documentclass[epj,nopacs,fleqn]{svjour}
\pdfoutput=1
\usepackage{graphicx,amsmath,amssymb,slashed,cite, verbatim,url}

\newcommand{\nn}{\nonumber}
\newcommand{\perc}{\%}
\newcommand\sss{\scriptscriptstyle}

\begin{document}

\title{Higgs characterisation at NLO in QCD: \\[3pt]
 CP properties of the top-quark Yukawa interaction} 

\author{
 Federico Demartin\inst{1},
 Fabio Maltoni\inst{1},
 Kentarou Mawatari\inst{2},
 Ben Page\inst{3},
 Marco Zaro\inst{4,5}
}

\institute{ 
 Centre for Cosmology, Particle Physics and Phenomenology (CP3),
 Universit\'e catholique de Louvain,\\
 B-1348 Louvain-la-Neuve, Belgium
 \and
 Theoretische Natuurkunde and IIHE/ELEM, Vrije Universiteit Brussel,
 and International Solvay Institutes,\\
 Pleinlaan 2, B-1050 Brussels, Belgium
 \and
 Departamento de F\'isica Te\'orica y del Cosmos y CAFPE,
 Universidad de Granada, Spain
 \and
 Sorbonne Universit\'es, UPMC Univ. Paris 06, UMR 7589, LPTHE, 
 F-75005, Paris, France
 \and
 CNRS, UMR 7589, LPTHE, F-75005, Paris, France
}

\abstract{
At the LHC the CP properties of the top-quark Yukawa interaction can be
probed through Higgs production in gluon fusion or in association with top quarks.
We consider the possibility for  both CP-even and CP-odd couplings to
the top quark to be present, and study CP-sensitive observables at
next-to-leading order (NLO) in QCD, including parton-shower effects.  
We show that the inclusion of NLO corrections sizeably reduces the
theoretical uncertainties, and confirm
that di-jet correlations in $H+2$ jet production through gluon fusion and
correlations of the top-quark decay products in $t\bar tH$ production can provide
sensitive probes of the CP nature of the Higgs interactions.
}

\date{}

\titlerunning{Higgs characterisation at NLO in QCD: 
 CP properties of the top-quark Yukawa interaction} 
\authorrunning{F.~Demartin et al.}

\maketitle


\vspace*{-12cm}
\noindent
\small{MCnet-14-21, CP3-14-59, LPN14-096}
\vspace*{10.5cm}

\section{Introduction}\label{sec:intro}

The top-quark Yukawa interaction has played a crucial role in the recent
discovery of the Higgs boson in the first run of the
LHC~\cite{Aad:2012tfa,Chatrchyan:2012ufa,Chatrchyan:2013lba,Aad:2013xqa}.
It is thanks to its large value that production in gluon fusion (GF),
which mostly proceeds through a top-quark loop in the Standard Model
(SM), has provided the necessary statistics for discovery already with a
modest integrated luminosity.
The wealth of production and decay channels available for a SM scalar with
a mass of about 125~GeV, has also made it possible to combine
information from different final-state
measurements~\cite{Zeppenfeld:2000td}.  
Global coupling
extractions~\cite{ATLAS-CONF-2014-009,Chatrchyan:2013lba} provide
indirect evidence that the Higgs boson couples to top quarks with a
strength in agreement with the SM expectations. 
Furthermore, the first exploratory searches of associated Higgs
production with a top-quark pair ($t\bar tH$), while not yet being
sensitive enough for an observation, already set an upper bound on the
strength of the interaction of 3--6 times the SM
expectation~\cite{Chatrchyan:2013yea,ATLAS-CONF-2014-011,CMS:2014jga}.  
In the coming run of the LHC at 13~TeV, $t\bar tH$ production will
certainly serve as a key channel to test the SM and explore new physics.  

While the path towards more and more precise determinations of the
strength of the Yukawa interaction with the top (and of the Higgs boson
couplings in general) is clear, the investigation of the structure and the properties of
such interaction is considerably more open.
One of the fundamental questions is whether the Higgs--top-quark coupling is
CP violating, {\it i.e.} the Higgs couples to both scalar and pseudoscalar fermion densities. 
In this context, it is important to stress that so far all experimental
determinations of the Higgs CP
properties~\cite{Aad:2013xqa,Chatrchyan:2013mxa} have been obtained
from the $H\to VV\to 4\ell$ decay mode and therefore only constrain the
$HVV$ interactions.  

Gathering information on the CP properties of the top-quark Yukawa interaction
is not an easy task. 
As there is no decay mode of the Higgs to or through top quarks that can be
effectively studied at the LHC, only Higgs production can be
considered. 
In addition, even though different couplings, either scalar,
pseudoscalar or mixed, have an impact on the production rates~\cite{Freitas:2012kw,Cheung:2013kla,Djouadi:2013qya} and  
can also be bound by indirect measurements~\cite{Brod:2013cka}, 
only specially designed observables can provide direct evidence of 
CP-violating effects at hadron colliders. 
In inclusive Higgs production, for example, at least two extra jets are
needed in the final state to be able to construct CP-sensitive
observables. 
These can probe the Higgs interaction with the top quark through GF (as well
as with $W$ and $Z$'s in vector boson fusion (VBF)). 
The $t\bar tH$ final state, on the other hand, provides many
CP-sensitive observables that can also be constructed from the daughters
of the top-quark decays. In fact, in $H+$jets and $t\bar tH$ production
information on the CP nature of the top-quark coupling is encoded in the
correlations between the jets and among the top--antitop decay products. 
This means that the choice of decay mode of the Higgs in the
corresponding analyses can be done freely and based on criteria other
than the requirement of a precise reconstruction of the Higgs resonance,
something that, in general, might not even be needed.  

In order to test the different hypotheses for the Higgs sector, the
approach based on an effective field theory (EFT) turns out to be
particularly suitable, given the fact that the current experimental data
do not show any clear sign of physics beyond the SM.
In such an approach, no new particle and symmetry is hypothesised 
on top of the SM ones.
This has the advantage of reducing the number of new parameters and
interactions compared to other approaches based only on Lorentz
symmetry, without losing the ability to describe in a model-independent
way the effects of any new physics we cannot directly access at the
current energies. 
Furthermore, the EFT approach can be systematically improved by
including higher-dimensional operators in the lagrangian on the one hand
(which are suppressed by higher powers of the scale $\Lambda$ where
new physics appears), and higher-order perturbative corrections on the
other hand. 

The aim of this work is to present how EFT predictions accurate to
next-to-leading order (NLO)
in QCD matched to a parton shower can be used to determine the CP
properties of the Higgs boson coupling to the top quark, through Higgs
production in association with jets or with a pair of top quarks.
To this aim we employ the {\it Higgs Characterisation} (HC) framework
originally proposed in ref.~\cite{Artoisenet:2013puc}, which follows the
general strategy outlined in ref.~\cite{Christensen:2009jx} and has
been recently applied to the VBF and VH
channels~\cite{Maltoni:2013sma}. 
In this respect, this work contributes to  the general effort of
providing NLO accurate tools and predictions to accomplish the most
general and accurate characterisation of Higgs interactions in the main
production modes at the LHC. 
Note that at variance with VBF and VH, $H+$jets and $t\bar tH$ are processes
mediated by QCD interactions at the Born level, hence higher-order
corrections are expected to be more important and certainly needed 
in analyses aiming at accurate and precise extractions of the
Higgs properties.   

First, we consider Higgs production in GF together with extra
jets, focusing on final states with at least two jets.
This process is not only a background to VBF, but it can also provide
complementary information on the Higgs boson coupling 
properties~\cite{Klamke:2007cu,Hagiwara:2009wt,Andersen:2010zx,Campanario:2010mi,Englert:2012ct,Englert:2012xt,Dolan:2014upa}.
In the heavy-top limit, the CP structure of the Higgs--top interaction is 
inherited by the effective Higgs--gluon 
vertices~\cite{Dawson:1990zj,Djouadi:1991tka,Kauffman:1993nv,Spira:1995rr,Kniehl:1995tn,Dawson:1998py}. 
Higgs plus two (three) jets through GF at LO has been computed in 
refs.~\cite{DelDuca:2001eu,DelDuca:2001fn} (refs.~\cite{Campanario:2013mga,Campanario:2014oua}), where the full top-mass
dependence was retained.  The results cited above show that the large
top-mass limit is a very good approximation as long as the transverse
momentum of the jets is not sensibly larger than the top mass and
justify the use of EFT approach for the Higgs--gluons interactions.
In the $m_t\to\infty$ limit, the resulting analytic expressions at NLO
for GF $Hjj$ production have been implemented in 
{\sc MCFM}~\cite{Campbell:2010cz}, which has been used by
{\sc Powheg Box}~\cite{Campbell:2012am} and 
{\sc Sherpa}~\cite{Hoeche:2014lxa} to obtain NLO results matched with
parton shower (NLO+PS).
Independent NLO+PS predictions in the {\sc Sherpa} package using 
{\sc GoSam}~\cite{vanDeurzen:2013rv} for the one-loop matrix elements
and in  {\sc MadGraph5\_aMC@NLO}~\cite{Alwall:2014hca}, which embodies 
{\sc MadFKS}~\cite{Frederix:2009yq} and {\sc MadLoop}~\cite{Hirschi:2011pa},
are also available.  
We note that all the above predictions are for the SM Higgs boson, 
{\it i.e.} the CP-even state, and $Hjj$ production for the CP-odd state
has been only available at LO, yet with the exact top-mass
dependence~\cite{Campanario:2010mi}.   
In this paper we present NLO results in the large top-mass limit for GF
production of a generic (mixed) scalar/pseudoscalar state in association
with one or two jets at the LHC, also matching to parton shower.   

Second, we study $t\bar tH$ production for arbitrary CP couplings,
including  NLO+PS effects.  
While NLO corrections in QCD for this process have been known for quite
some time~\cite{Beenakker:2001rj,Dawson:2002tg}, the NLO+PS prediction
has been done only recently, for both CP eigenstates, $0^+$ and
$0^-$, in {\sc aMC@NLO}~\cite{Frederix:2011zi} and in the
{\sc Powheg Box}~\cite{Garzelli:2011vp} for the CP-even case only. 
The spin-correlation effects of the top--antitop decay products have also
been studied at the NLO+PS level with the help of  
{\sc MadSpin}~\cite{Artoisenet:2012st,Biswas:2014hwa}.  
Weak and electroweak corrections have also been reported recently in 
refs.~\cite{Frixione:2014qaa} and \cite{Yu:2014cka}, respectively.
The phenomenology of a  CP-mixed Higgs coupling to the top quark at the
LHC has been
studied at LO in ref.~\cite{Ellis:2013yxa}.
In addition to the case where the Higgs has definite CP quantum numbers,
here we consider the more general case of a CP-mixed particle ($0^\pm$)
including NLO in QCD, parton-shower effects and spin-sucorrelated decays.  

The paper is organised as follows.
In the next section we recall the effective lagrangian employed for a
generic spin-0 resonance and define sample scenarios used to determine
the CP properties of the Higgs boson.  
We also briefly describe our setup for the computation of NLO
corrections in QCD together with matching to parton shower.
In Sect.~\ref{sec:hjj} we present results of $H+$jets in GF, focusing
on the $H+2$ jet production. We also make a comparison with VBF
production with dedicated kinematical cuts.  
In Sect.~\ref{sec:tth} we illustrate the $t\bar tH$ production channel.
In Sect.~\ref{sec:summary} we briefly summarise our findings and 
in Appendix~\ref{app:eftnlo} we present the Feynman rules, the UV and
the $R_2$ counterterms necessary to NLO computations for GF in the
heavy-top-quark limit.

\section{Setup}\label{sec:setup}

In this section, we summarise our setup. We start from the definition of
the effective lagrangian, pass to the identification of suitable
benchmark scenarios, and finally to event generation at NLO in QCD
accuracy, including parton-shower effects.

\subsection{Effective lagrangian and benchmark scenarios}

The most robust approach to build an effective lagrangian is 
to employ all the SM symmetries, {\it i.e.} start from a linearly realised electroweak symmetry
 and systematically write all  higher-dimensional operators, organised in terms of increasing dimensions.
The complete basis at dimension six has been known for a long
time~\cite{Buchmuller:1985jz,Grzadkowski:2010es} and recently 
reconsidered in more detail in the context of the Higgs boson;
see {\it e.g.},~\cite{Giudice:2007fh,Buchalla:2012qq,Contino:2013kra}. This approach has been 
followed in the {\sc FeynRules}~\cite{Alloul:2013bka} implementation of ref.~\cite{Alloul:2013naa}, 
where the effective lagrangian is written in terms of fields above the electroweak symmetry breaking (EWSB) scale and then expressed in terms of gauge eigenstates.

In ref.~\cite{Artoisenet:2013puc} we have
followed an alternative approach (and yet fully equivalent in the context of the
phenomenological applications of this paper, as explicitly seen in tables~1 and 3 of ref.~\cite{Alloul:2013naa}) and implemented
the EFT lagrangian starting from the mass eigenstates, so below the
EWSB scale, and for various spin--parity assignments  ($X(J^P)$ with
$J^P=0^{\pm},1^{\pm},2^+$).  We have also used {\sc FeynRules}, whose
output in the {\sc UFO} format~\cite{Degrande:2011ua,deAquino:2011ub}
can be directly passed to {\sc MadGraph5\_aMC@NLO} \cite{Alwall:2014hca}.  We stress that this procedure is fully automatic for computations at LO, while  at NLO 
the UFO model has to be supplemented with suitable counterterms, as will
be recalled in Sect.~\ref{sec:NLO}, a procedure that in this work has been performed by hand. 

The  term of interest in the effective lagrangian can be written as (see eq.~(2.2) in ref.~\cite{Artoisenet:2013puc}): 
\begin{align}
 {\cal L}_0^t 
   = -\bar\psi_t\big(
         c_{\alpha}\kappa_{\sss Htt}g_{\sss Htt} 
       +i s_{\alpha}\kappa_{\sss Att}g_{\sss Att}\, \gamma_5 \big)
      \psi_t\, X_0 \,,
\label{eq:0ff}
\end{align}
where $X_0$ labels the scalar boson, $c_{\alpha}\equiv\cos\alpha$ and
$s_{\alpha}\equiv\sin\alpha$ can be thought of as
``CP mixing'' parameters, $\kappa_{\sss Htt,Att}$ are the dimensionless
real coupling parameters, and
$g_{\sss Htt}=g_{\sss Att}=m_t/v\,(=y_t/\sqrt{2})$, with $v\!\sim\! 246$~GeV.  
While obviously redundant (only two independent real quantities are
needed to parametrise the most general CP-violating interaction), 
this parametrisation has several practical advantages, among which the possibility of  easily 
interpolating  between the CP-even ($c_{\alpha}=1,s_{\alpha}=0$) and
CP-odd  ($c_{\alpha}=0,s_{\alpha}=1$) assignments
as well as recovering the SM case by the dimensionless and dimensionful
coupling parameters $\kappa_i$ and $g_{\sss Xyy'}$.

\begin{table}
\center
\begin{tabular}{lccc}
\hline
 $g_{\sss Xyy'}$ & $gg$ & $\gamma\gamma$ & $Z\gamma$  \\[0.5mm]
\hline\\[-3mm]
 $X=H$ & $-\alpha_s/3\pi v$ & $47\alpha_{\rm EM}/18\pi v$ 
         & $ C (94 c^2_W-13)/9\pi v$  \\[1mm]
 $X=A$ & $\alpha_s/2\pi v$ & $4\alpha_{\rm EM}/3\pi v$ 
         & $2 C (8c^2_W-5)/3\pi v$ 
 \\[0.5mm] 
\hline
\end{tabular}
\caption{Loop-induced couplings $g_{\sss Xyy'}$ in the lagrangian~\eqref{L_loop}.
 $c_W=\cos\theta_W$ and 
 $C=\sqrt{\frac{\alpha_{\sss\rm EM}G_F m_Z^2}{8\sqrt{2}\pi}}$.} 
\label{tab:gXaa}
\end{table}

\begin{table}
\center
\begin{tabular}{lll}
 \hline
 parameter & description \\
 \hline\\[-3mm]
 $\Lambda$ [GeV] & cutoff scale \\[0.5mm]
 $c_{\alpha}\,(\equiv \cos\alpha$) & mixing between $0^+$ and
         $0^-$ \\[0.5mm]
 $\kappa_i$ & dimensionless coupling parameter \\[0.5mm]
 \hline
\end{tabular}
\caption{HC model parameters.}
\label{tab:param}
\end{table}

The Higgs interaction with the top-quarks induces a (non-decoupling) effective
couplings to photons, gluons, and photon-$Z$
gauge bosons through a top-quark loop. In the HC framework,
the effective lagrangian for such loop-induced interactions 
with vector bosons reads (eq.~(2.4) in ref.~\cite{Artoisenet:2013puc}):
\begin{align}
 {\cal L}_0^{\rm loop} &=\bigg\{ 
  -\frac{1}{4}\big[c_{\alpha}\kappa_{\sss Hgg}g_{\sss Hgg} \,
 G_{\mu\nu}^aG^{a,\mu\nu} \nn\\
   &\hspace*{4.2em}+s_{\alpha}\kappa_{\sss Agg}g_{\sss Agg}\,G_{\mu\nu}^a\widetilde G^{a,\mu\nu} \big] \nn\\
   &\hspace*{2.1em} -\frac{1}{4}\big[c_{\alpha}\kappa_{\sss H\gamma\gamma}
  g_{\sss H\gamma\gamma} \, A_{\mu\nu}A^{\mu\nu} \nn\\
   &\hspace*{4.2em}+s_{\alpha}\kappa_{\sss A\gamma\gamma}g_{ \sss A\gamma\gamma}\,
  A_{\mu\nu}\widetilde A^{\mu\nu}
  \big] \nn\\
   &\hspace*{2.1em} -\frac{1}{2}\big[c_{\alpha}\kappa_{\sss HZ\gamma}g_{\sss HZ\gamma} \, 
  Z_{\mu\nu}A^{\mu\nu} \nn\\
   &\hspace*{4.2em}+s_{\alpha}\kappa_{\sss AZ\gamma}g_{\sss AZ\gamma}\,Z_{\mu\nu}\widetilde A^{\mu\nu} \big] 
 \bigg\} X_0\,,
\label{L_loop}
\end{align}
where the (reduced) field strength tensors are defined as
\begin{align}
 G_{\mu\nu}^a &=\partial_{\mu}^{}G_{\nu}^a-\partial_{\nu}^{}G_{\mu}^a
  +g_sf^{abc}G_{\mu}^bG_{\nu}^c\,, \\
 V_{\mu\nu} &=\partial_{\mu}V_{\nu}-\partial_{\nu}V_{\mu}\quad 
  (V=A,Z,W^{\pm})\,,
\end{align}
and the dual tensor is
\begin{align}
 \widetilde V_{\mu\nu}
 =\frac{1}{2}\epsilon_{\mu\nu\rho\sigma}V^{\rho\sigma}\,.
\end{align}
We note that the $X_0$--gluon lagrangian provides not only the $ggX_0$,
but also the $gggX_0$ and $ggggX_0$ effective vertices, see Appendix~\ref{app:eftnlo}.%
\footnote{The CP-odd case does not have the $ggggX_0$ vertex due to the
anti-symmetric nature of the interaction.}
For the $X_0\gamma\gamma$ and $X_0Z\gamma$ interactions, in addition to
the top-quark loop, a $W$-boson loop contributes for the CP-even case and in fact dominates. 
As a result, these processes are less sensitive to the CP properties of
the top Yukawa coupling. 
The dimensionful loop-induced couplings $g_{\sss Xyy'}$ are shown in
table~\ref{tab:gXaa}. In the following, we focus only on the gluonic operators in
eq.~\eqref{L_loop}.  
As mentioned in the introduction, the EFT prediction can be improved by
including higher-dimensional operators, and this can be achieved
rather easily in our framework by adding, e.g. the dimension-seven 
Higgs--gluon lagrangian~\cite{Harlander:2013oja}, into the {\sc HC}
model.   
Finally, we remind the reader that in the HC lagrangian the loop-induced $X_0ZZ$
and $X_0WW$ interactions are parametrized by the cutoff $\Lambda$, since
those are sub-leading contribution to the SM tree-level interaction; see
eq.~\eqref{eq:0vv} below. 
    
In order to compare GF and VBF in the $Hjj$
channel, we also write the effective lagrangian for the interactions with
massive gauge bosons (eq.~(2.4) in ref.~\cite{Artoisenet:2013puc}): 
\begin{align}
 &{\cal L}_0^{Z,W} =\bigg\{
  c_{\alpha} \kappa_{\sss{\rm SM}} \, \Big[\frac{1}{2}g_{\sss HZZ}\, Z_\mu Z^\mu 
                           + g_{\sss HWW}\, W^+_\mu W^{-\mu}\Big] \nn\\
  &\hspace*{1.15em} -\frac{1}{4}\frac{1}{\Lambda}\big[c_{\alpha}\kappa_{\sss HZZ} \, Z_{\mu\nu}Z^{\mu\nu}
        +s_{\alpha}\kappa_{\sss AZZ}\,Z_{\mu\nu}\widetilde Z^{\mu\nu} \big] \nn\\
  &\hspace*{1.15em} -\frac{1}{2}\frac{1}{\Lambda}\big[c_{\alpha}\kappa_{\sss HWW} \, W^+_{\mu\nu}W^{-\mu\nu}
        +s_{\alpha}\kappa_{\sss AWW}\,W^+_{\mu\nu}\widetilde W^{-\mu\nu}\big] \nn\\ 
  &\hspace*{1.15em} -\frac{1}{\Lambda}c_{\alpha} 
    \big[ 
          \kappa_{\sss H\partial Z} \, Z_{\nu}\partial_{\mu}Z^{\mu\nu} \nn\\
  &\hspace*{4.5em}     
 + \big( \kappa_{\sss H\partial W} W_{\nu}^+\partial_{\mu}W^{-\mu\nu}+h.c.\big)
 \big]
 \bigg\} X_0\,,
 \label{eq:0vv}
\end{align}
where $g_{\sss HZZ}=2m_Z^2/v$ and $g_{\sss HWW}=2m_W^2/v$ are the SM
couplings, and $\Lambda$ is the cutoff scale.  
The HC model parameters are summarised in table~\ref{tab:param}.

\begin{table}
\center
\begin{tabular}{ll}
\hline \\[-3mm]
 scenario for GF/$t\bar tH$ & HC parameter choice \\
\hline \\[-3mm]
 $0^+$(SM) &  $\kappa_{\sss Hgg/Htt}=1\ (c_{\alpha}=1)$\\[0.5mm]
 $0^-$ &  $\kappa_{\sss Agg/Att}=1\ (c_{\alpha}=0)$\\[0.5mm]
 $0^{\pm}$ & $\kappa_{\sss Hgg,Agg/Htt,Att}=1\ (c_{\alpha}=1/\sqrt{2})$\\[0.5mm]
\hline 
\end{tabular}
\caption{Benchmark scenarios for GF/$t\bar tH$.}
\label{tab:GFscenarios}
\end{table}

\begin{table}
\center
\begin{tabular}{ll}
\hline\\[-3mm]
 scenario for VBF & HC parameter choice \\
\hline\\[-3mm]
 $0^+$(SM) &  $\kappa_{\sss SM}=1\ (c_{\alpha}=1)$\\[0.5mm]
 $0^+$(HD) &  $\kappa_{\sss HZZ,HWW}=1\ (c_{\alpha}=1)$\\[0.5mm]
 $0^-$(HD) &  $\kappa_{\sss AZZ,AWW}=1\ (c_{\alpha}=0)$\\[0.5mm]
 $0^{\pm}$(HD) & $\kappa_{\sss HZZ,HWW,AZZ,AWW}=1\ (c_{\alpha}=1/\sqrt{2})$\\[0.5mm]
\hline 
\end{tabular}
\caption{Benchmark scenarios for VBF used for comparison with Higgs production in GF.}
\label{tab:VBFscenarios}
\end{table}

In table~\ref{tab:GFscenarios} we list the representative scenarios that 
we later use for illustration.  
The first scenario, which we label $0^+$(SM), corresponds to the SM, with the
couplings to fermions as described by eq.~\eqref{eq:0ff}, and the
effective couplings to gluons as described by the corresponding gluonic 
operators in eq.~\eqref{L_loop}. 
The second scenario, which we label $0^-$, corresponds to a pure pseudoscalar
state.  The third scenario, $0^\pm$, describes a CP-mixed case, where
the spin-0 boson is a scalar/pseudoscalar state in equal proportions. 

To compare between $H+2$ jets in GF and in VBF, we collect in
table~\ref{tab:VBFscenarios} some of the new physics scenarios
considered in the previous HC paper~\cite{Maltoni:2013sma}. 
The first scenario corresponds to the SM.
The second scenario, $0^+$(HD), represents a scalar state interacting 
with the weak bosons in a custodial invariant way 
through the higher-dimensional (HD) operators of eq.~\eqref{eq:0vv}
corresponding to $\kappa_{\sss HZZ, HWW}$. 
The third scenario, $0^-$(HD), is the analogous of a pure pseudoscalar
state, while the fourth scenario is representative of a CP-mixed case,
with equal contributions from the scalar and pseudoscalar components.

\subsection{NLO corrections matched with parton shower}
\label{sec:NLO}

{\sc MadGraph5\_aMC@NLO} is designed to perform
automatic computations of tree-level and NLO differential cross
sections, including the possibility of matching LO and NLO calculations to
parton showers via the MC@NLO method~\cite{Frixione:2002ik}, and also to merge LO~\cite{Alwall:2008qv} and 
NLO~\cite{Frederix:2012ps} samples that differ in parton multiplicities. 
Currently, NLO computations are restricted to QCD corrections.
They can be achieved fully automatically in the SM. 
Recently, the computation of ultraviolet (UV) and $R_2$ counterterms, 
the latter being originally necessary to compute one-loop amplitudes with the 
{\sc CutTools}~\cite{Ossola:2007ax} implementation of the OPP
integrand-reduction method \cite{Ossola:2006us},  was
automated for any renormalisable theory~\cite{Degrande:2014vpa}.

The UV and $R_2$ counterterms for QCD one-loop amplitudes in the SM were presented
in~\cite{Draggiotis:2009yb} and have been available in
{\sc MadGraph5\_aMC@NLO} for some time.  The corresponding  terms for 
effective interactions between the SM Higgs and gluons 
were presented in~\cite{Page:2013xla}.  Here, we have derived them for the 
pseudoscalar case, listed in Appendix~\ref{app:eftnlo}, and coded by hand in a UFO model named \texttt{HC\_NLO\_X0}. The resulting model is publicly available online in the {\sc FeynRules} repository~\cite{FR-HC:Online}.

\subsection{Simulation parameters}\label{sec:param}

We generate events for the LHC with centre-of-mass
(CM) energies $\sqrt{s}=8$ and $13$~TeV, and we set the $X_0$ resonance
mass to $m_{X_0}=125$~GeV.  We take the heavy-top-quark limit for GF, while we 
set the top-quark mass to $m_{t}=173$~GeV in $t\bar tX_0$ production. 

Parton distribution functions (PDFs) are evaluated by using the
NNPDF2.3 (LO/NLO) parametrisation~\cite{Ball:2012cx} through the LHAPDF
interface~\cite{Whalley:2005nh}.  
For NLO predictions, the PDF uncertainty is computed together with the
uncertainty in the strong coupling constant $\alpha_s(m_Z)$ as described
in~\cite{Demartin:2010er}.
We assume the strong coupling constant to be  distributed as a gaussian
around the value  
\begin{align}
 \alpha_s^{(\rm{NLO})}(m_Z) = 0.1190 \pm 0.0012~(68\,\perc~\mathrm{C.L.})\,,
\label{eq:asnlo}
\end{align}
where the confidence interval is taken accordingly to 
the PDF4-LHC recommendation~\cite{Alekhin:2011sk,Botje:2011sn}.
At the present time there is no PDF set that allows the correct
assessment of
the PDF+$\alpha_s$ uncertainty at LO. 
Therefore, for LO predictions we compute the sole PDF uncertainty, with
the strong coupling at the $m_Z$ scale fixed to $\alpha_s^{(\rm{LO})}(m_Z) = 0.130$~\cite{Ball:2011uy,Martin:2009iq}.

Central values $\mu_0$ for the renormalisation and factorisation scales
$\mu_{R,F} $ are set to 
\begin{align}
 \mu_0^{(\rm{GF})} = H_T/2
\label{eq:mu0GF} 
\end{align}
for $X_0$(+jets) production in the GF channel,
\begin{align}
 \mu_0^{(\rm{VBF})} = m_W
\label{eq:mu0VBF} 
\end{align}
for $X_0jj$ production in the VBF channel, and
\begin{align}
 \mu_0^{(t\bar tH)} = \sqrt[3]{m_T(t)\,m_T(\bar t)\,m_T(X_0)}
\label{eq:mu0Htt}
\end{align}
for $t\bar tX_0$ production, where $m_T\equiv\sqrt{m^2+p_T^2}$ is the
transverse mass of a particle, and $H_T$ is the sum of the transverse
masses of the particles in the final state.    
Uncertainties coming from missing higher orders are estimated 
varying $\mu_R$ and $\mu_F$, independently, by a factor 2 around $\mu_0$, 
\begin{align}
 1/2<\mu_{R,F}/\mu_{\rm 0}<2\,.
\end{align}

We note here that scale and PDF uncertainties are evaluated
automatically  at no extra computing cost via a reweighting technique~\cite{Frederix:2011ss}.
In addition, such information is available on an event-by-event basis
and therefore uncertainty bands can be plotted for any observables of
interest.   We define the total theoretical uncertainty of an observable
as the linear sum of two terms: the PDF+$\alpha_s$ uncertainty on the one hand, and the overall 
scale dependence on the other.

For parton showering and hadronisation we employ 
{\sc HERWIG6}\cite{Corcella:2000bw}. We recall that matching and merging
to {\sc HERWIG++}\cite{Bahr:2008pv}, 
{\sc Pythia6}~\cite{Sjostrand:2006za} (virtuality ordered, or $p_T$ 
ordered for processes with no final-state radiation) and
{\sc Pythia8}~\cite{Sjostrand:2007gs} are also available.
Jets are reconstructed employing the 
anti-$k_T$  algorithm~\cite{Cacciari:2008gp} as implemented in 
{\sc FastJet}~\cite{Cacciari:2011ma}, with distance parameter $R=0.4$
(both for jets in $H+$jets production and for $b$-tagged jets
coming from top decays in $t\bar tH$ production) and
\begin{align}
 p_T(j)>30~{\rm GeV}\,,\quad |\eta(j)|<4.5\,.
\label{eq:mincutsHj}
\end{align}
%

\section{Gluon fusion production with jets}\label{sec:hjj}

\begin{table*}
\center
\begin{tabular}{r|lllll}
 \hline\\[-3mm]
  scenario \hspace{4.2em}
  & $\sigma_{\rm LO}$~(pb) 
  & $\sigma_{\rm NLO}$~(pb) & $K$ 
  & $\sigma_{\rm NLO+PS}$~(pb) & $R$\\[0.5mm]
 \hline\\[-2.5mm]
             $0^+$
             & 4.002(4)~${}^{+46.8}_{-29.6}$\,{\scriptsize $\pm3.3\perc$}
             & 5.484(7)~${}^{+17.0}_{-16.8}$\,{\scriptsize $\pm1.2\perc$}
	     & 1.37
             & 4.618~${}^{+21.8}_{-18.8}$\,{\scriptsize $\pm1.2\perc$} 
             & 0.84
 \\[1mm]
 LHC 8 TeV \enskip \quad $0^-$
             & 9.009(9)~${}^{+46.8}_{-29.6}$\,{\scriptsize $\pm3.3\perc$}
             & 12.34(2)~${}^{+17.1}_{-16.8}$\,{\scriptsize $\pm1.2\perc$} 
	     & 1.37
             & 10.38~${}^{+21.7}_{-18.8}$\,{\scriptsize $\pm1.2\perc$}
             & 0.84
 \\[1mm]
             $0^\pm$
             & 6.511(6)~${}^{+46.8}_{-29.6}$\,{\scriptsize $\pm3.3\perc$} 
             & 8.860(14)~${}^{+16.9}_{-16.8}$\,{\scriptsize $\pm1.2\perc$} 
	     & 1.36
             & 7.474~${}^{+21.7}_{-18.8}$\,{\scriptsize $\pm1.2\perc$} 
	     & 0.84
 \\[1mm]
 \hline\\[-2.5mm]
             $0^+$
             & 10.67(1)~${}^{+41.7}_{-27.5}$\,{\scriptsize $\pm2.6\perc$} 
             & 14.09(2)~${}^{+16.2}_{-14.9}$\,{\scriptsize $\pm1.1\perc$} 
	     & 1.32
             & 12.08~${}^{+19.8}_{-16.7}$\,{\scriptsize $\pm1.0\perc$} 
             & 0.86
 \\[1mm]
 LHC 13 TeV \quad $0^-$
             & 24.01(2)~${}^{+41.7}_{-27.5}$\,{\scriptsize $\pm2.6\perc$} 
             & 31.67(6)~${}^{+16.2}_{-14.9}$\,{\scriptsize $\pm1.1\perc$} 
	     & 1.32
             & 27.14~${}^{+20.3}_{-16.4}$\,{\scriptsize $\pm1.0\perc$}  
	     & 0.86
 \\[1mm]
             $0^\pm$
             & 17.36(2)~${}^{+41.7}_{-27.5}$\,{\scriptsize $\pm2.6\perc$} 
             & 22.83(3)~${}^{+16.2}_{-14.9}$\,{\scriptsize $\pm1.1\perc$} 
	     & 1.32
             & 19.59~${}^{+19.5}_{-16.6}$\,{\scriptsize $\pm1.0\perc$}  
             & 0.86
 \\[1mm]
 \hline
\end{tabular}
\caption{LO and NLO cross sections and corresponding $K$ factors for 
 $pp\to X_0+1$ jet (GF channel) at the 8- and 13-TeV LHC, for the three
 scenarios defined in table~\ref{tab:GFscenarios}.
 The integration error in the last digit(s) (in parentheses), and the
 fractional scale (left) and PDF(+$\alpha_s$) (right) uncertainties are
 also reported. 
 In addition to fixed-order results, the PS-matched NLO cross sections and the ratios $R\equiv\sigma_{\rm NLO+PS}/\sigma_{\rm NLO}$ are also shown.} 
\label{tab:xsecHj}
\end{table*}

\begin{table*}
\center
\begin{tabular}{r|lllll}
 \hline\\[-3mm]
  scenario \hspace{4.2em}
  & $\sigma_{\rm LO}$~(pb) 
  & $\sigma_{\rm NLO}$~(pb) & $K$
  & $\sigma_{\rm NLO+PS}$~(pb) & $R$ \\[0.5mm]
 \hline\\[-2.5mm]
             $0^+$
             & 1.351(1)~${}^{+67.1}_{-36.8}$\,{\scriptsize $\pm4.3\perc$} 
             & 1.702(6)~${}^{+19.7}_{-20.8}$\,{\scriptsize $\pm1.7\perc$} 
	     & 1.26
             & 1.276~${}^{+29.4}_{-23.9}$\,{\scriptsize $\pm1.7\perc$}
	     & 0.75
 \\[1mm]
 LHC 8 TeV \enskip \quad $0^-$
             & 2.951(3)~${}^{+67.2}_{-36.8}$\,{\scriptsize $\pm4.4\perc$} 
             & 3.660(15)~${}^{+19.1}_{-20.6}$\,{\scriptsize $\pm1.7\perc$} 
	     & 1.24
             & 2.755~${}^{+29.8}_{-24.1}$\,{\scriptsize $\pm1.8\perc$}
	     & 0.75
 \\[1mm]
             $0^\pm$
             & 2.142(2)~${}^{+67.1}_{-36.8}$\,{\scriptsize $\pm4.4\perc$} 
             & 2.687(10)~${}^{+19.6}_{-20.8}$\,{\scriptsize $\pm1.7\perc$} 
	     & 1.25
             & 2.022~${}^{+29.7}_{-24.1}$\,{\scriptsize $\pm1.8\perc$}  
             & 0.75
 \\[1mm]
 \hline\\[-2.5mm]
             $0^+$
             & 4.265(4)~${}^{+61.5}_{-34.9}$\,{\scriptsize $\pm3.3\perc$} 
             & 5.092(23)~${}^{+15.4}_{-17.9}$\,{\scriptsize $\pm1.2\perc$} 
	     & 1.19
             & 4.025~${}^{+23.9}_{-21.3}$\,{\scriptsize $\pm1.2\perc$}
	     & 0.79
 \\[1mm]
 LHC 13 TeV \quad $0^-$
             & 9.304(9)~${}^{+61.6}_{-34.9}$\,{\scriptsize $\pm3.4\perc$} 
             & 11.29(4)~${}^{+16.0}_{-18.2}$\,{\scriptsize $\pm1.2\perc$} 
	     & 1.21
             & 8.701~${}^{+24.6}_{-21.6}$\,{\scriptsize $\pm1.3\perc$} 
             & 0.77
 \\[1mm]
             $0^\pm$
             & 6.775(6)~${}^{+61.5}_{-34.9}$\,{\scriptsize $\pm3.3\perc$} 
             & 8.055(35)~${}^{+15.8}_{-18.2}$\,{\scriptsize $\pm1.2\perc$} 
	     & 1.19
             & 6.414~${}^{+24.4}_{-21.5}$\,{\scriptsize $\pm1.2\perc$}  
             & 0.80 
 \\[1mm]
 \hline
\end{tabular}
\caption{Same as table~\ref{tab:xsecHj}, but for $pp\to X_0+2$ jets (GF).}
\label{tab:xsecHjj}
\end{table*}

In {\sc MadGraph5\_aMC@NLO} the code and the events for $X_0$ plus two jets in the
GF channel can be automatically generated by issuing the following
commands (note the \texttt{/ t} syntax to forbid diagrams containing
top loops):   
\begin{verbatim}
 > import model HC_NLO_X0-heft
 > generate p p > x0 j j / t [QCD]
 > output
 > launch
\end{verbatim}
where the {\tt -heft} suffix in the model name refers to the corresponding model restriction.
As a result, all the amplitudes featuring the Higgs--gluon effective
vertices in the heavy-top limit are generated, including corrections up
to NLO in QCD. Analogous commands can be issued to generate events 
for $X_0$ plus zero and one jet at NLO. 
The NLO computation for $Hjjj$ in GF has been recently achieved interfacing
{\sc Sherpa} with {\sc GoSam}~\cite{Cullen:2013saa}.
We note that {\sc MadGraph5\_aMC@NLO} provides the FxFx 
merging~\cite{Frederix:2012ps} to combine several NLO+PS samples, which
differ by final-state multiplicities, and NLO merged Higgs production
in GF was discussed in refs.~\cite{Frederix:2012ps,Alwall:2014hca}.  

As mentioned above, since our interest is geared towards QCD effects in
production distributions, we do not include Higgs decays in our
studies. We stress, however, that decays (as predicted in the HC model) can be
efficiently included at the partonic event level by employing {\sc MadSpin}~\cite{Artoisenet:2012st}, 
before passing the short-distance events to a parton-shower program.

\subsection{Total rates}

We start by showing results for total cross sections for Higgs plus jet
production in GF, not only for $H+2$ jets but also for $H+1$ jet as a
reference.   
We remark here that as GF is the dominant Higgs production mechanism, 
enormous theoretical efforts to achieve more precise
computation have been made over the last decade and we refer to the
reports by the LHC Higgs Cross Section Working Group 
(LHCHXSWG)~\cite{Dittmaier:2011ti,Dittmaier:2012vm,Heinemeyer:2013tqa} for 
more details. We note that a first calculation of Higgs plus one jet at NNLO ($gg$
only and in the EFT) has been reported in ref.~\cite{Boughezal:2013uia}. 

Table~\ref{tab:xsecHj} collects the LO and NLO total
cross sections and the corresponding $K$ factors for $pp\to X_0 j$ at
the 8- and 13-TeV LHC, together with uncertainties, for the three
scenarios defined in table~\ref{tab:GFscenarios}.
The acceptance cuts in eq.~\eqref{eq:mincutsHj} are imposed. 

Requiring the presence of jets in the final state entails  imposing cuts
at the generation level as well as after event generation in the case of
NLO+PS simulation. 
We have checked that the cuts at the generation level
were loose enough not to affect the NLO+PS rates and distributions. 
Since reconstructed jets after parton shower and hadronisation can be
different from the fixed-order parton jets, the parton-shower matched
cross section can be different from the fixed-order prediction.

The figure in parentheses is the integration error in the last digit(s).
The first uncertainty (in percent) corresponds to the envelope obtained
by varying independently the renormalisation and factorisation scales by
a factor 2 around the central value, $\mu_0=H_T/2$.
The second one corresponds to the PDF($+\alpha_s$) uncertainty.
As mentioned in sect.~\ref{sec:param}, the full PDF+$\alpha_s$
uncertainty is available only at NLO.
It is known that PDF and $\alpha_s$ uncertainties are comparable for GF  
at NLO~\cite{Demartin:2010er}, thus we take them both  into
account. 
We can see that both the scale dependence and PDF+$\alpha_s$
uncertainties are independent of the scenarios, and as expected they are
significantly reduced going from LO to NLO.
It is also evident that the residual scale dependence is the dominant source
of uncertainty in the GF channel.  
We also note that $\sigma(0^-)$ is larger than $\sigma(0^+)$ by a factor of 2.25 at LO
(and to a good approximation even at NLO) due to the different coupling normalisation 
(see table~\ref{tab:gXaa}), and $\sigma(0^{\pm})$ is equal to the average of
$\sigma(0^+)$ and $\sigma(0^-)$. This means that there are no interference effects 
in the total rates for this process.

In addition to the fixed-order results, we also show the NLO cross
sections matched with parton shower ($\sigma_{\rm NLO+PS}$) in the
table. The ratios to the fixed-order NLO rates, $R\equiv\sigma_{\rm NLO+PS}/\sigma_{\rm NLO}$ 
are shown in the last column. These ratios are smaller than one, as extra radiation generated by the
parton shower tends to spread the energy of the original hard partons,
affecting the spectrum of the jets and leading to more events which fail
to pass the cuts.  The survival rate after shower slightly increases as increasing the
collision energy.   We note that the ratios can slightly depend on the parton-shower
programs~\cite{Frixione:2013mta}, and these differences shall be
considered as matching systematics. Another effect of the parton shower that we observe 
is a slightly increased scale dependence in the results, compared to the corresponding
fixed-order predictions.  

Table~\ref{tab:xsecHjj} presents results for $pp\to X_0+2$ jets.
The features of the cross sections and uncertainties are qualitatively
similar to the 1-jet case in table~\ref{tab:xsecHj}, while
rather different quantitatively. 
As one increases the number of extra jets, the cross section becomes
smaller (as expected, yet mildly) and the $K$ factors are also reduced.
On the other hand, the scale dependence increases, especially in the LO
results, as more powers of $\alpha_s$ enter the matrix elements. 
Once again, the $K$ factors do not depend on the scenarios.
We note that the LO ratio $\sigma(0^-)/\sigma(0^+)$ slightly deviates
from 2.25 because of the missing $ggggA$ vertex as well as the different
helicity structure of the
amplitudes~\cite{Kauffman:1998yg}.  

\subsection{Distributions}

In the previous section we have seen that
if the strength of the scalar and pseudoscalar couplings in the Higgs--top-quark interaction 
is similar ({\it i.e.}  $\kappa_{\sss Htt}g_{\sss Htt}\sim\kappa_{\sss Att}g_{\sss Att}$ in 
eq.~\eqref{eq:0ff}), the total Higgs production rate in GF is sensitive 
to the CP mixing of the Higgs boson.  We now turn to distributions,
where GF jet--jet
correlations are known tools to determine the Higgs CP 
properties~\cite{Klamke:2007cu,Hagiwara:2009wt,Andersen:2010zx,Campanario:2010mi,Englert:2012ct,Englert:2012xt,D
olan:2014upa}.
In the following, all the distributions will be shown for the 13-TeV LHC. 
For these studies, we require the presence of at least two reconstructed
jets in the final states. The jets are ordered by the transverse momenta.

\begin{figure}
 \center 
 \includegraphics[width=1.0\columnwidth]{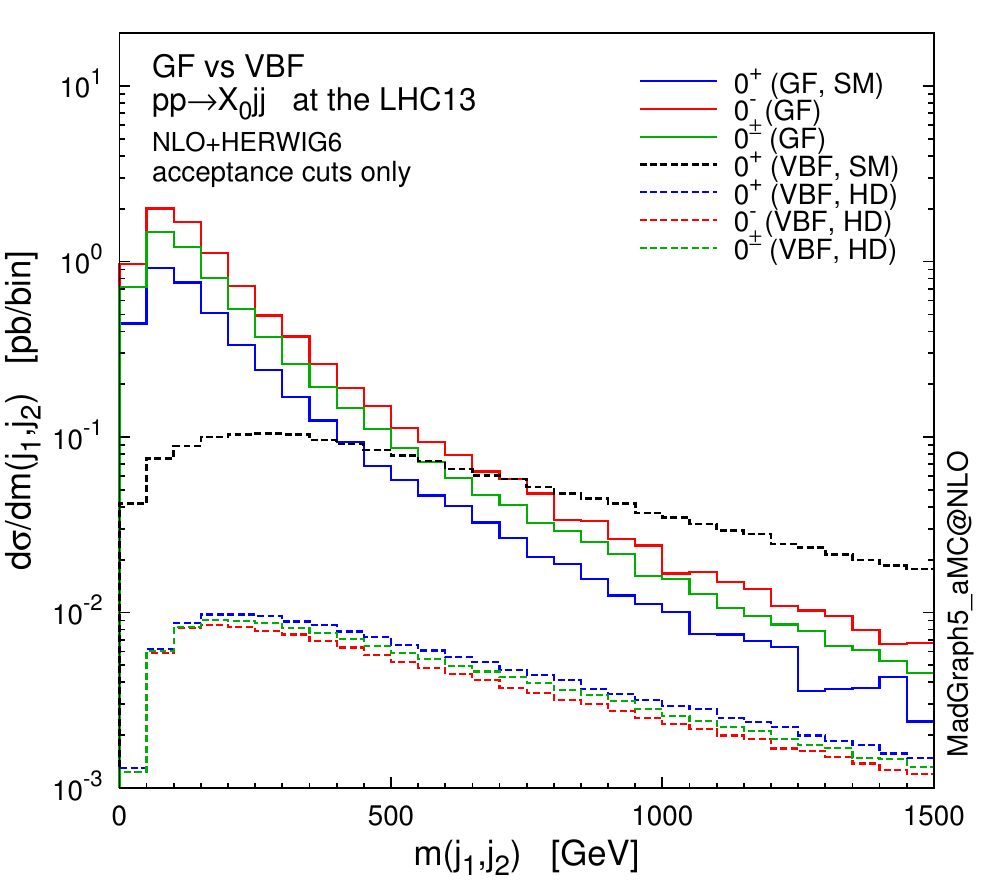}
\caption{Distribution of the invariant mass of the two leading jets in 
 $pp\to X_0jj$ through GF (solid lines) and VBF (dashed) at the 13-TeV LHC.
 The different hypotheses are defined in tables~\ref{tab:GFscenarios}
 and \ref{tab:VBFscenarios}.} 
\label{fig:mjj}
\end{figure}

We start by showing the invariant mass distribution $m_{jj}$ of the two
leading jets in fig.~\ref{fig:mjj}, where GF and VBF are compared for the various scenarios defined
in tables~\ref{tab:GFscenarios} and \ref{tab:VBFscenarios}. 
For the VBF HD scenarios we fix the cutoff scale to $\Lambda=1$~TeV.
GF is dominant in the small di-jet mass region, while VBF tends to
produce a jet pair with higher invariant mass~\cite{DelDuca:2001fn}.   
This is because, for $Hjj$ production in GF, the $gg$ and $qg$
initial states are dominant, and hence the Higgs can be radiated off the
initial or final gluon legs, leading to more central jets with the
acceptance cuts only.
For the VBF process, on the other hand, the Higgs boson is produced through
the $t,u$-channel weak-boson fusion, leading to forward hard jets. 
Based on this fact, we usually require a minimum $m_{jj}$ as a VBF cut in
order to minimise the GF contribution to extract the VBF information.
The shapes of the $m_{jj}$ spectra are similar among the different CP
scenarios within the same channel.
This means that, apart from the difference between GF and VBF, the
invariant mass cut acts in a similar way on every CP scenario in a
given channel; more details for the VBF case can be found  
in ref.~\cite{Maltoni:2013sma}. 

Looking at the subprocesses contributing to $X_0+2$ jets is instructive. 
The $qq \to X_0 qq$ subprocess features VBF-like $t$-channel gluon exchange diagram and 
is not affected by the $m_{jj}$ cut, since the jets tend to be produced in the
forward region, similarly to the weak-boson case~\cite{Englert:2012xt}.
Moreover, even for the $gg$ and $qg$ induced subprocesses, the
$t$-channel contribution becomes relatively important by imposing the 
invariant mass cut.  In other words, the VBF cut maximises the contributions featuring gluons in the $t$-channel, 
which are the most sensitive to the CP properties of $X_0$ also in the GF case~\cite{Hagiwara:2009wt}.
To illustrate how the CP-sensitive observables change with the VBF cut,
on top of the acceptance cuts, we impose an invariant mass cut as
\begin{align}
 m(j_1,j_2)>250,\ 500~{\rm GeV} \,.
\label{eq:mjjcut} 
\end{align}
We do not require a minimum rapidity separation, although this is
another common VBF cut, since $\Delta\eta_{jj}$ itself is an observable
sensitive to the CP properties of $X_0$~\cite{Englert:2012xt,Djouadi:2013yb}.

\begin{figure*}
 \center 
 \includegraphics[width=0.325\textwidth]{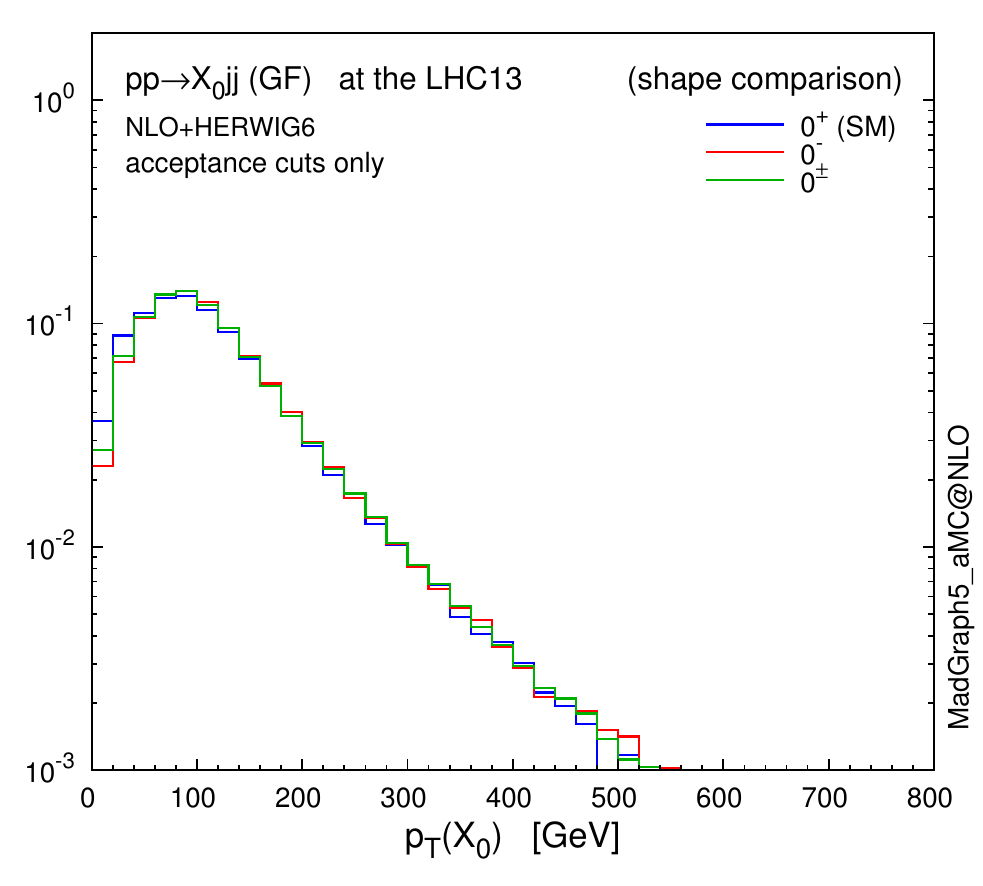}
 \includegraphics[width=0.325\textwidth]{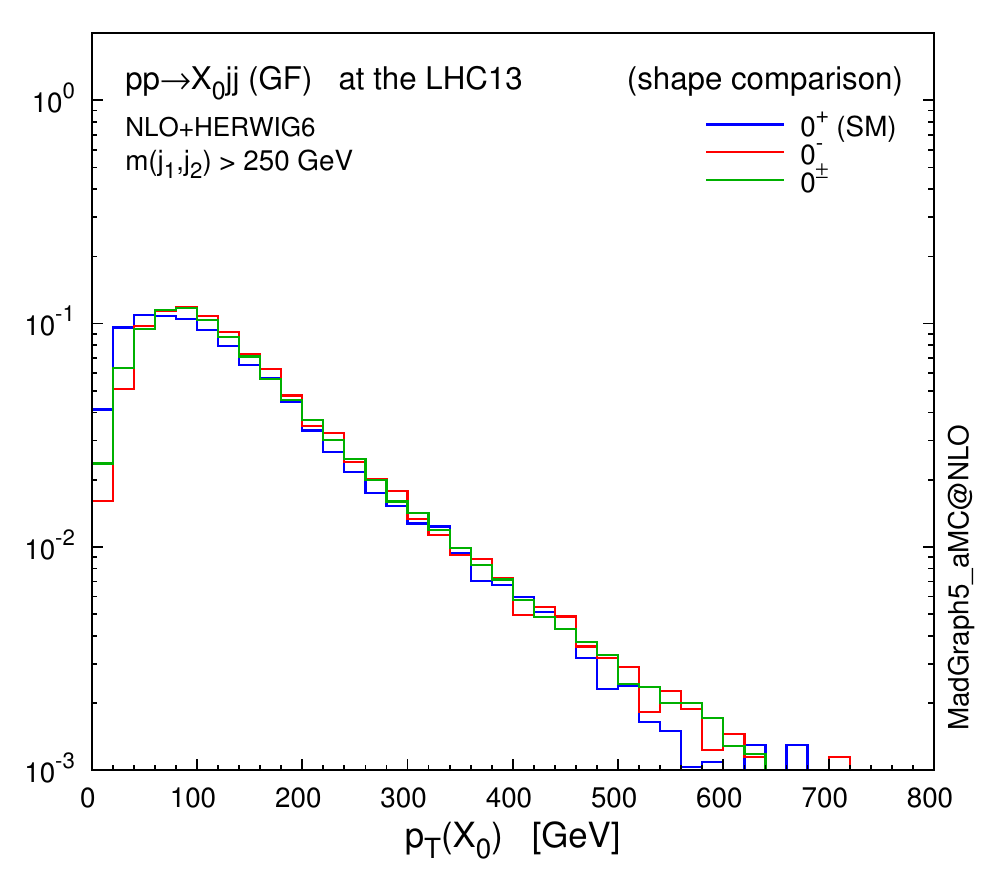}
 \includegraphics[width=0.325\textwidth]{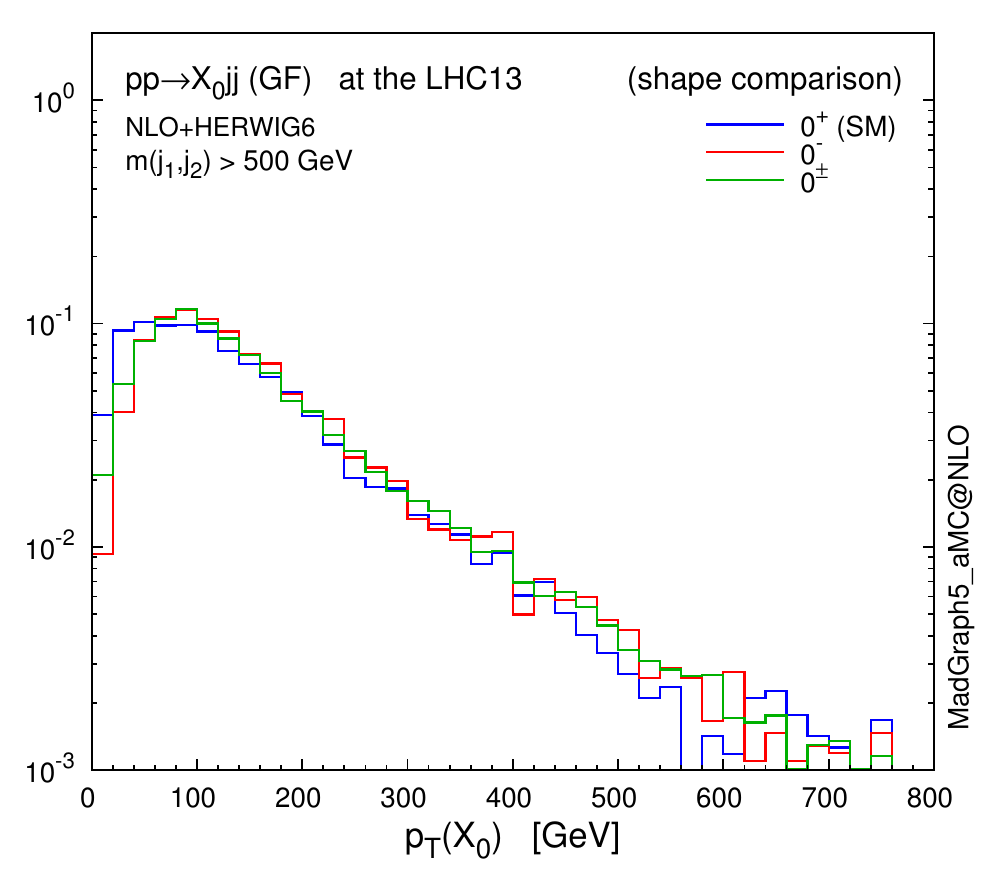}
 \includegraphics[width=0.325\textwidth]{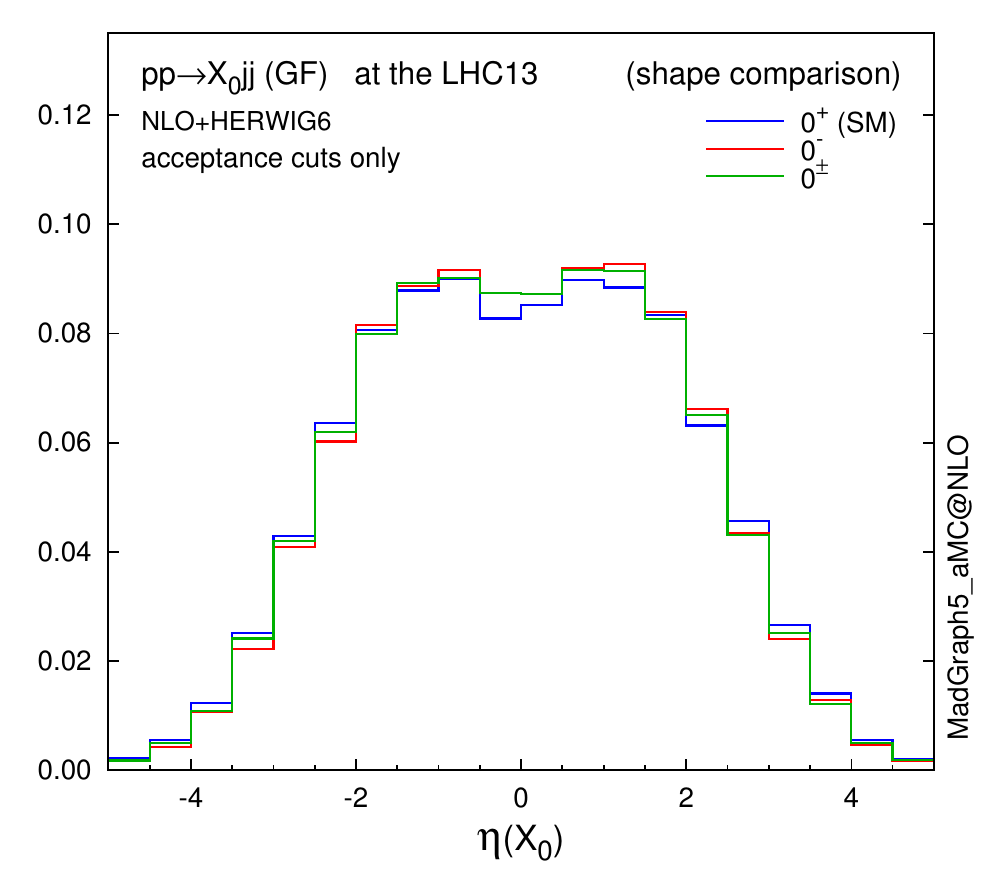}
 \includegraphics[width=0.325\textwidth]{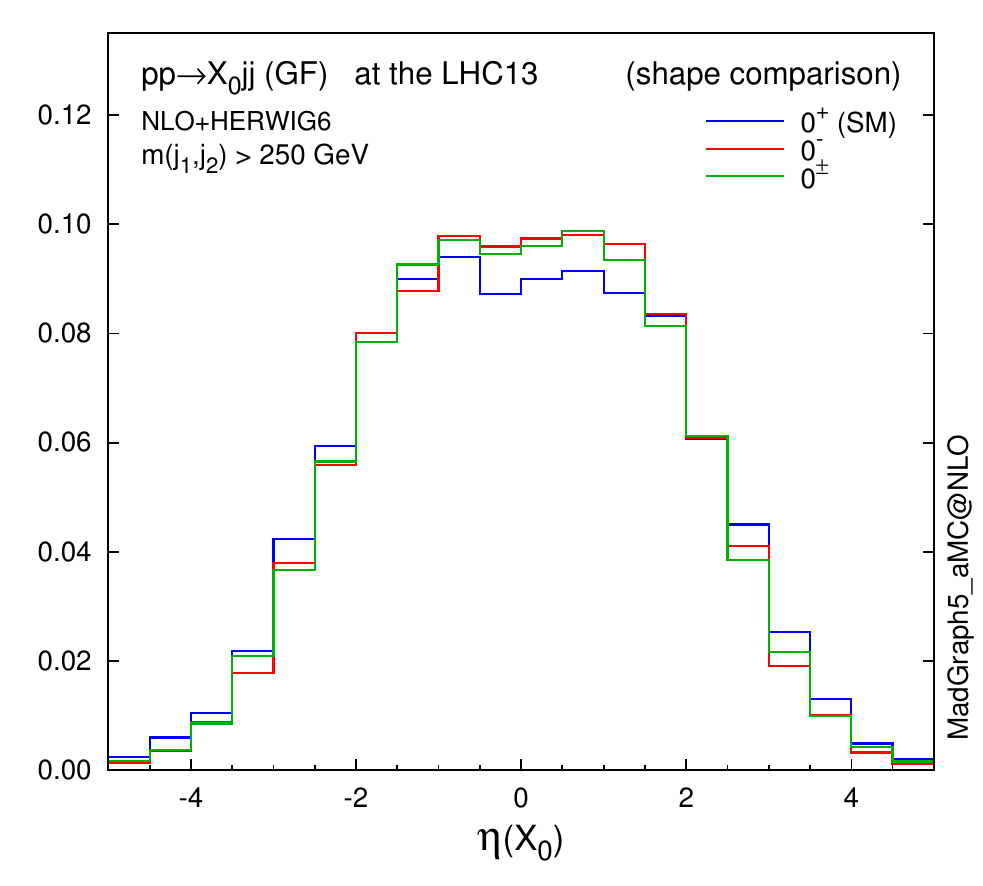}
 \includegraphics[width=0.325\textwidth]{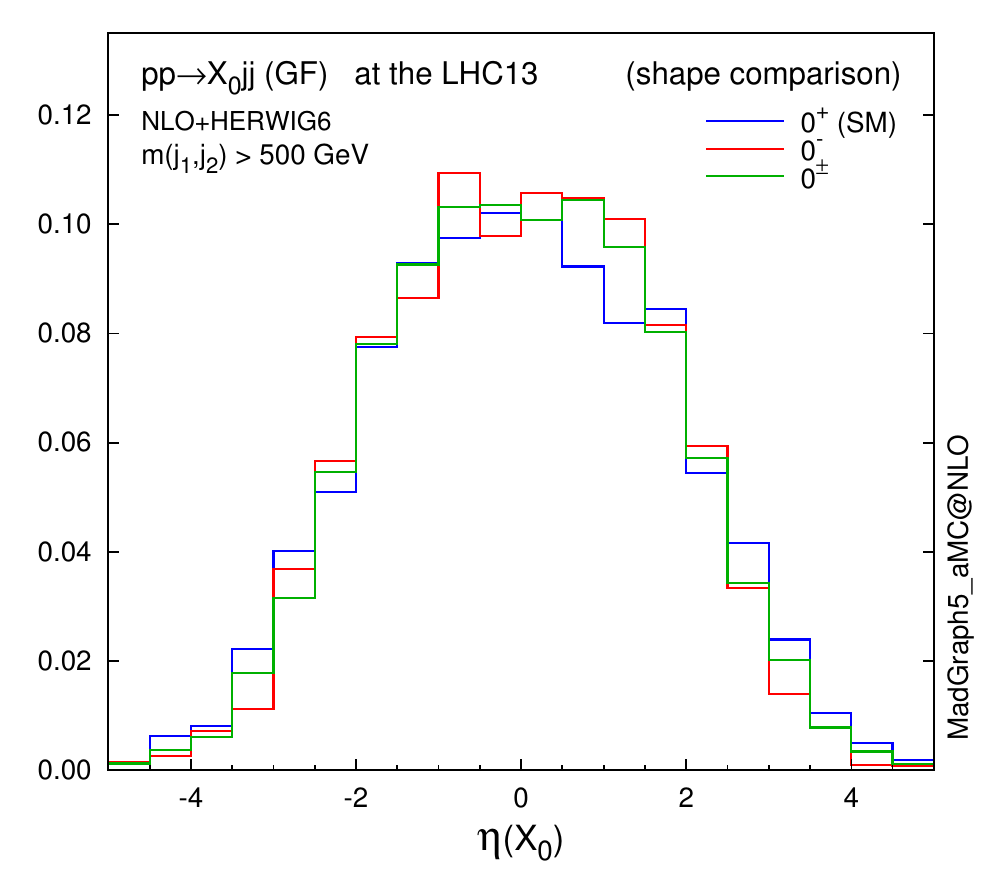}
 \caption{Normalized distributions (shape comparison) in $p_T$ and $\eta$ of the resonance
 $X_0$, with the acceptance cuts for jets (left), plus
 $m(j_1,j_2)>250$~GeV (centre) and $500$~GeV (right).
 The three spin-0 hypotheses are defined in table~\ref{tab:GFscenarios}.}  
\label{fig:gf_x}
\end{figure*} 

\begin{figure*}
 \center 
 \includegraphics[width=0.325\textwidth]{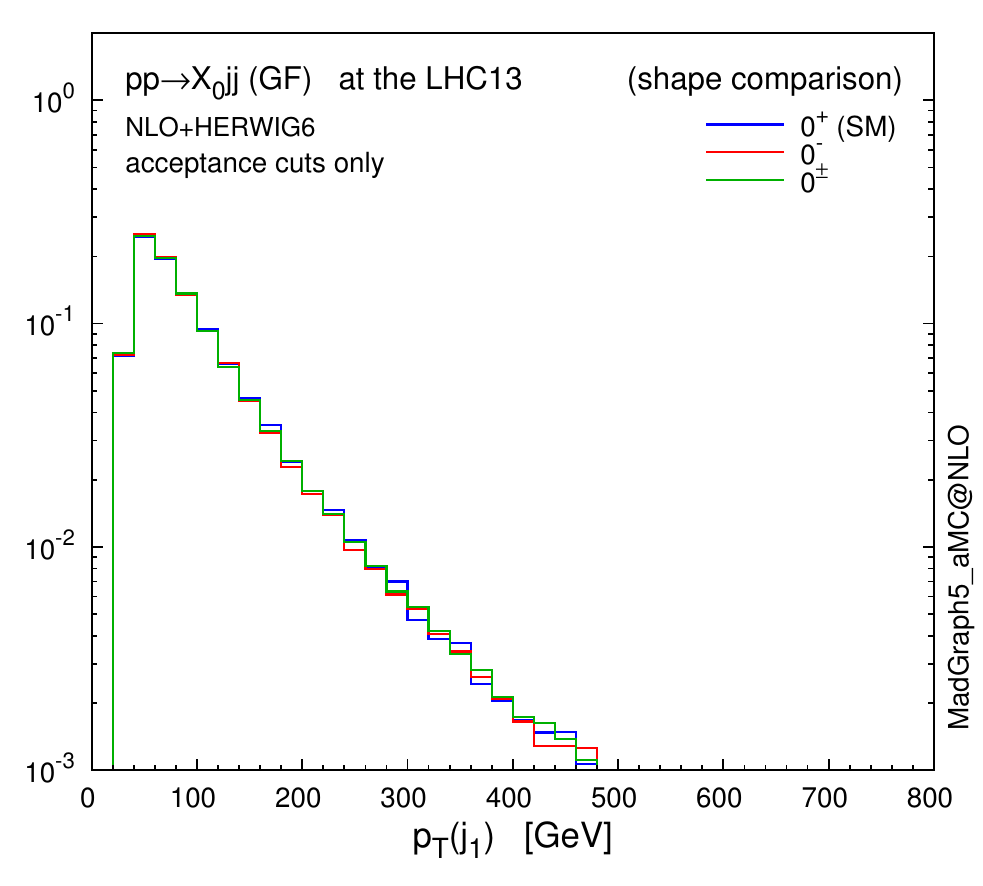}
 \includegraphics[width=0.325\textwidth]{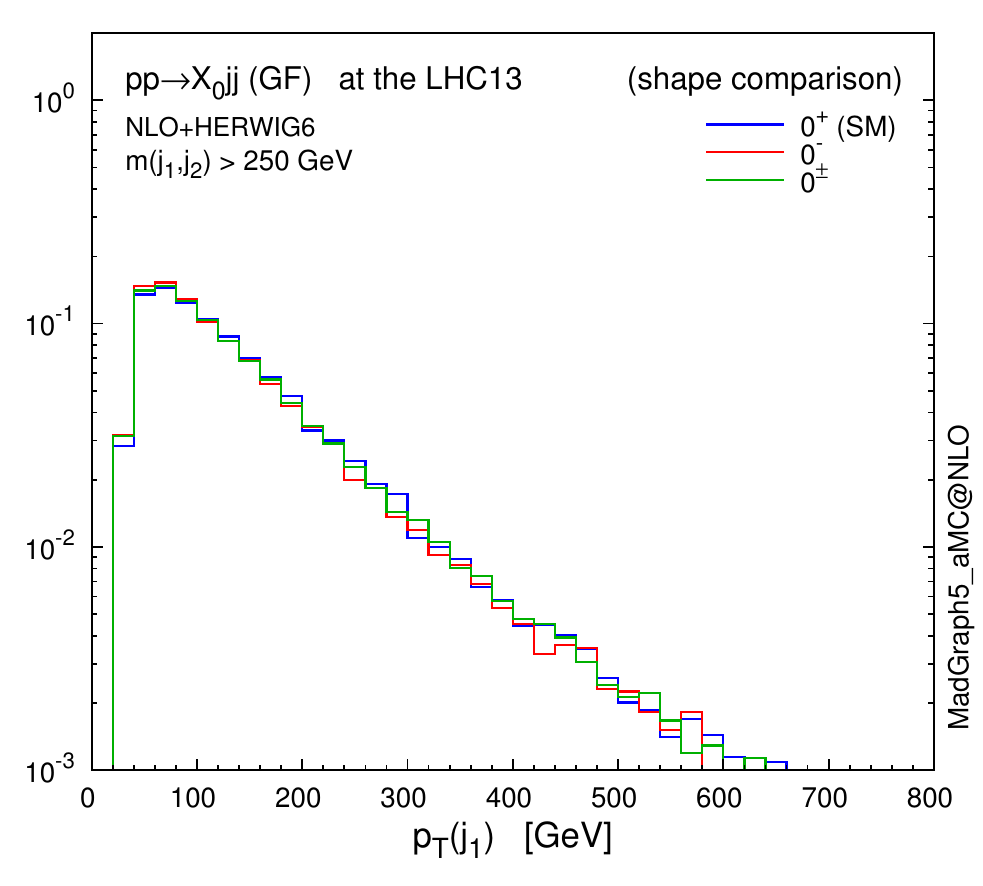}
 \includegraphics[width=0.325\textwidth]{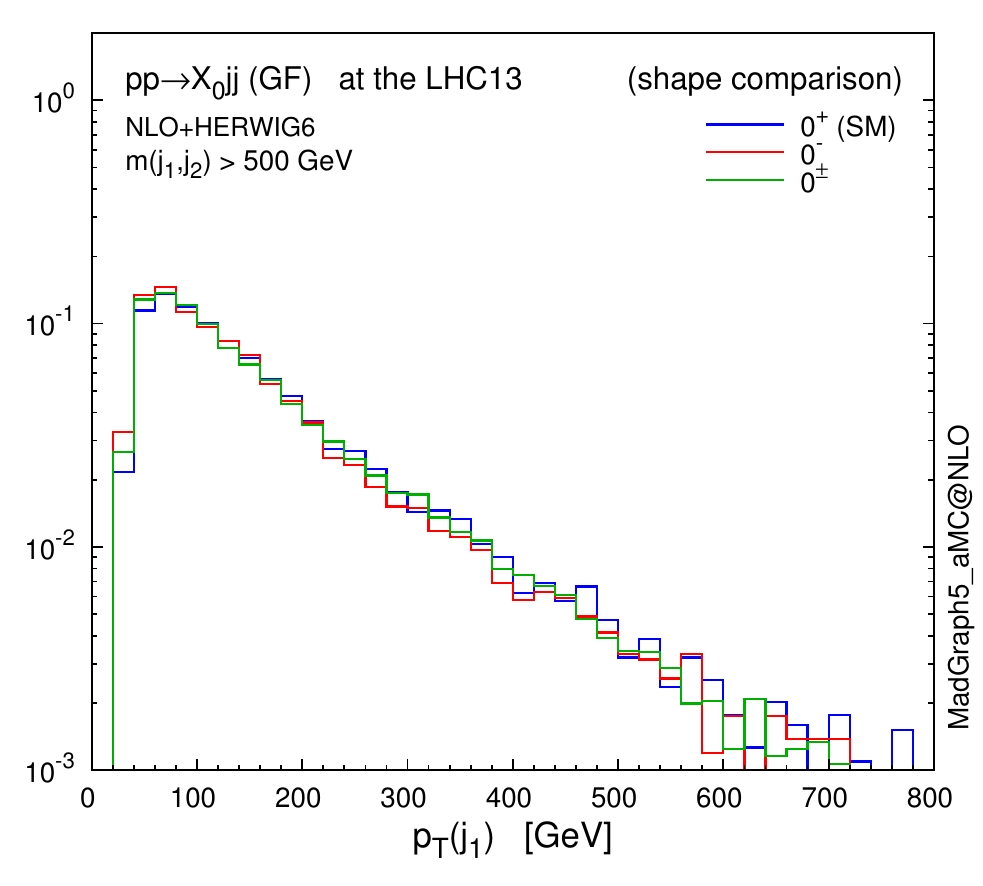}
 \includegraphics[width=0.325\textwidth]{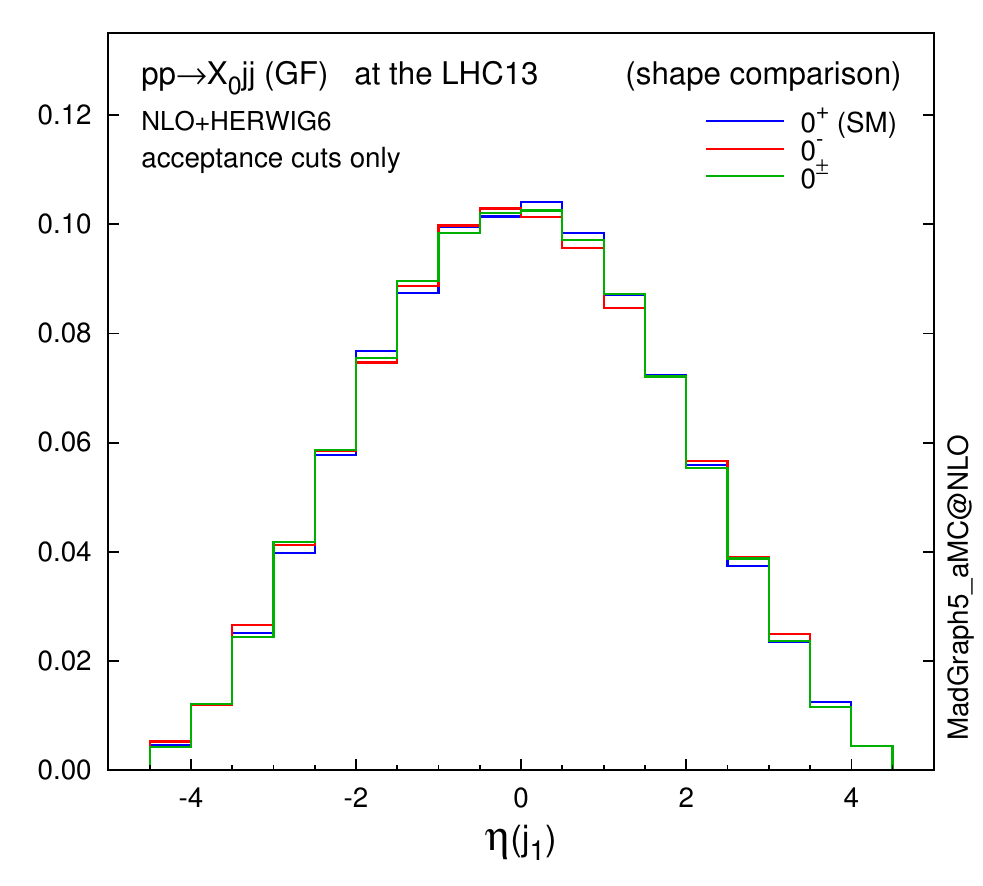}
 \includegraphics[width=0.325\textwidth]{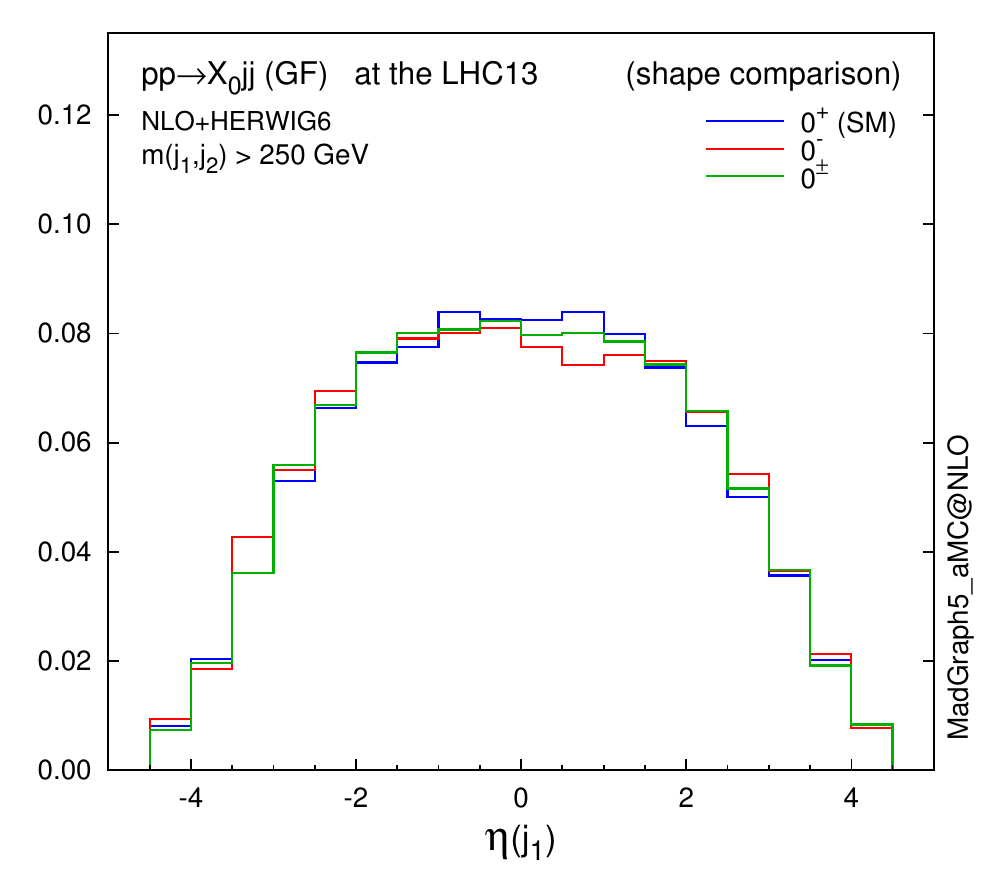}
 \includegraphics[width=0.325\textwidth]{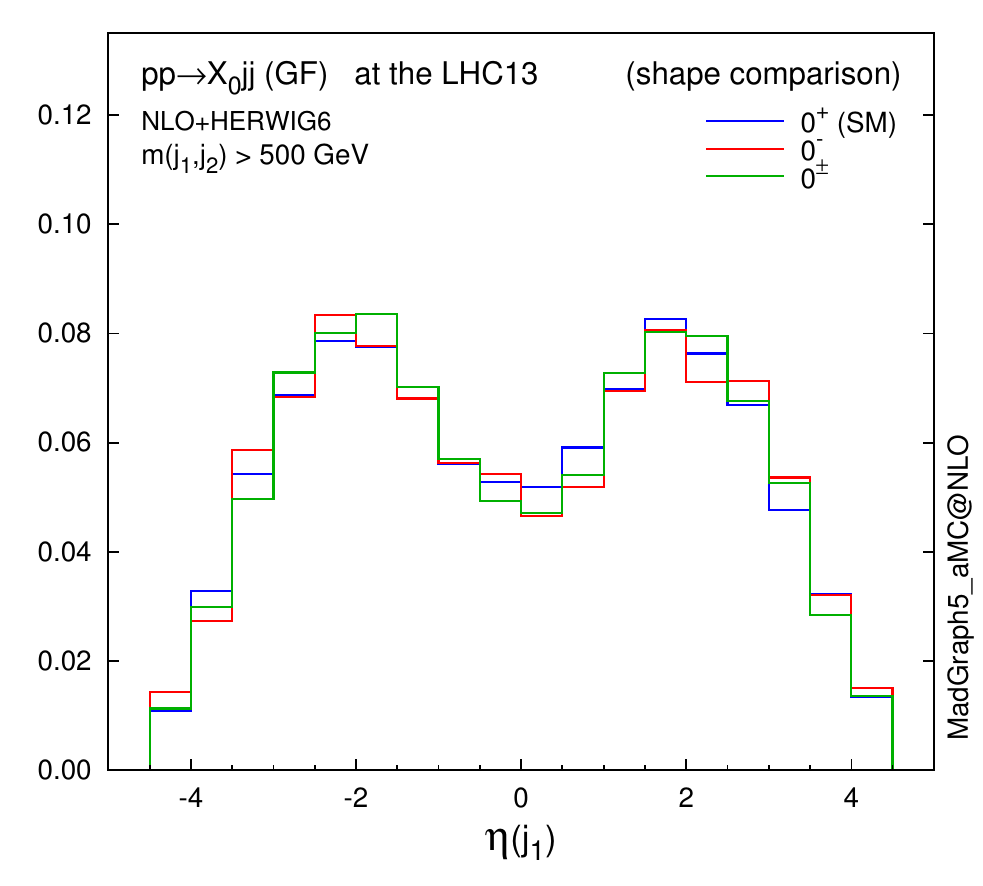}
 \caption{Same as fig.~\ref{fig:gf_x}, but for the leading jet.}
\label{fig:gf_j1}
\end{figure*} 

\begin{figure*}
 \center 
 \includegraphics[width=0.325\textwidth]{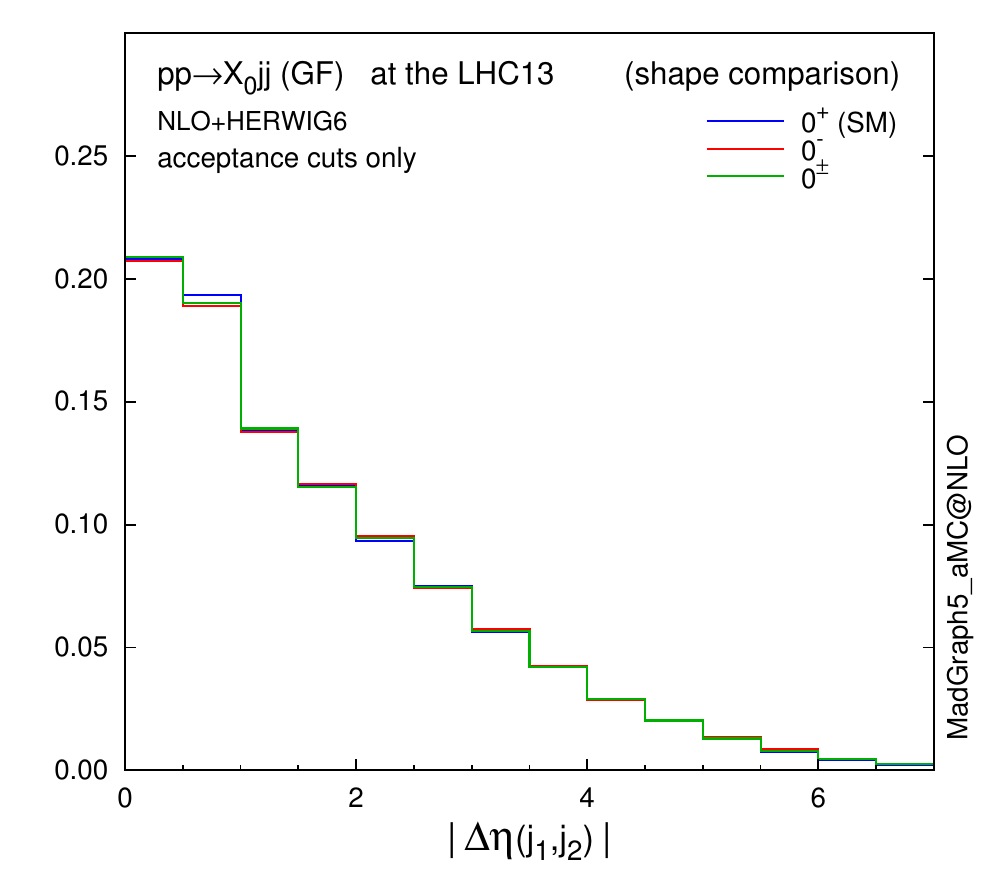}
 \includegraphics[width=0.325\textwidth]{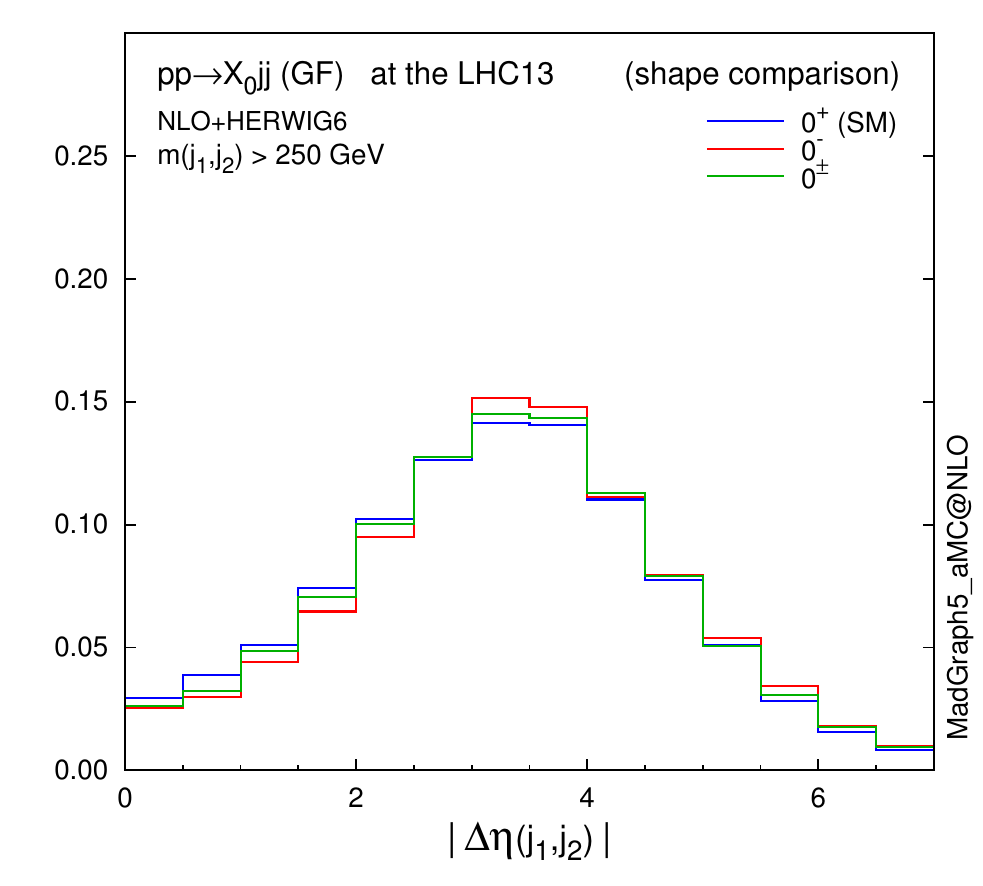}
 \includegraphics[width=0.325\textwidth]{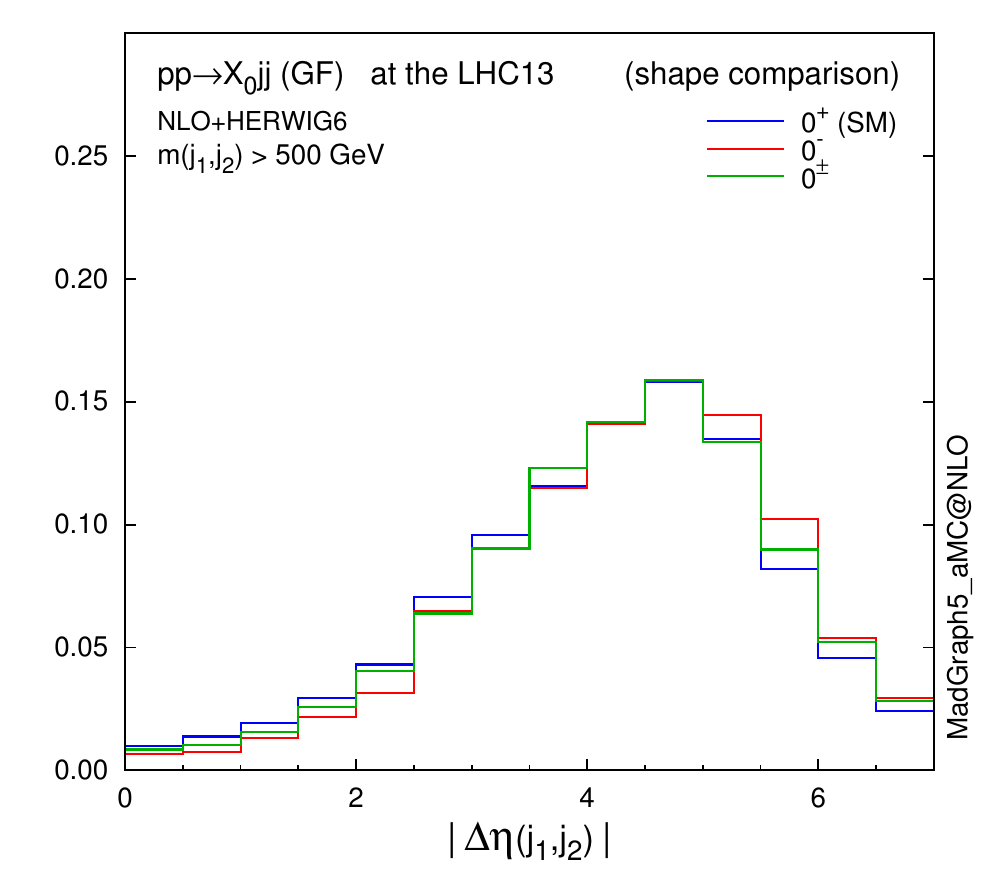}
 \includegraphics[width=0.325\textwidth]{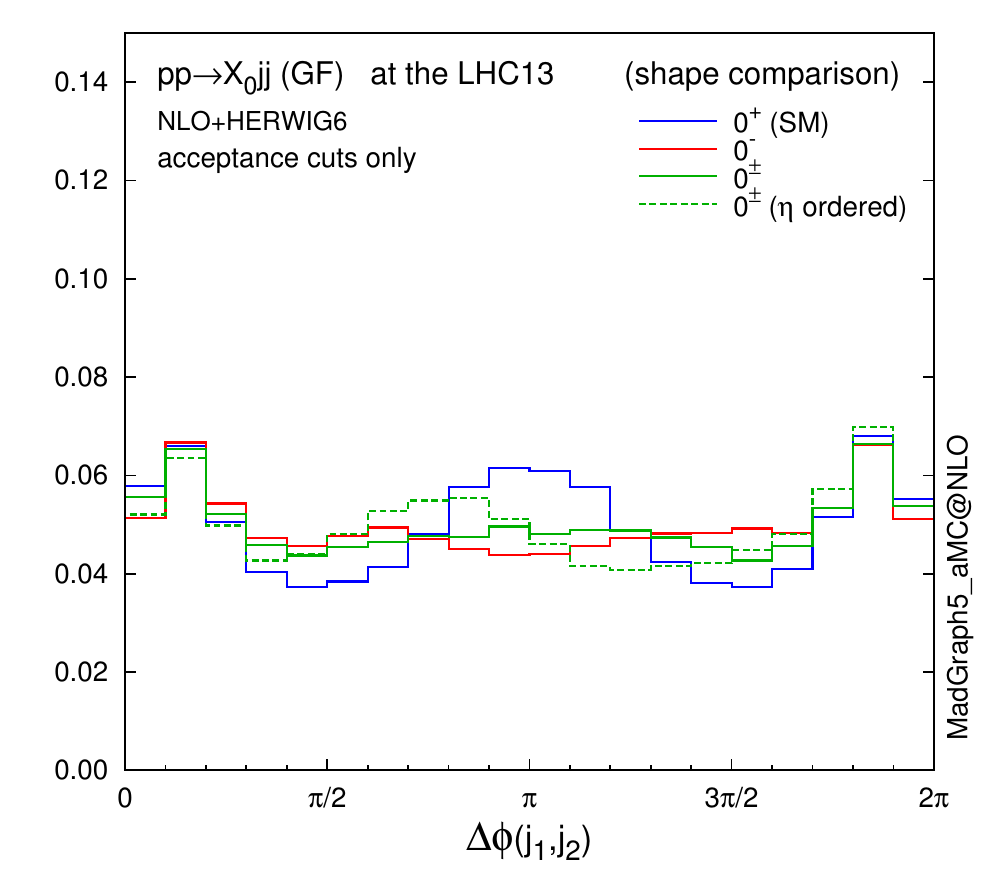}
 \includegraphics[width=0.325\textwidth]{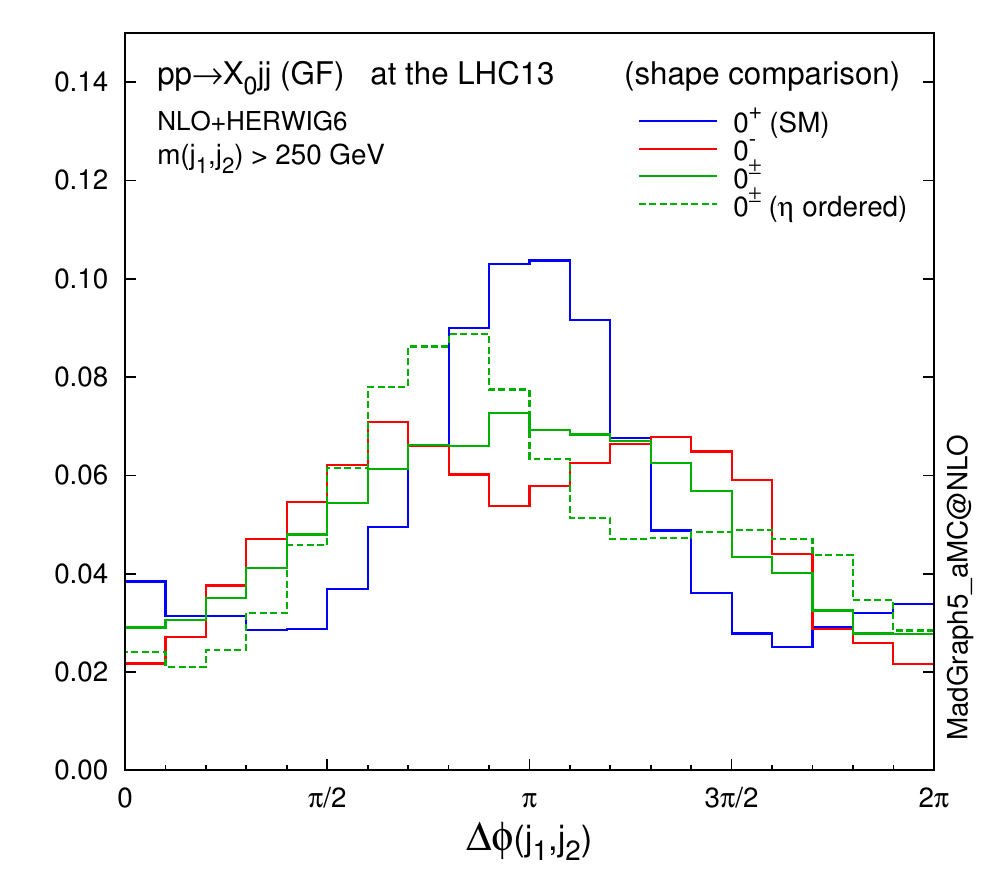}
 \includegraphics[width=0.325\textwidth]{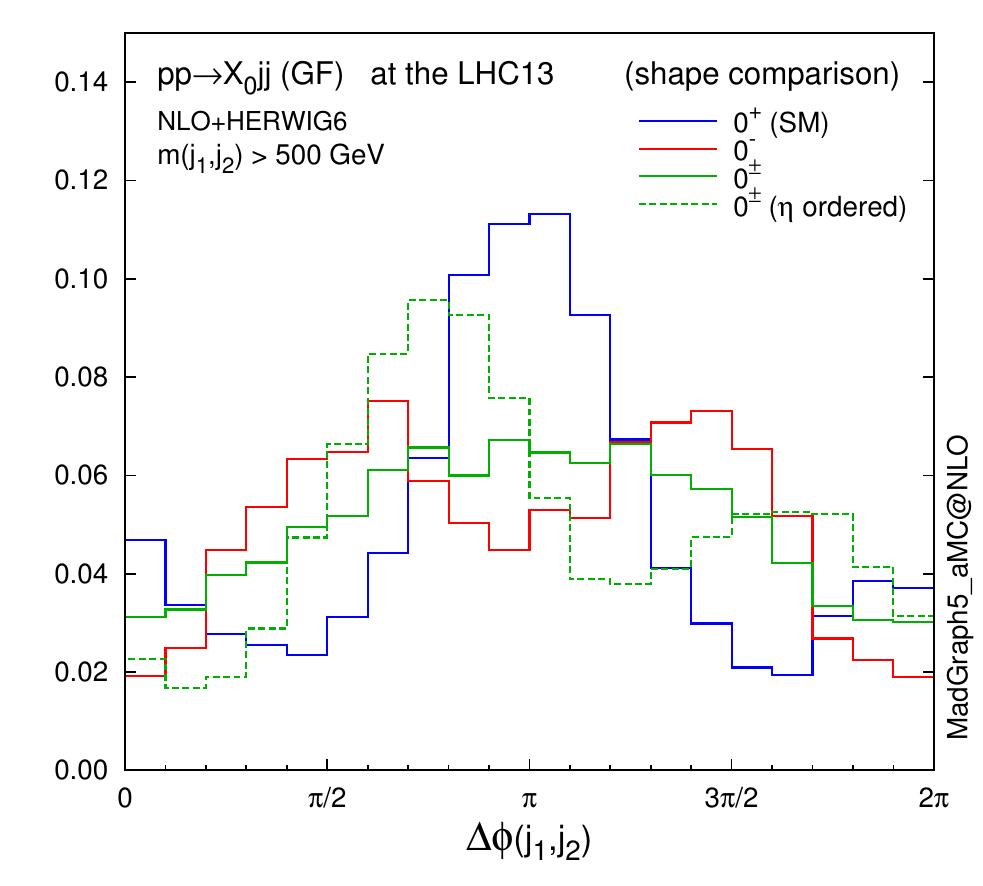}
 \caption{Same as fig.~\ref{fig:gf_x}, but for $\Delta\eta$ and $\Delta\phi$
 distributions between the two tagging jets.
 For $\Delta\phi$, the distribution with the additional $\eta$ jet
 ordering is also shown by a dashed line for the $0^{\pm}$ case.} 
\label{fig:gf_j1j2}
\end{figure*} 

Figs.~\ref{fig:gf_x} and \ref{fig:gf_j1} show the effect of
the invariant mass cut on the $p_T$ and $\eta$ distributions for
the resonance $X_0$ and the leading jet. Imposing larger $m_{jj}$ cuts leads to harder transverse momenta 
for both the $X_0$ and the jets; as a result, the $X_0$ is produced more centrally, 
while the jets are shifted to the forward regions and the difference in the low $p_T(X_0)$ region 
between the various CP scenarios becomes more pronounced. This behaviour is due to the
fact that at larger $m_{jj}$   topologies featuring the emission of the
Higgs boson by
a gluon exchanged in the $t$-channel are enhanced, similarly to the typical VBF topology.

A possible concern is to what extent the  EFT approach is valid. 
In fact the heavy-top-quark effective lagrangian in eq.~\eqref{L_loop} is  a
good approximation for single light-Higgs production. 
The EFT closely reproduces the $m_{jj}$ spectrum of the loop
computation even in the very high invariant mass region~\cite{DelDuca:2001fn}. 
However, this approximation fails when the transverse momenta of the
jets are larger than the top mass~\cite{DelDuca:2001eu},
overestimating the exact prediction for the $p_T(j_{1})>m_t$
region. Since the events are generated predominantly in the small $p_T(j_1)$
region, we choose not to apply any rejection of events with large $p_T$
in the following analyses.

\begin{figure*}
 \center 
 \includegraphics[width=0.32\textwidth]{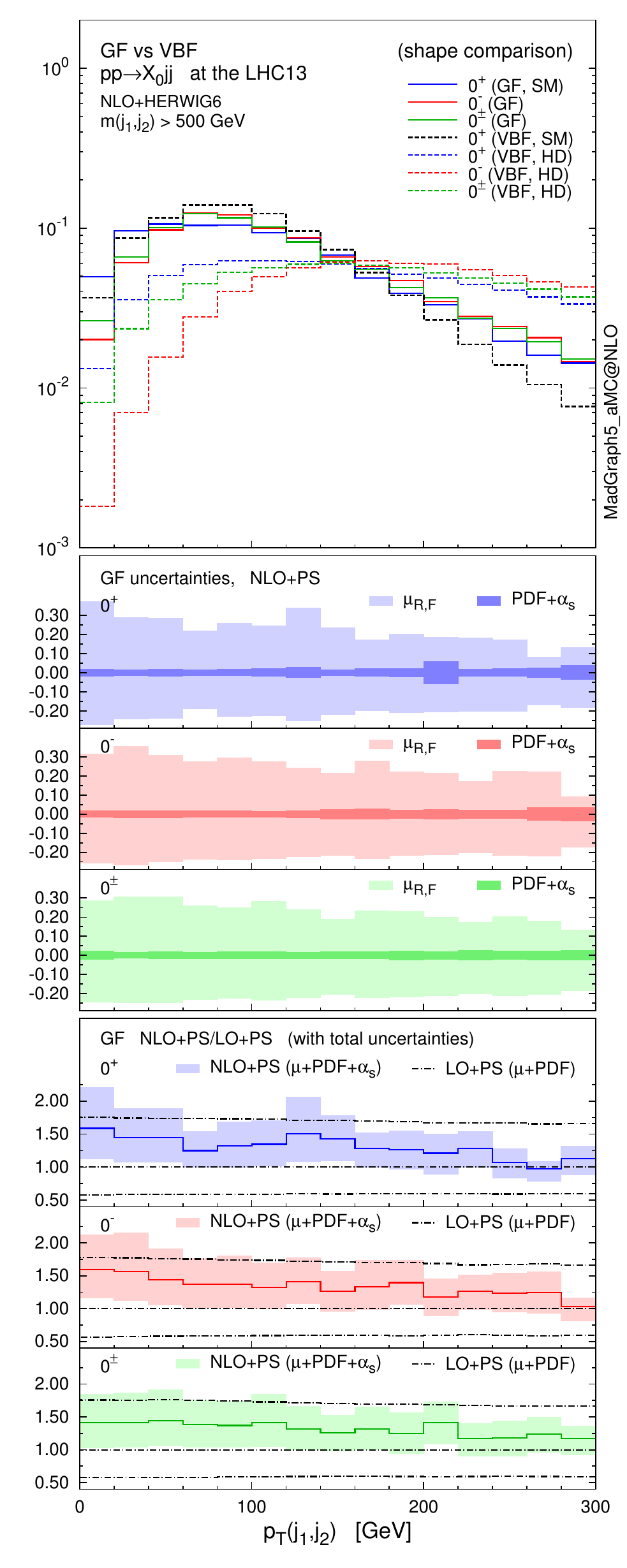}
 \includegraphics[width=0.32\textwidth]{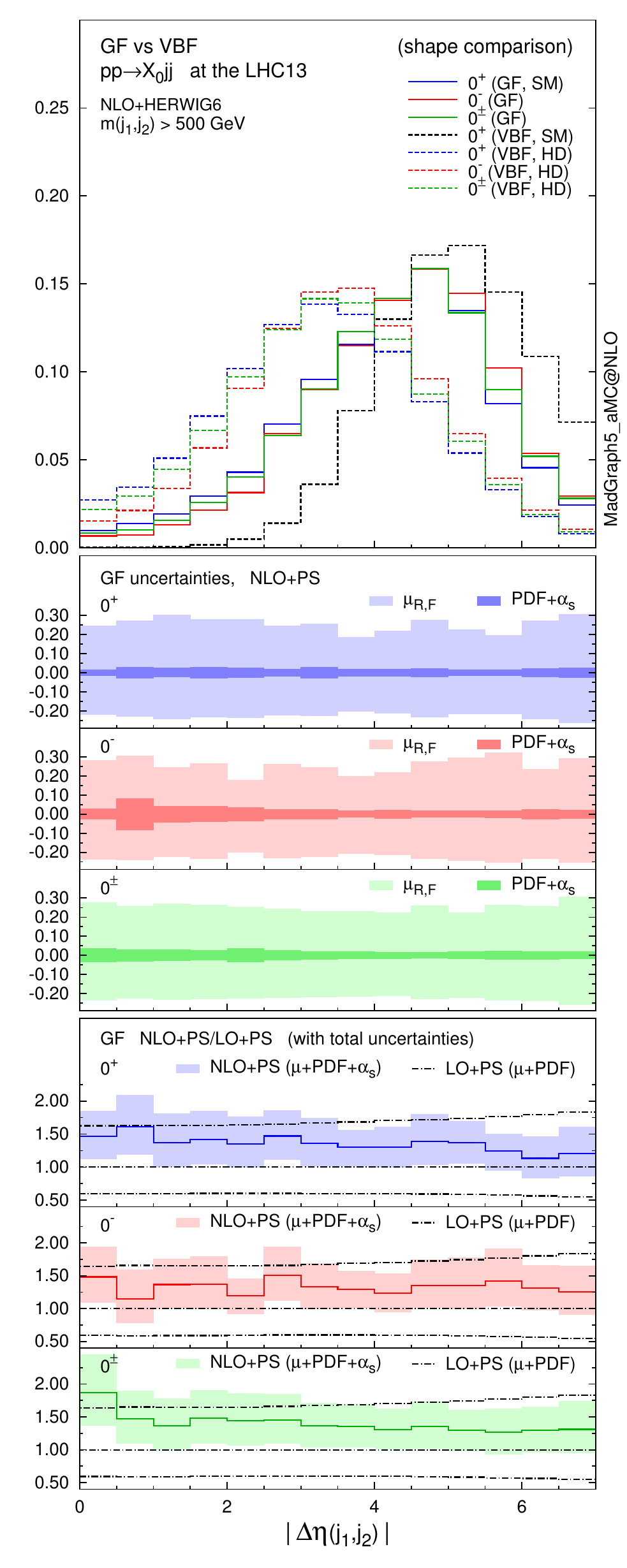}
 \includegraphics[width=0.32\textwidth]{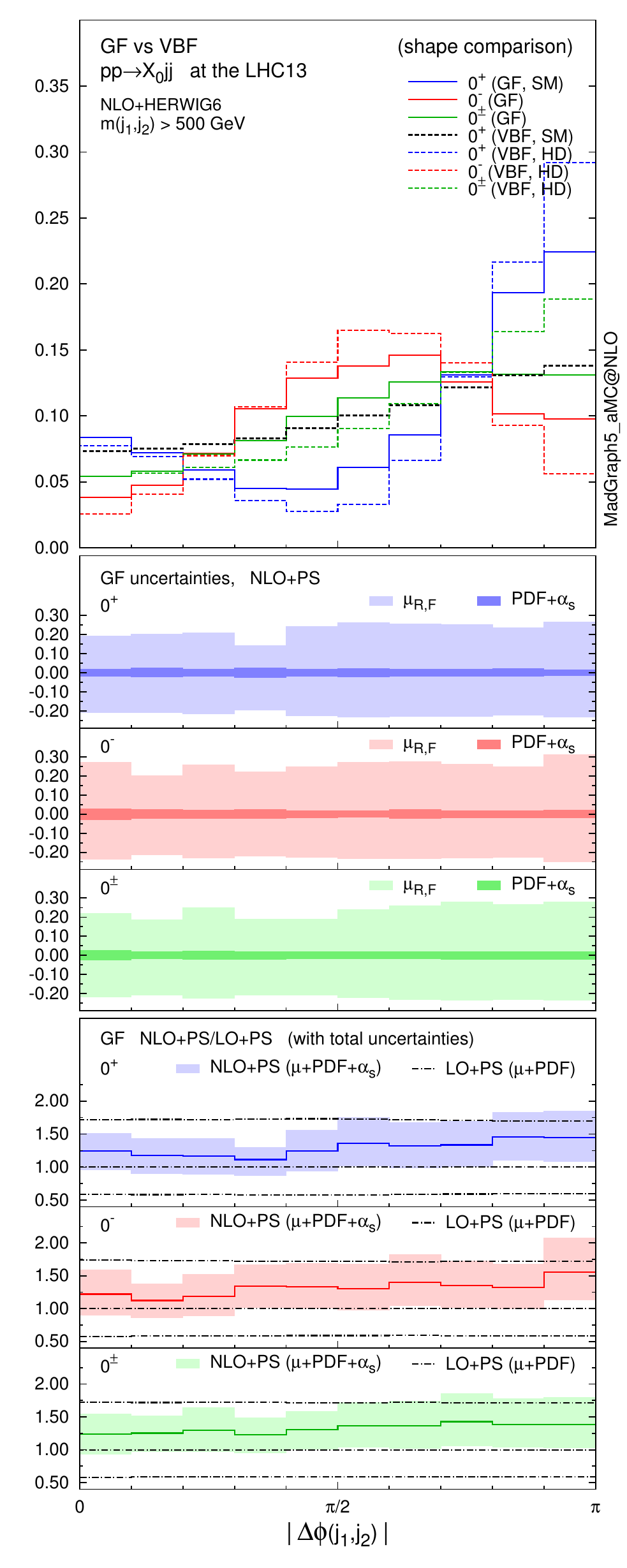}
 \caption{Normalized distributions (shape comparison) of the $p_T$ of the di-jet system
 (left), the rapidity (centre) and azimuthal (right) difference between
 the two tagging jets for $pp\to X_0jj$ in GF (solid lines) and VBF
 (dashed) at the 13-TeV LHC, where the acceptance cuts plus the
 $m_{jj}>500$~GeV cut are applied. 
 For each GF scenario, the middle panels show the scale and
 PDF$+\alpha_s$ uncertainties, while the bottom ones give the ratio of
 NLO+PS to LO+PS results with the total uncertainties.} 
\label{fig:Hjj}
\end{figure*}

The most sensitive observables for the CP nature of the Higgs
boson couplings to the top quark in this channel are di-jet correlations, shown in
fig.~\ref{fig:gf_j1j2}.  As already seen in fig.~\ref{fig:gf_j1}, the invariant mass cut
effectively suppresses the central jet activity, although the different CP scenarios in the rapidity separation 
$\Delta\eta_{jj}\equiv\eta(j_1)-\eta(j_2)$ can be hardly distinguished.
On the other hand, the azimuthal angle between the two jets 
$\Delta\phi_{jj}\equiv\phi(j_1)-\phi(j_2)$ is well known to be very sensitive 
to the CP mixing and our results confirm that this is indeed the case also at NLO
(for a LO vs. NLO comparison see fig.~\ref{fig:Hjj} in the following).  

A remarkable observation is that the $\Delta\phi_{jj}$ distribution
is more sensitive to the CP-mixed state, when the two leading jets
(ordered by $p_T$) are reordered in pseudorapidity%
\footnote{The definition is analogous to eq.~(4.1) in
 ref.~\cite{Klamke:2007cu}.}
(dashed green), compared to the distribution with the usual $p_T$ jet
ordering (solid green).
This is especially true for the maximal mixing scenario, which we
consider here, since with just $p_T$ ordering the $\pi/4$ phase shift,
generated by quantum interference between the CP-even and -odd
components, is cancelled between $+\Delta\phi_{jj}$ and
$-\Delta\phi_{jj}$~\cite{Klamke:2007cu}. Indeed, the distribution for $0^{\pm}$ without $\eta$ ordering is just the weighted average of the $0^+$ and $0^-$ cases.

\begin{table}
\center
\begin{tabular}{r|rrr}
 \hline\\[-3mm]
  $m_{jj}>$ \hspace{4.2em} & 250 GeV & 500 GeV & 500 GeV  \\[0.3mm]
   &  &  &  + jet veto \\[0.3mm]
 \hline\\[-3mm]
             $0^+$       
             & 22.7 \perc
             & 6.6 \perc
             & 5.0 \perc \\[0.3mm]
 LHC 8 TeV \enskip\quad  $0^-$
             & 21.4 \perc
             & 5.7 \perc
             & 4.5 \perc \\[0.3mm]
             $0^\pm$
             & 21.5 \perc
             & 6.2 \perc
             & 4.6 \perc \\[0.3mm]
 \hline\\[-3mm]
             $0^+$       
             & 26.3 \perc
             & 9.0 \perc
             & 6.4 \perc \\[0.3mm]
 LHC 13 TeV \quad $0^-$
             & 25.4 \perc
             & 8.6 \perc
             & 6.2 \perc \\[0.3mm]
             $0^\pm$
             & 25.6 \perc
             & 8.6 \perc
             & 6.2 \perc \\[0.3mm]
 \hline
\end{tabular}
\caption{Selection efficiencies with different di-jet invariant mass cuts for $pp\to X_0jj$.
 A jet veto defined in~\eqref{jetveto} is also applied in the last column.}
\label{tab:jetveto}
\end{table}

The NLO computation allows also to investigate the effect of applying a veto on additional jets in the event,
a  procedure that is  known to suppress the central QCD activities and to enhance the VBF
signal~\cite{Cox:2010ug,Gerwick:2011tm}. We implement it by vetoing
 events containing a third jet laying in pseudorapidity between the forward and backward tagging jets,
\begin{align}
 \min \big\{ \eta(j_1), \eta(j_2) \big\} < \eta(j_{\rm veto}) < 
 \max \big\{ \eta(j_1), \eta(j_2) \big\} \,.
\label{jetveto}
\end{align}
Table~\ref{tab:jetveto} collects the selection efficiencies on the NLO+PS
samples after $m_{jj}>250$~GeV and 500~GeV cuts, and $m_{jj}>500$~GeV
plus the central jet veto, with respect to the acceptance cuts only.  
As already seen in fig.~\ref{fig:mjj}, the efficiencies are very similar
among the different scenarios.
The additional jet veto could be useful to enhance the sensitivity to
CP-mixing, especially for the 13-TeV run. 
Indeed we have checked that the size of the variation  in the
$\Delta\phi_{jj}$ distribution in fig.~\ref{fig:gf_j1j2} becomes
slightly larger.  The related jet binning uncertainties have been discussed in detail 
in  ref.~\cite{Gangal:2013nxa}.

Finally, we discuss the theoretical uncertainties for the CP-sensitive
observables.  Figure~\ref{fig:Hjj} displays, from left to right, normalised
distributions of the $p_T$ of the di-jet system (which is equivalent to
$p_T(X_0)$ only at LO), the pseudorapidity and the azimuthal difference
between the two tagging jets for $pp\to X_0+2$ jets in GF (solid lines) at
the 13-TeV LHC.
The acceptance cuts and the invariant mass cut 
$m_{jj}>500$~GeV are imposed.
The middle panels show the scale and PDF$+\alpha_s$ uncertainties for
each scenario, while the bottom ones give the ratio of NLO+PS to LO+PS
results with the total theoretical uncertainties. 
The total uncertainty is defined as the linear sum of the scale and
PDF$+\alpha_s$ uncertainties. 
The scale uncertainty is dominant, as observed in
table~\ref{tab:xsecHjj}, and both the scale and PDF$+\alpha_s$
uncertainties change very mildly over the phase space.
In all cases NLO corrections are relevant and cannot be described by an 
overall $K$ factor.

In the main panel, we also draw a comparison with the VBF contributions
(dashed lines).
The $p_T(j_1,j_2)$ and $\Delta\eta(j_1,j_2)$ distributions show that in
the SM VBF case the Higgs boson is produced more centrally
while the tagging jets are more  forward than in GF production.
For the three HD VBF cases, conversely, the jets are more central.
We recall that the type of operators are the same both for the GF and
the HD VBF, {\it i.e.} the dimension-five operators $X_0V_{\mu\nu}V^{\mu\nu}$
and $X_0V_{\mu\nu}\widetilde V^{\mu\nu}$. 

We track down the slight difference between GF and HD VBF in $\Delta\eta_{jj}$
to  the presence of the mass of the $t$-channel vector boson, {\it i.e.} massless
gluons vs. massive weak bosons.  On the other hand, the slightly weaker modulation for $\Delta\phi_{jj}$ 
in GF is due to the presence of the $gg$ and $qg$ initiated
contributions~\cite{Hagiwara:2009wt,Englert:2012xt}. 
We note that the interference between GF and VBF can be safely 
neglected~\cite{Andersen:2007mp,Bredenstein:2008tm}.

\section{Associated production with a top-quark pair}\label{sec:tth}

The code and events for $t\bar tX_0$ hadroproduction can be
automatically generated by issuing the following commands in {\sc MadGraph5\_aMC@NLO}:
\begin{verbatim}
 > import model HC_NLO_X0
 > generate p p > x0 t t~ [QCD]
 > output
 > launch
\end{verbatim}
The top quark decays are subsequently performed starting from the event file 
(in the Les Houches format~\cite{Alwall:2006yp}) by {\sc MadSpin}~\cite{Artoisenet:2012st} following 
a procedure~\cite{Frixione:2007zp} that keeps intact production and decay spin correlations. 

\subsection{Total rates}
 
In table~\ref{tab:xsec_tth} we show results for total cross sections 
at LO and NLO accuracy and the corresponding $K$ factors at  8 and
13-TeV LHC for the three scenarios defined in table~\ref{tab:GFscenarios}. 
The uncertainties correspond respectively to i) the integration error on the last digit(s), reported in
parentheses,  
ii) the envelope obtained by independently varying the renormalisation and
factorisation scales by a factor 2 around the central value given in eq.~\eqref{eq:mu0Htt},
and   iii) the PDF+$\alpha_s$ uncertainty (only PDF uncertainty for LO).

\begin{table*}
\center
\begin{tabular}{r|lllll}
 \hline\\[-3mm]
  scenario \hspace{4.2em}
  & $\sigma_{\rm LO}$~(fb) 
  & $\sigma_{\rm NLO}$~(fb) & $K$ 
  & $\sigma_{\rm NLO+PS}^{\rm dilep}$~(fb) & $R$\\[0.5mm]
 \hline\\[-2.5mm]
             $0^+$
             & 130.3(1)~${}^{+36.8}_{-24.6}$\,{\scriptsize $\pm5.9\perc$}
             & 134.9(2)~${}^{+3.2}_{-8.3}$\,{\scriptsize $\pm3.0\perc$}
	     & 1.04
             & 3.088~${}^{+3.1}_{-8.4}$\,{\scriptsize $\pm2.8\perc$} 
             & $ 2.29 \times 10^{-2}$
 \\[1mm]
 LHC 8 TeV \enskip \quad $0^-$
             & 44.49(4)~${}^{+42.5}_{-27.6}$\,{\scriptsize $\pm10.3\perc$}
             & 47.07(6)~${}^{+6.5}_{-11.5}$\,{\scriptsize $\pm4.9\perc$} 
	     & 1.06
             & 1.019~${}^{+5.5}_{-11.0}$\,{\scriptsize $\pm4.3\perc$}
             & $ 2.16 \times 10^{-2}$
 \\[1mm]
             $0^\pm$
             & 87.44(8)~${}^{+38.2}_{-25.4}$\,{\scriptsize $\pm6.9\perc$} 
             & 90.93(12)~${}^{+3.9}_{-9.1}$\,{\scriptsize $\pm3.4\perc$} 
	     & 1.04
             & 2.052~${}^{+3.6}_{-9.0}$\,{\scriptsize $\pm3.2\perc$} 
	     & $ 2.26 \times 10^{-2}$
 \\[1mm]
 \hline\\[-2.5mm]
             $0^+$
             & 468.6(4)~${}^{+32.8}_{-22.8}$\,{\scriptsize $\pm4.5\perc$} 
             & 525.1(7)~${}^{+5.7}_{-8.7}$\,{\scriptsize $\pm2.1\perc$} 
	     & 1.12
             & 11.52~${}^{+5.5}_{-8.7}$\,{\scriptsize $\pm2.0\perc$} 
             & $ 2.19 \times 10^{-2}$
 \\[1mm]
 LHC 13 TeV \quad $0^-$
             & 196.8(2)~${}^{+37.1}_{-25.2}$\,{\scriptsize $\pm7.5\perc$} 
             & 224.3(3)~${}^{+6.8}_{-10.5}$\,{\scriptsize $\pm3.2\perc$} 
	     & 1.14
             & 4.488~${}^{+5.6}_{-9.8}$\,{\scriptsize $\pm2.8\perc$}  
	     & $ 2.00 \times 10^{-2}$
 \\[1mm]
             $0^\pm$
             & 332.4(3)~${}^{+34.0}_{-23.5}$\,{\scriptsize $\pm5.4\perc$} 
             & 374.1(5)~${}^{+6.0}_{-9.3}$\,{\scriptsize $\pm2.5\perc$} 
	     & 1.13
             & 8.022~${}^{+5.4}_{-8.9}$\,{\scriptsize $\pm2.2\perc$}  
             & $ 2.14 \times 10^{-2}$
 \\[1mm]
 \hline
\end{tabular}
\caption{LO and NLO cross sections and corresponding $K$ factors
 for $pp\to t\bar tX_0$ at the 8- and 13-TeV LHC, for the three
 scenarios defined in table~\ref{tab:GFscenarios}.
 The integration error in the last digit(s) (in parentheses), and the
 fractional scale (left) and PDF(+$\alpha_s$) (right) uncertainties are
 also reported. 
 In addition to the fixed-order results, the PS-matched NLO cross
 sections for the di-leptonic decay channel
 $\sigma_{\rm NLO+PS}^{\rm dilep}$ and the ratios
 $R\equiv\sigma_{\rm NLO+PS}^{\rm dilep}/\sigma_{\rm NLO}$
 are also shown, where the acceptance cuts in 
 eqs.~\eqref{eq:mincutstth_leptons} and \eqref{eq:mincutstth_bjets} are
 applied.}
\label{tab:xsec_tth}
\end{table*}

At variance with the GF process, the production rate for the
pseudoscalar case is smaller than that for the scalar case.
Such a difference is proportional to  the top-quark mass, as the amplitudes for the scalar and pseudoscalar interactions
are identical in the limit where the Yukawa coupling is kept  constant and the quark mass is neglected.  
In $pp$ collisions at the LHC energies 
the contribution of the $gg$ initial  state is dominant over $q\bar q$ annihilation for all the scenarios.
It is rather interesting to observe, however, that for a CP-odd scalar $q\bar q$
annihilation contributes at LO to just 16\,\% (10\,\%) of the total cross section
at 8 (13)~TeV, compared to around 40\,\% (30\,\%) of the SM-like CP-even case. 
This difference is such that the CP-odd case exhibits slightly larger scale and PDF uncertainties.
Once again, we note that the scale dependence is larger than the
PDF$+\alpha_s$ uncertainty (though not by as much as in GF $H+$jets),
and that all the uncertainties are significantly reduced going from LO
to NLO, as expected. Increasing the collision energy from 8 to 13~TeV enhances 
the cross sections by about a factor 4 while the $K$ factors only slightly increase.
As in the GF case, $\sigma(0^{\pm})$ is equal to the average of
$\sigma(0^+)$ and $\sigma(0^-)$. We have verified explicitly that at the
LO the interference between amplitudes corresponding to different parity interactions is exactly zero.  At NLO, the interference at the amplitude level is nonzero, yet the total rates do sum up to each of the parity-definite contributions. 

To investigate the spin correlations effects among the decay products from the top and
antitop quarks, we present results for the di-leptonic decay channel of the
top pair, $t\to b\ell^+\nu_\ell$ and 
$\bar t\to\bar b\ell^-\bar{\nu}_\ell$ with $\ell=e,\mu$.
We require two leptons and two $b$-tagged jets that pass the acceptance 
cuts, respectively,
\begin{align} 
 p_T(\ell)>20~{\rm GeV}\,,\quad |\eta(\ell)|<2.5\,,
\label{eq:mincutstth_leptons}
\end{align}
and
\begin{align}
 p_T(j_b)>30~{\rm GeV}\,,\quad |\eta(j_b)|<2.5\,.
\label{eq:mincutstth_bjets}
\end{align}
It is known that dedicated top and Higgs reconstruction are crucial 
in order to obtain the significant $t\bar tH$ signal over the
background, at least for the dominant $H\to b\bar b$ decay channel. 
Several proposals have been put forward from using multivariate analysis, e.g., matrix element method~\cite{Artoisenet:2013vfa} to jet substructure/boosted techniques~\cite{Butterworth:2008iy,Plehn:2009rk,Buckley:2013auc,Bramante:2014gda}.
In this work we are mainly concerned in checking what observables can be sensitive to CP effects and
do not consider either backgrounds or reconstruction issues. However, we will consider how CP-sensitive
observables are affected by the requirement of a large transverse
momentum for the Higgs, i.e. a ``boosted Higgs''.

In table~\ref{tab:xsec_tth}, we also report the PS-matched  NLO cross
sections for the di-leptonic decay channel and the corresponding ratios to the
fixed-order NLO prediction,
$R\equiv\sigma_{\rm NLO+PS}^{\rm dilep}/\sigma_{\rm NLO}$, where 
acceptance cuts (assuming 100\,\% $b$-tag and lepton efficiencies) are taken into  
account.  Accounting for  the branching fraction of the di-lepton mode, 
$(0.213)^2\sim 0.045$, the ratios show that parton shower and the cuts lead to a decrease of about a factor 2 in
the cross section. Increasing the CM energy results in the slightly smaller $R$ ratios.

\subsection{Distributions}

\begin{figure}
 \center 
 \includegraphics[width=1.0\columnwidth]{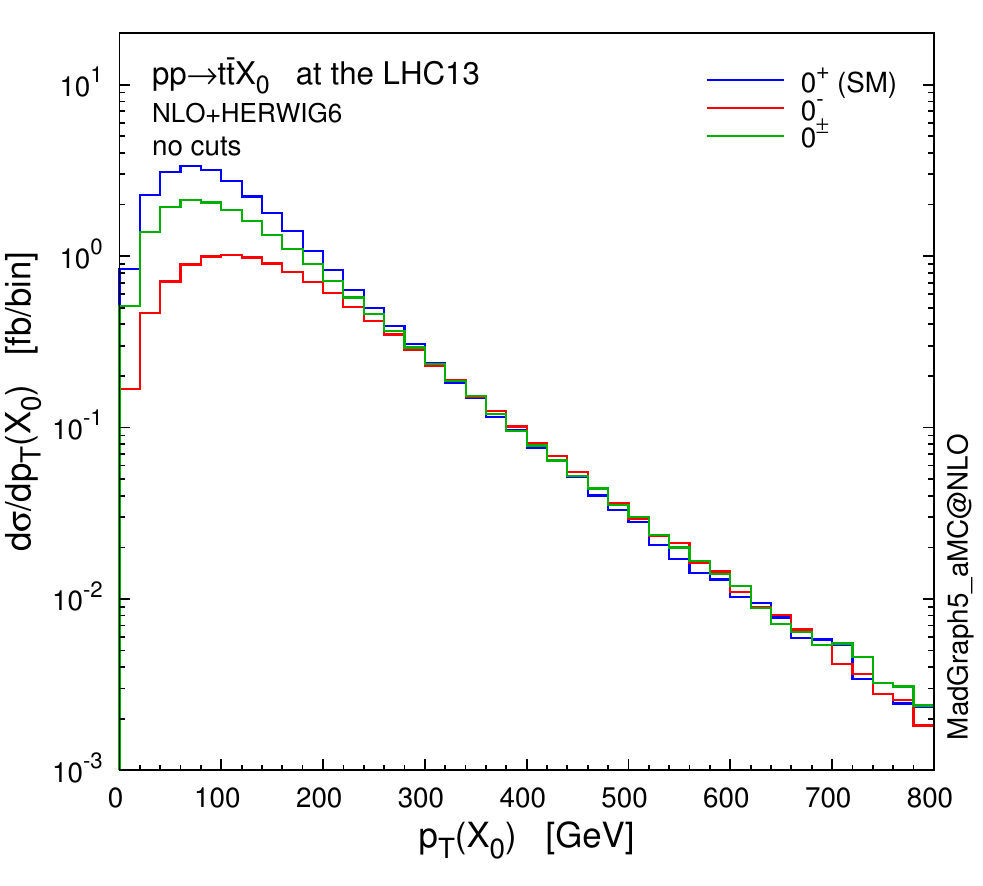}
 \caption{Distribution of the transverse momentum of $X_0$ in
 $pp\to t\bar tX_0$ at the 13-TeV LHC.
 The different hypotheses are defined in table~\ref{tab:GFscenarios}.} 
\label{fig:tth_diffxs}
\end{figure} 

\begin{figure*}
 \center 
 \includegraphics[width=0.24\textwidth]{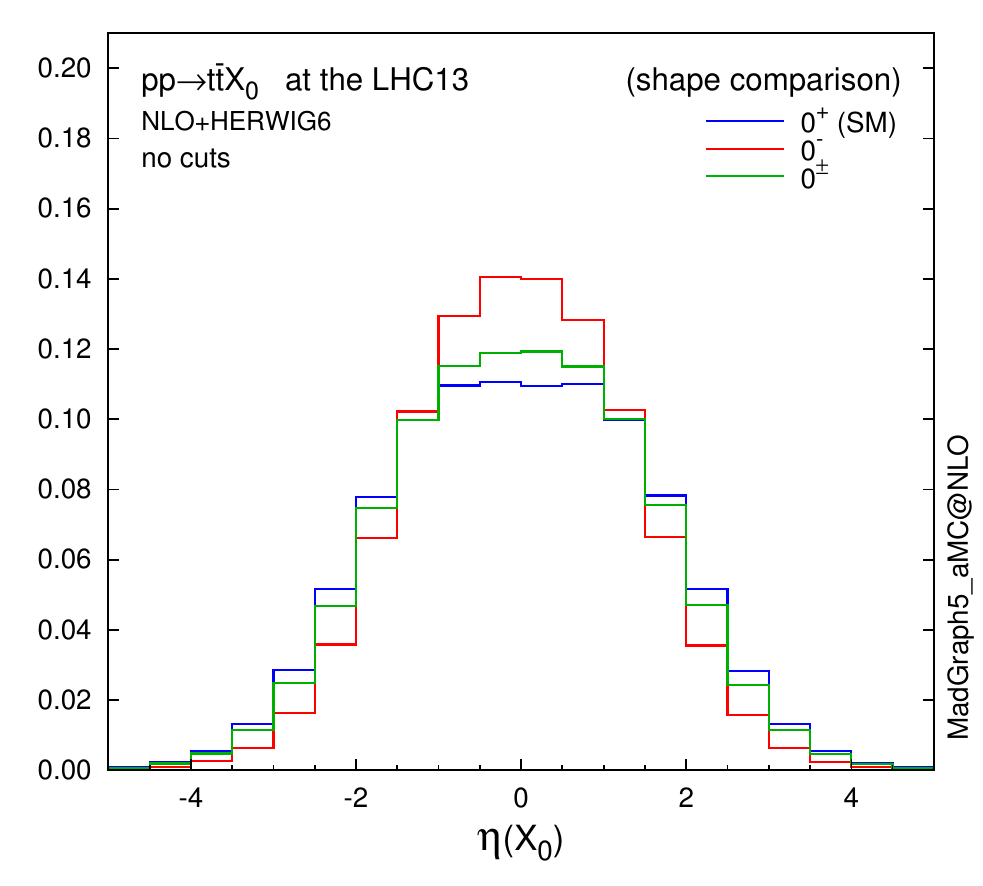}
 \includegraphics[width=0.24\textwidth]{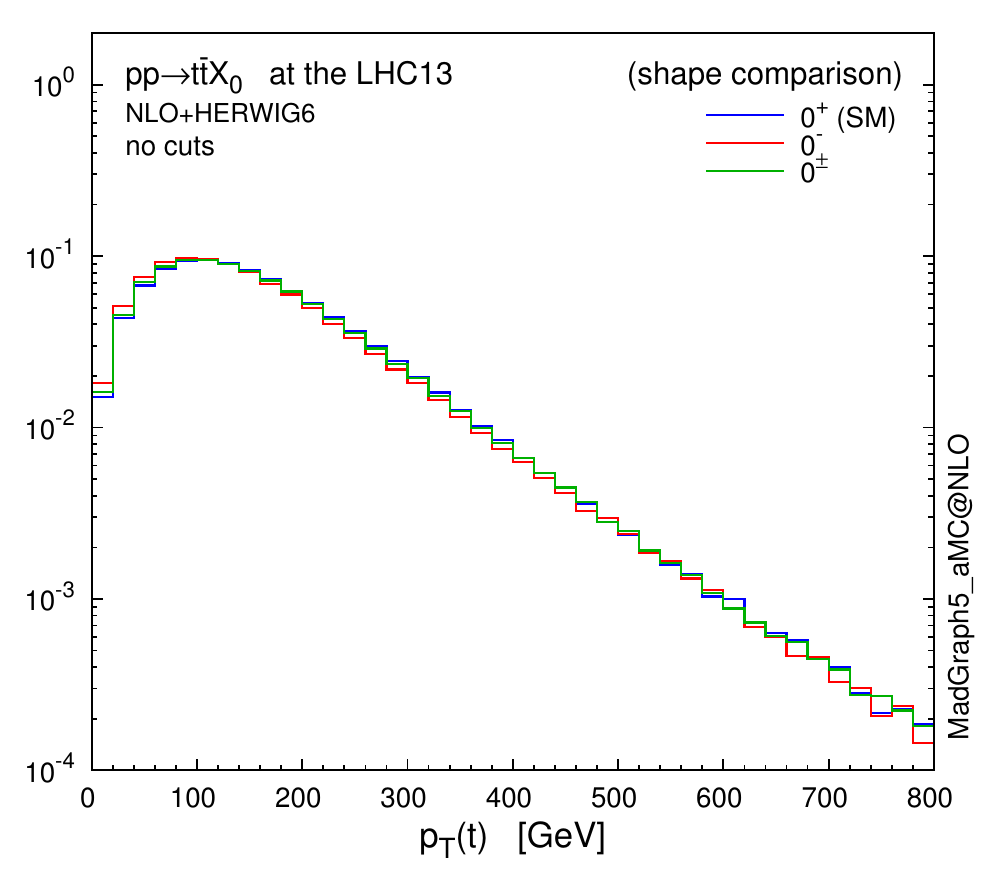}
 \includegraphics[width=0.24\textwidth]{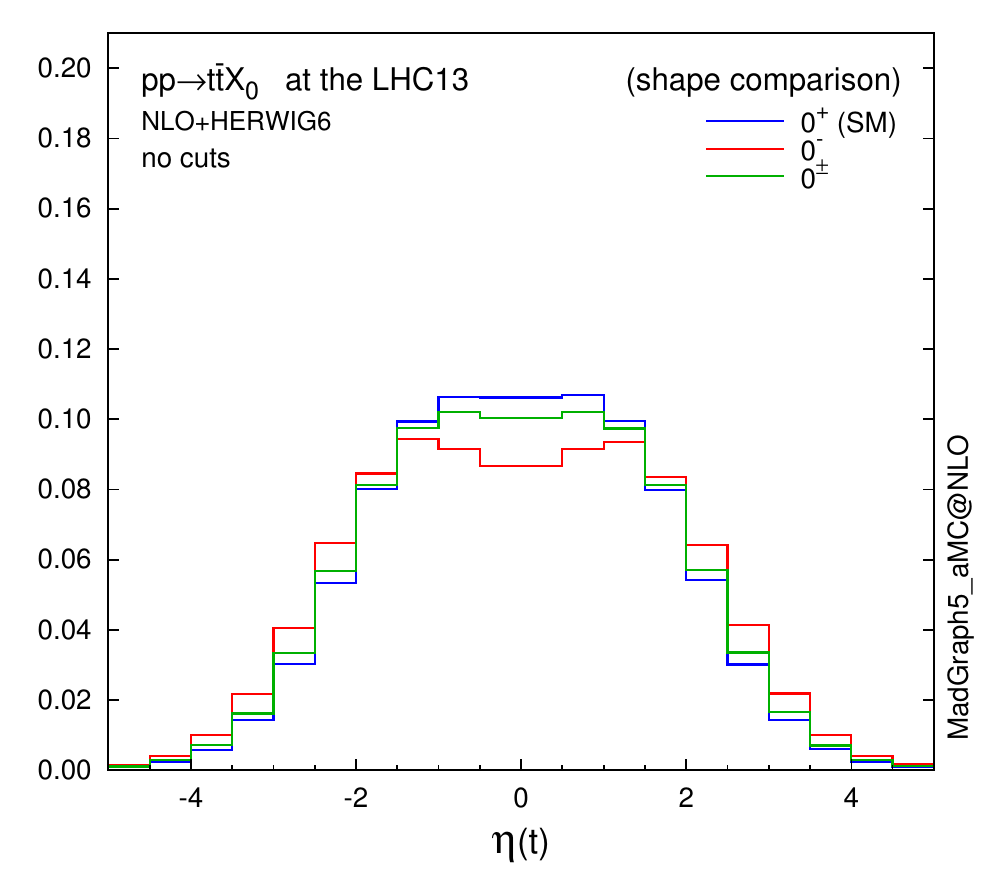}
 \includegraphics[width=0.24\textwidth]{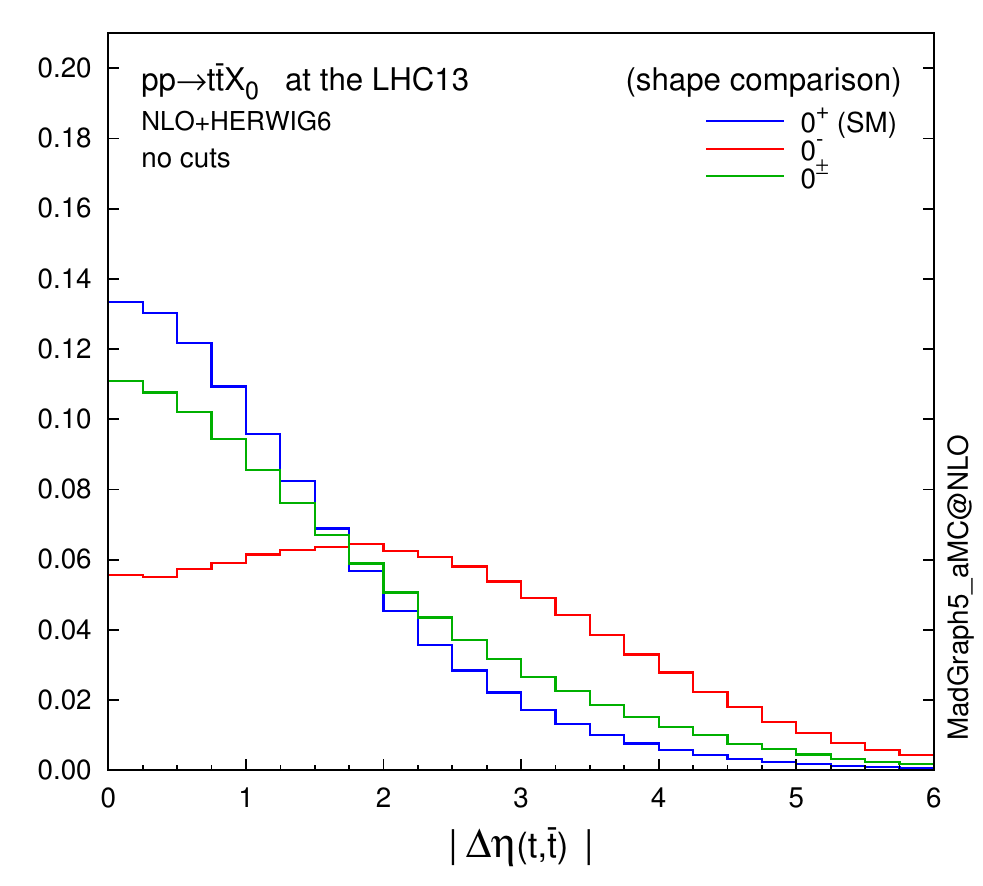}
 \includegraphics[width=0.24\textwidth]{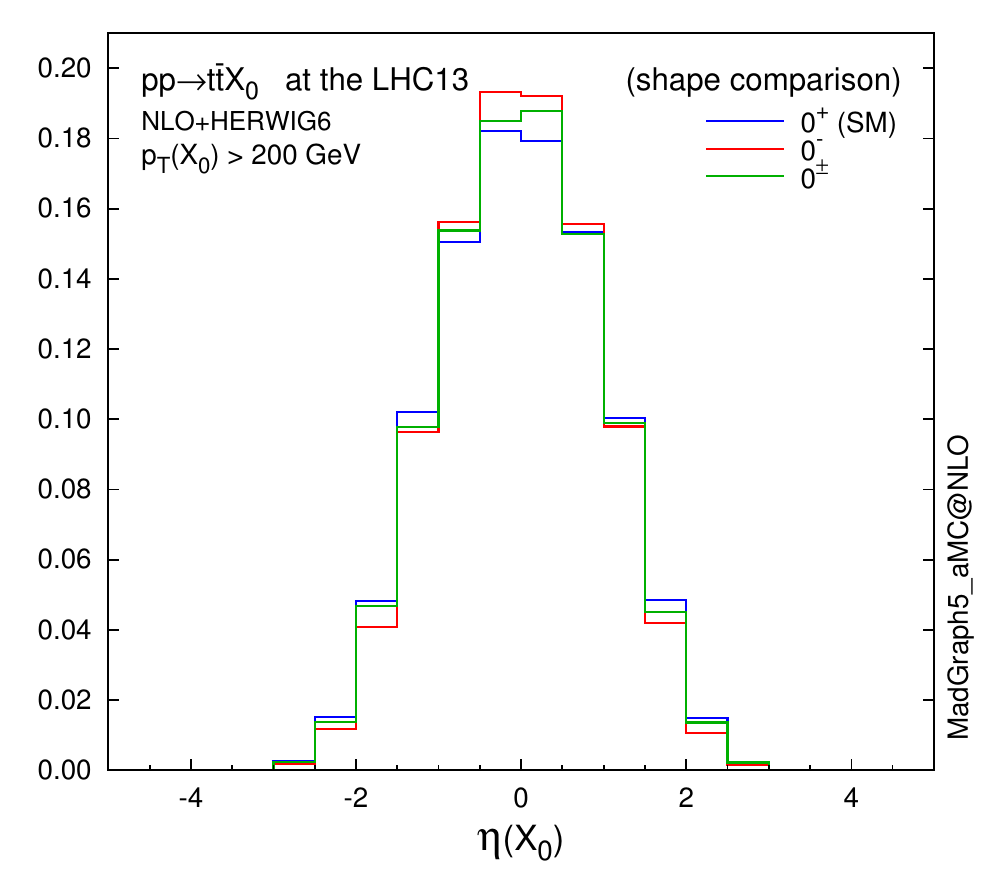}
 \includegraphics[width=0.24\textwidth]{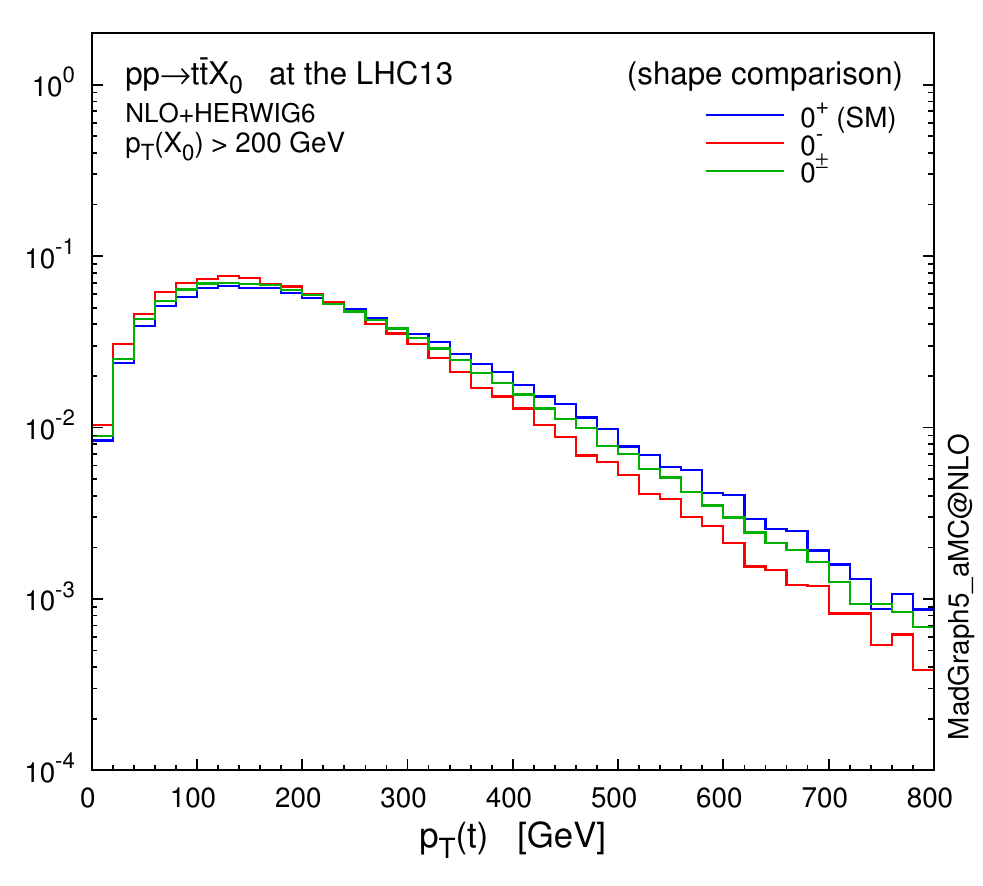}
 \includegraphics[width=0.24\textwidth]{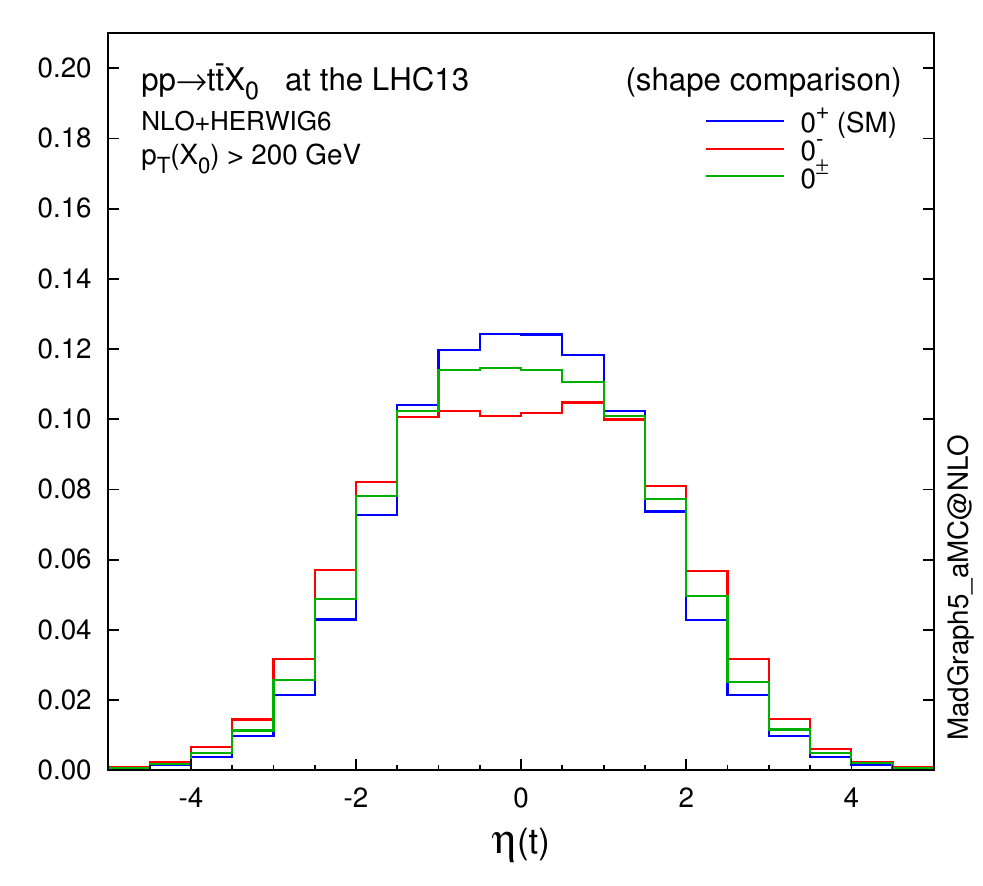}
 \includegraphics[width=0.24\textwidth]{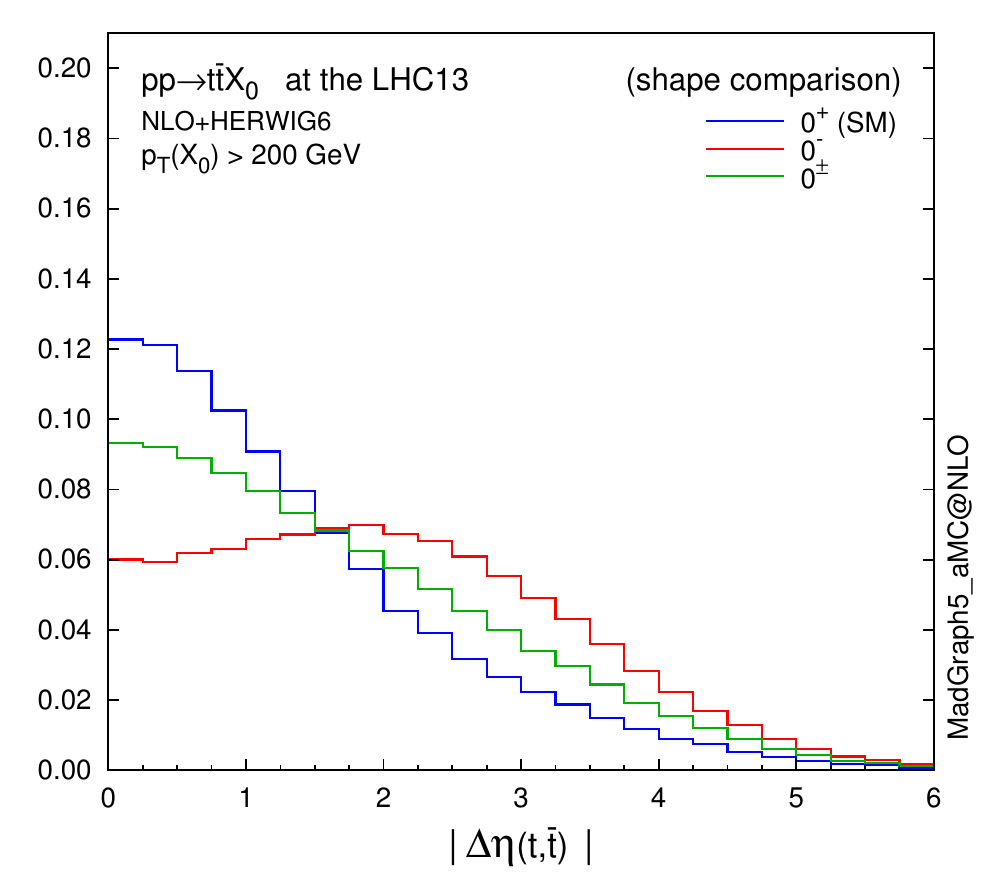}
\caption{Normalized distributions (shape comparison) without cuts (top), while with the
 $p_{T}(X_0)>200$~GeV cut (bottom).
 The three spin-0 hypotheses are defined in table~\ref{tab:GFscenarios}.}  
\label{fig:tth1}
\end{figure*} 

\begin{figure*}
 \center 
 \includegraphics[width=0.24\textwidth]{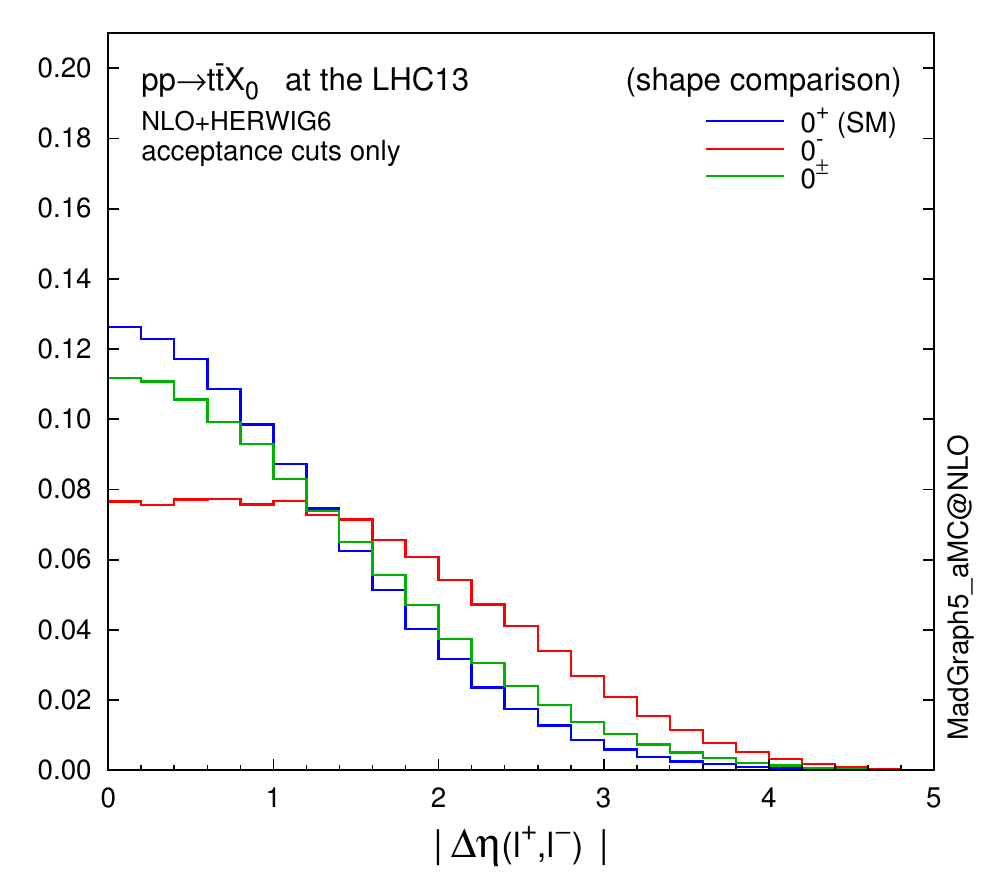}
 \includegraphics[width=0.24\textwidth]{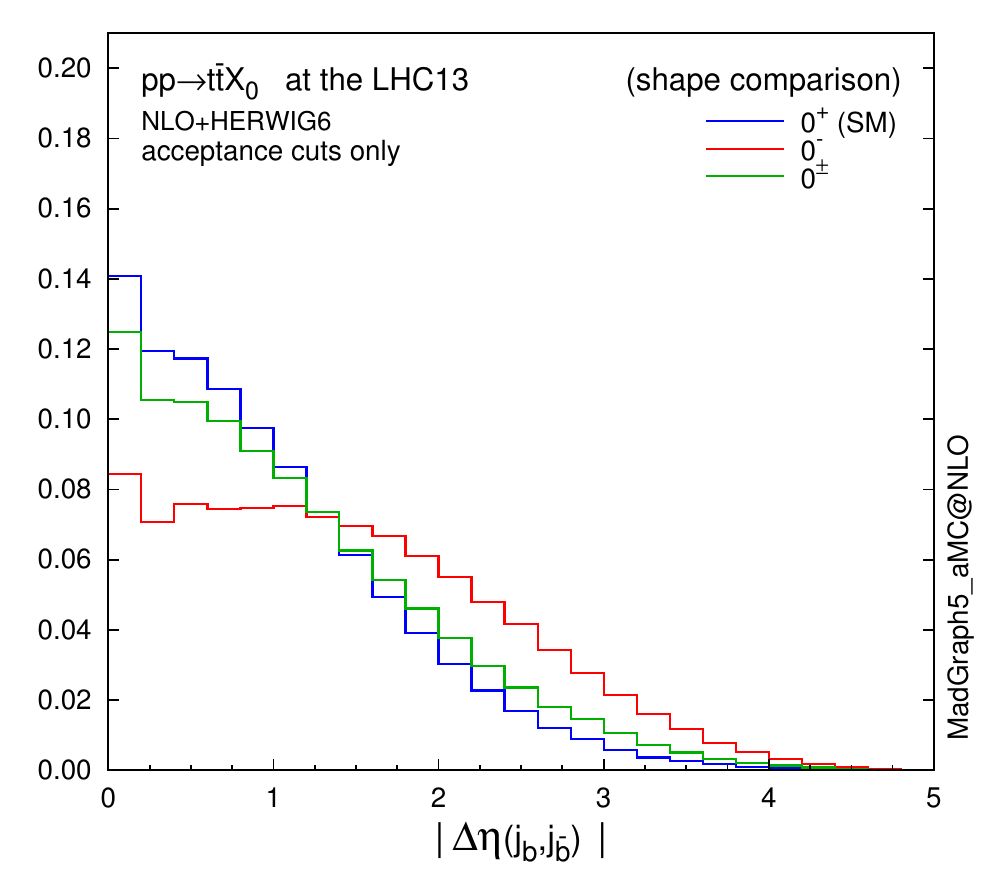}
 \includegraphics[width=0.24\textwidth]{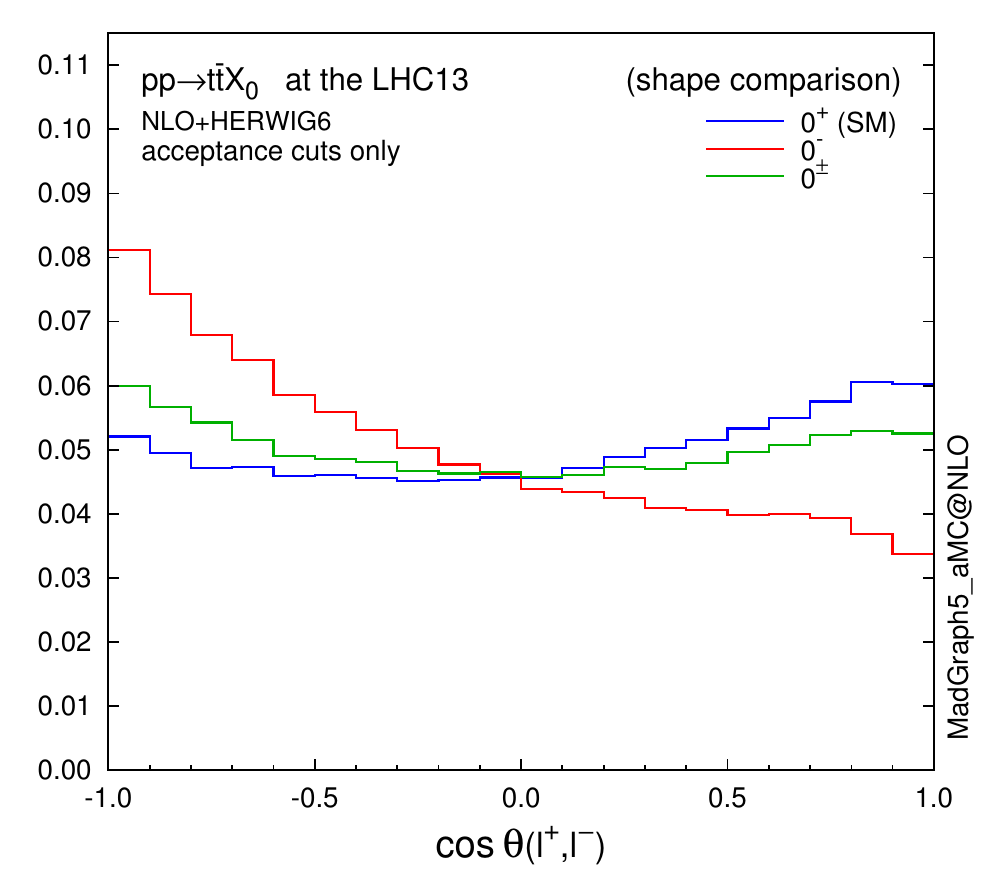}
 \includegraphics[width=0.24\textwidth]{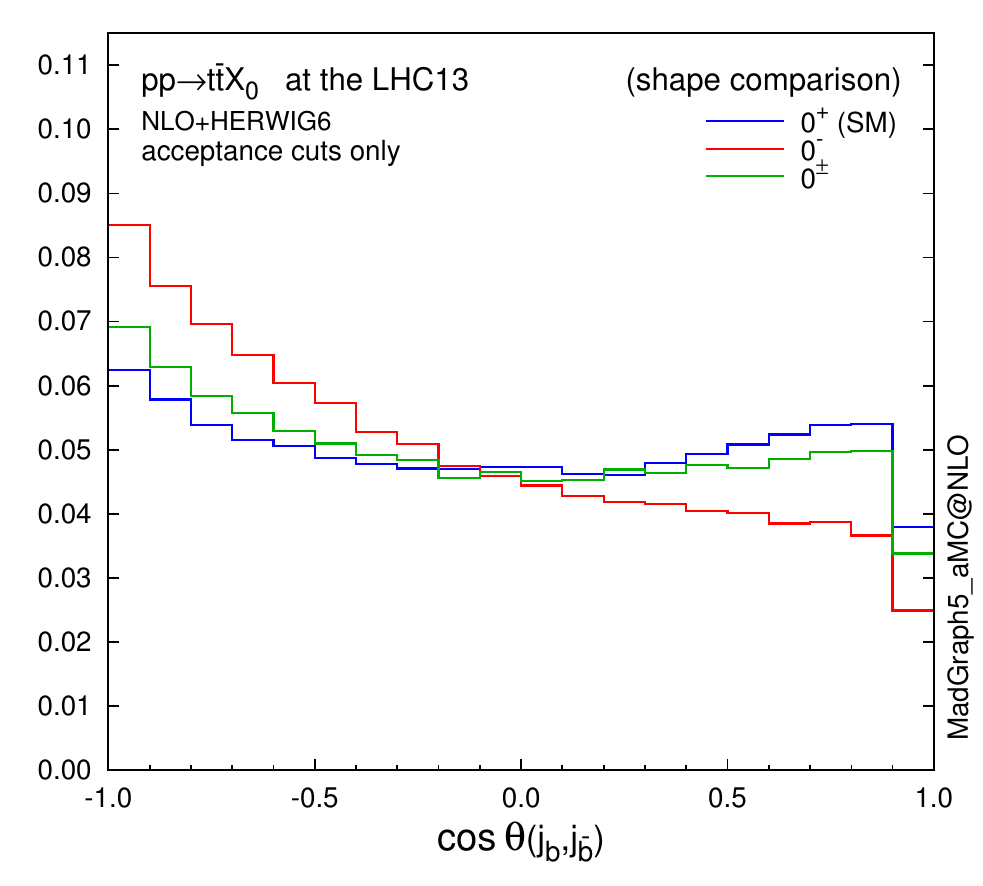}
 \includegraphics[width=0.24\textwidth]{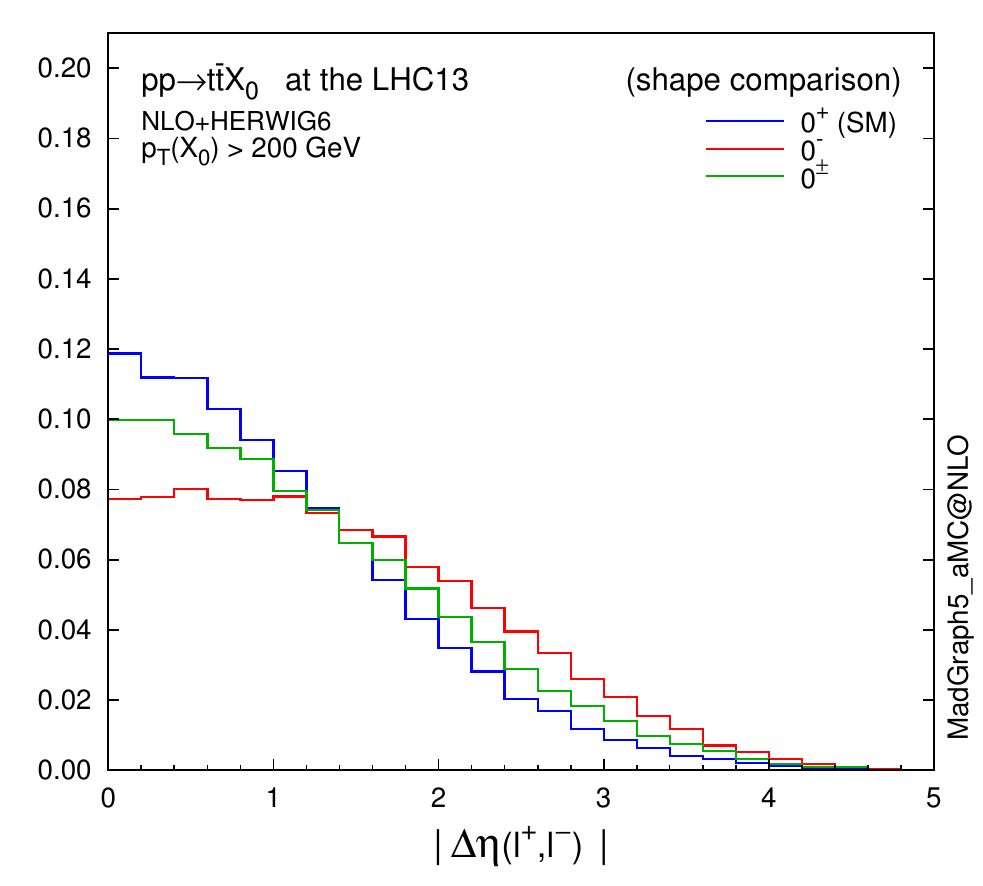}
 \includegraphics[width=0.24\textwidth]{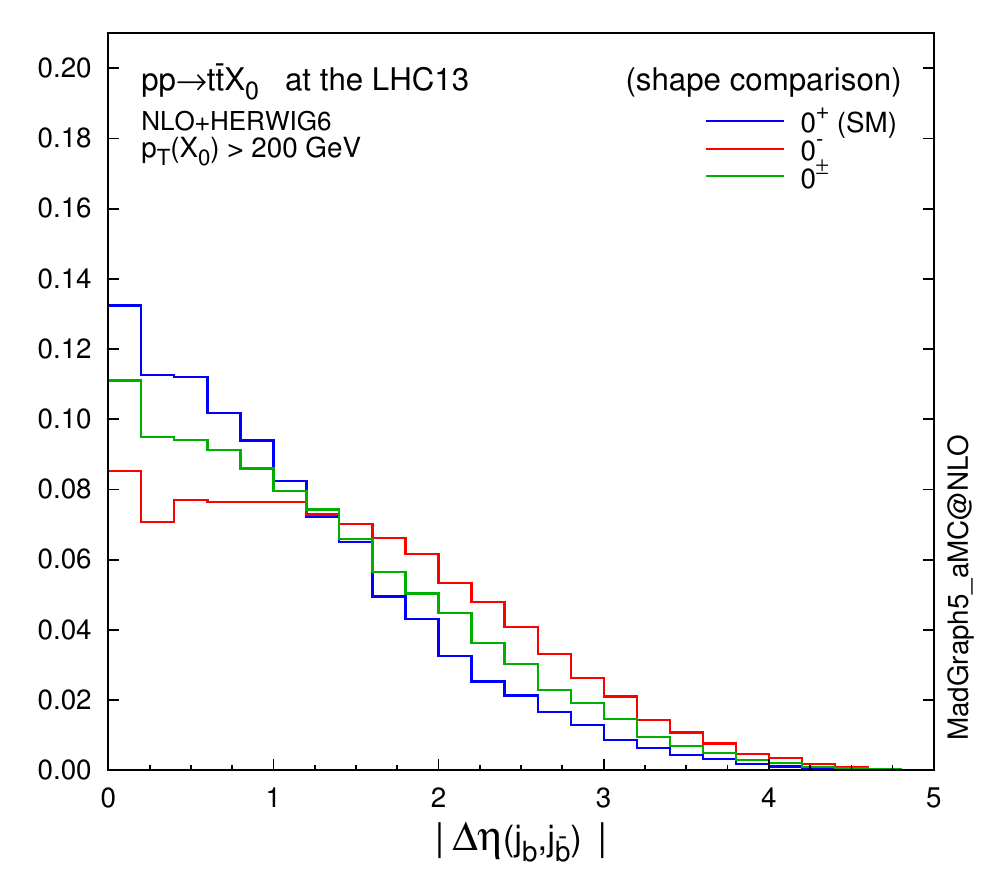}
 \includegraphics[width=0.24\textwidth]{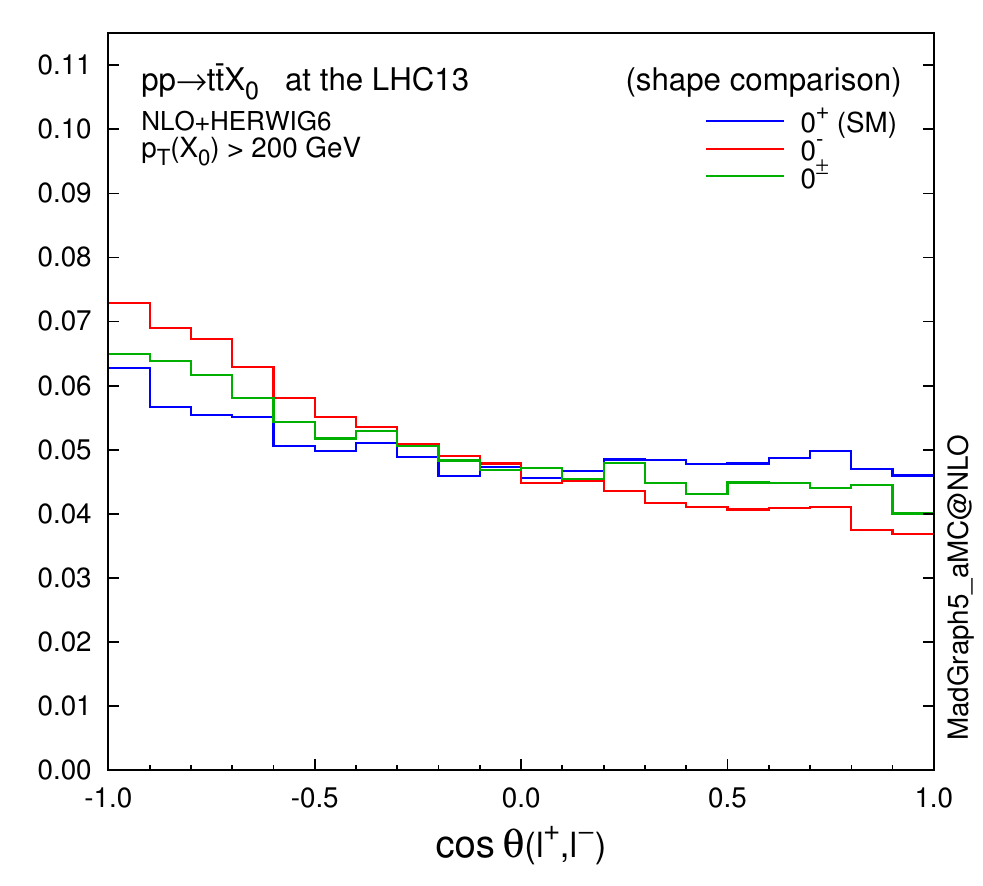}
 \includegraphics[width=0.24\textwidth]{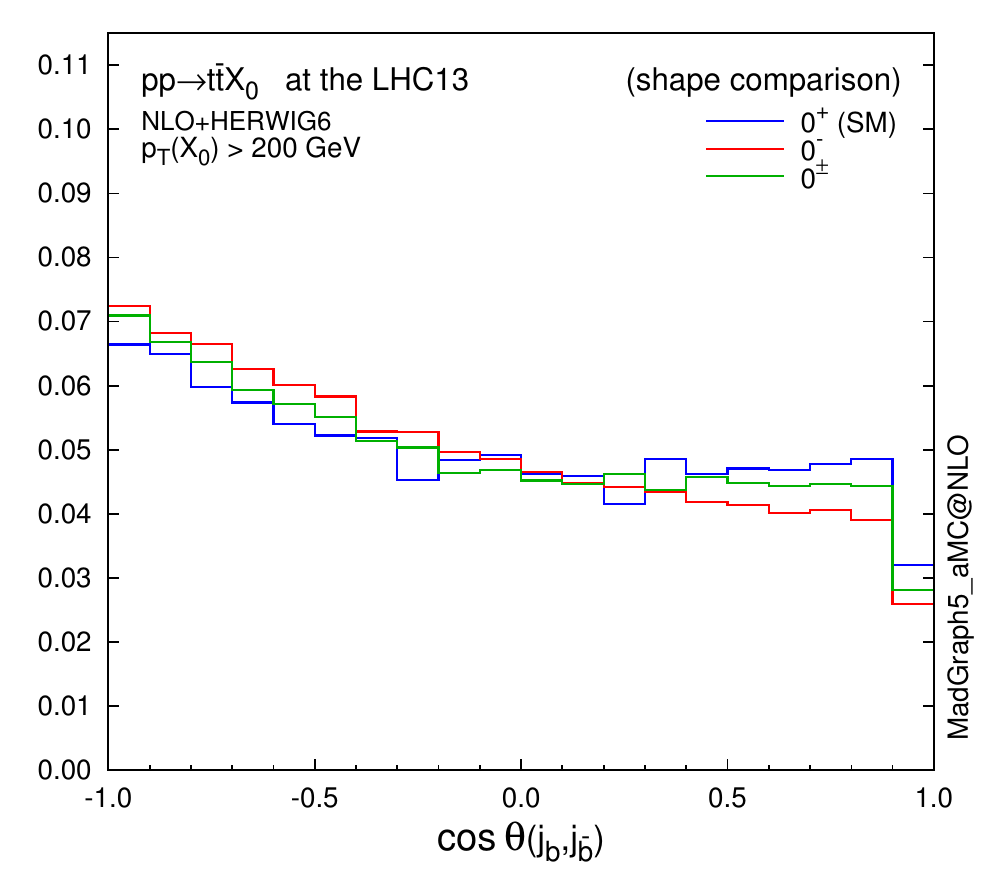}
\caption{Normalized distributions (shape comparison) for the correlations between
 the top-quark decay products with the acceptance cuts (top) plus the
 $p_{T}(X_0)>200$~GeV cut (bottom).}
\label{fig:tth2}
\end{figure*} 

In fig.~\ref{fig:tth_diffxs} we show differential cross sections for
$t\bar tX_0$ production at the 13-TeV LHC as a function of the
transverse momentum of the resonance $p_T(X_0)$.
As one can see, the difference between the various scenarios is
significant in the low-$p_T$ region, while the high-$p_T$ tail of the distributions,
featuring exactly the same shape, are not sensitive to the CP mixing~\cite{Frederix:2011zi}.
It is also interesting to see that our normalisation choice, $g_{\sss Htt}=g_{\sss Att}=m_t/v\,(=y_t/\sqrt{2})$
leads to exactly the same rates at high $p_T$ independently of the mixing parameter $\alpha$. This is a known
feature of scalar radiation from a heavy quark at high
$p_T$~\cite{Dawson:1997im,Dittmaier:2000tc,Beenakker:2001rj}. This
raises the important question whether  boosted analyses can be sensitive
to CP properties of the
Higgs--top-quark coupling, which we address below. 

Figure~\ref{fig:tth1} shows some other  relevant distributions in the $ t \bar t X_0$ final state, without and
with the $p_T(X_0)>200$~GeV cut:
the pseudorapidity distribution of $X_0$, 
the top-quark transverse momentum and pseudorapidity, and the 
pseudorapidity distance between the top and antitop quarks
$\Delta\eta(t,\bar t)\equiv\eta(t)-\eta(\bar t)$. 
Compared to the SM, a CP-odd $X_0$ tends to be produced more centrally, 
while  the accompanying top quarks are  more forward.  
The most sensitive distribution  to CP mixing is the rapidity difference between 
the top and antitop. This observable is hardly affected by the
$p_T(X_0)>200$~GeV cut, thus the correlations among the top--antitop
decay products provide a good CP-discriminating power also in the
boosted regime. 

\begin{figure*}
 \center 
 \includegraphics[width=0.322\textwidth]{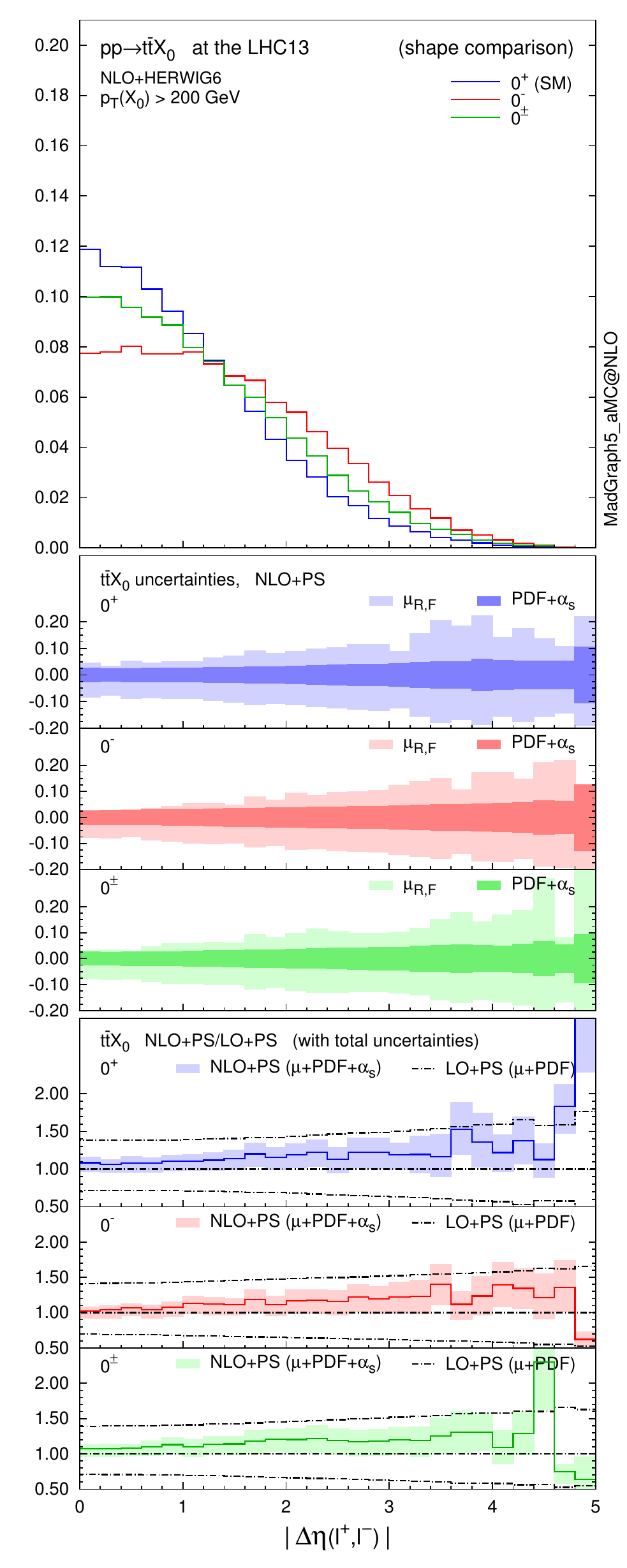}
 \includegraphics[width=0.322\textwidth]{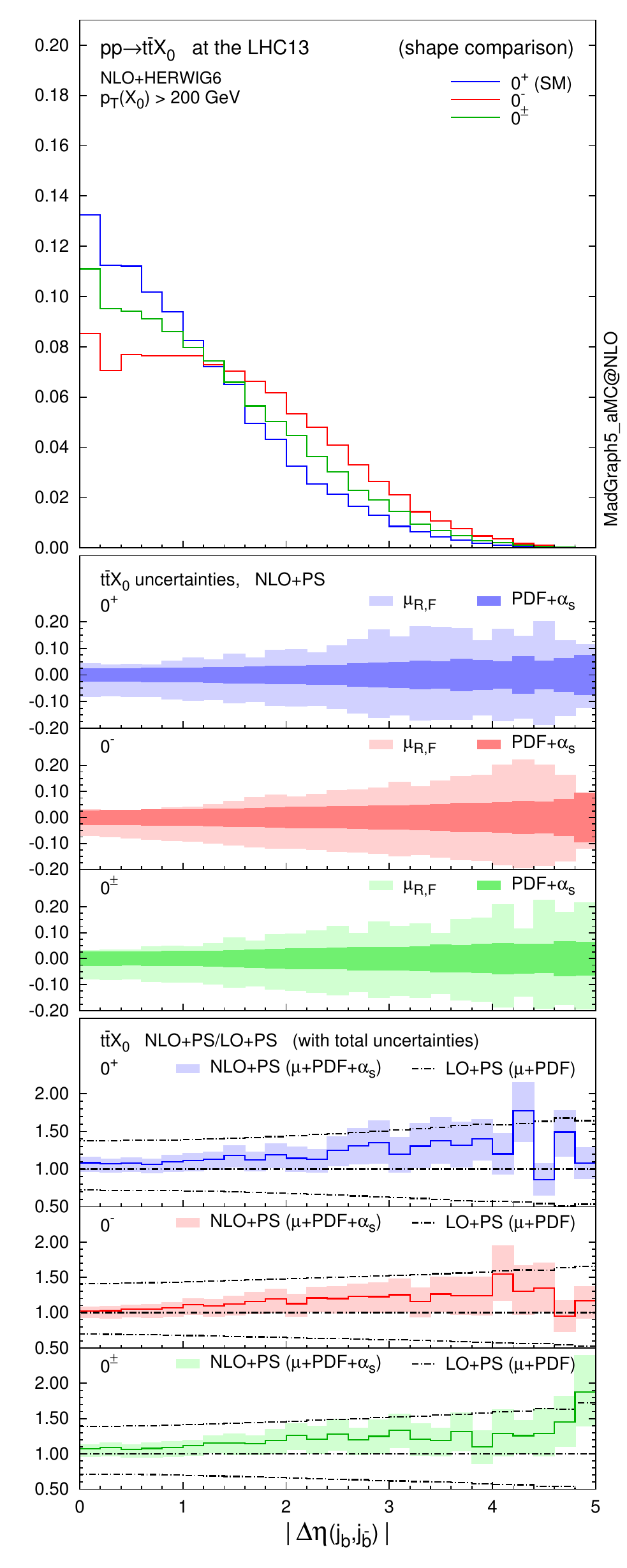}
 \includegraphics[width=0.322\textwidth]{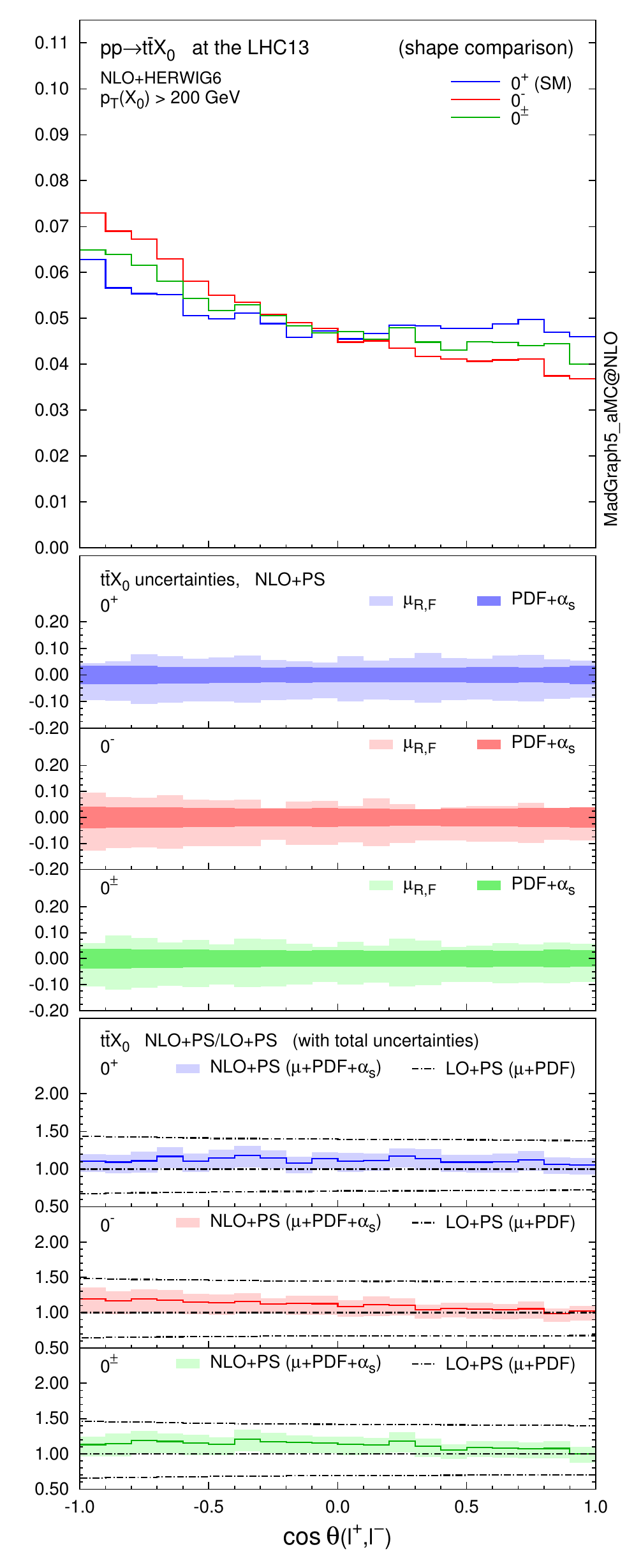}
 \caption{Normalized distributions (shape comparison) of the rapidity separation between
 the leptons (left) and the $b$-jets (centre), and the opening angle
 between the leptons (right), for $pp\to t\bar tX_0$ at the 13-TeV 
 LHC, where the acceptance cuts plus $p_T(X_0)>200$~GeV are applied. 
 For each scenario,
 the middle panels show the scale and PDF$+\alpha_s$ uncertainties,
 while the bottom ones give the ratio of NLO+PS to LO+PS results, with
 total uncertainties.} 
\label{fig:tth_scale}
\end{figure*} 

In fig.~\ref{fig:tth2}, we show the correlations between the top decay
products (in the di-leptonic channel). As expected from the $\Delta\eta_{t\bar t}$ distribution,
$\Delta\eta_{\ell\bar\ell}$ and $\Delta\eta_{b\bar b}$ are almost insensitive
to the $p_T(X_0)$ cut, while the angles between the leptons and between the $b$-jets
are significantly affected by the boost.
The angular observables in different frames have been studied in 
ref.~\cite{Biswas:2014hwa}.
We note that, although we only consider the fully leptonic channel here,
there is no limitation to study the semi-leptonic and fully hadronic
channels by using {\sc MadSpin}.

Finally, we discuss the theoretical uncertainties. 
Figure~\ref{fig:tth_scale} displays, from left to right, the rapidity
distance between the leptons ($\Delta \eta_{\ell\bar\ell}$) 
and between the $b$-tagged jets ($\Delta \eta_{b\bar b}$), and the
opening angle between the leptons ($\cos \theta_{\ell\bar\ell}$),
where the acceptance cuts in eqs.~\eqref{eq:mincutstth_leptons} 
and \eqref{eq:mincutstth_bjets} plus the $p_T(X_0)>200$~GeV cut are
applied. 
The middle panels show the uncertainties due to the scale variation and the 
PDF+$\alpha_s$  for each scenario,  while the bottom ones give the ratio of NLO+PS to LO+PS results, 
each one with its total uncertainty band.
We can see that, depending on the observable considered,
the NLO corrections and the corresponding uncertainties can change significantly over the phase space.
As in the $H$+jets case, NLO corrections are significant for all the observables,
considerably reduce the theoretical uncertainty, and cannot be described by
an overall $K$ factor.

\section{Summary}\label{sec:summary}

In this work we have presented for the first time 
results at NLO in QCD, including parton-shower effects, for the hadroproduction of a
spin-0 particle with CP-mixed  coupling to the top quark, in gluon-fusion plus one and two jets and 
in association with a top-quark pair.  Our results are obtained in a fully automatic 
way through the implementation of the relevant interactions
in {\sc FeynRules} and then performing event generation in the {\sc MadGraph5\_aMC@NLO} framework.

We have presented illustrative distributions obtained by interfacing NLO
parton-level events to the {\sc HERWIG6} parton shower. NLO corrections
improve the predictions of total cross sections by reducing 
PDF+$\alpha_s$ uncertainty and scale dependence. 
In addition, our simulations show that NLO+PS effects need
to be accounted for to make accurate predictions on the kinematical
distributions of the final-state objects, such as the Higgs boson, the jets
and the top decay products. 

We have confirmed that di-jet correlations in Higgs plus two jet production, 
in particular the azimuthal difference between the jets,
are sensitive probes of the CP mixing of the Higgs.
In associated production with a top pair, we have shown that many correlations 
between the top and antitop decay products can be sensitive to the CP nature of 
the Higgs. In particular, the pseudorapidity separation between the leptons
or between the $b$-jets is a promising observable when analysing events 
with a Higgs boson at high transverse momentum.
The quantitative determination of the CP mixing has been done for the GF channel at LO in
ref.~\cite{Dolan:2014upa}, while the LO parton-level analysis has been
done for the $t\bar tH$ channel including $tH$ and $\bar tH$ in 
ref.~\cite{Ellis:2013yxa}. 
The estimation of the impact of the NLO+PS corrections as well as
detector effects is desired and will be reported elsewhere.

As a final remark, we note that in this work we have only addressed the issue of the
CP properties of the flavour-diagonal Higgs--top-quark interactions, which can
be parametrised in full generality
as in eq.~\eqref{eq:0ff}. At the dimension-six level, however, other operators appear that lead to effective 
three-point and four-point Higgs--top-quark interactions of different type~\cite{AguilarSaavedra:2009mx,Zhang:2010dr,Zhang:2012cd,Degrande:2010kt,Degrande:2012gr}, including flavour changing neutral ones~\cite{AguilarSaavedra:2009mx,Zhang:2013xya,Zhang:2014rja}, which can also be studied in the same production channels as discussed here,{ \it i.e.} $H$+jets and $t\bar t H$. Work in promoting predictions for these processes to NLO accuracy in QCD is in progress.

\section*{Acknowledgments}

We would like to thank the Higgs Cross Section Working Group for the encouragement in pursuing the Higgs Characterisation project.  We are thankful to Pierre Artoisenet, Stefano Carrazza, Stefano Forte,
Rikkert Frederix, Valentin Hirschi, Olivier Mattelaer and Tiziano Peraro
for their support during the preparation of this work and to Stefano Frixione for many useful discussions and comments
on the manuscript.  We thank Roberto Pittau for his participation to the initial stages of this project.

This work has been performed in the framework of the ERC grant 291377
``LHCtheory: Theoretical predictions and analyses of LHC physics:
advancing the precision frontier'' 
and
of the FP7 Marie Curie Initial Training Network MCnetITN (PITN-GA-2012-315877).
It is also supported in part by the 
Belgian Federal Science Policy Office through the Interuniversity Attraction Pole P7/37. 
The work of FD and FM is supported by the IISN ``MadGraph'' convention
4.4511.10 and the IISN ``Fundamental interactions'' convention 4.4517.08.
KM is supported in part by the Strategic Research Program ``High Energy
Physics'' and the Research Council of the Vrije Universiteit Brussel.  
The work of MZ is partially supported by the Research Executive Agency
(REA) of the European Union under the Grant Agreement Number 
PITN-GA-2010-264564 (LHCPhenoNet) and by the ILP LABEX (ANR-10-LABX-63), 
in turn supported by  French state funds managed by the ANR within the ``Investissements d'Avenir''
programme under reference ANR-11-IDEX-0004-02.

\appendix
\section{Feynman rules, UV and $\boldsymbol R_2$ terms for gluon-fusion
 Higgs production at NLO QCD} 
\label{app:eftnlo}

In this appendix we present the Feynman rules, UV and $R_2$ terms 
necessary for NLO-QCD automatic computations,
for gluon fusion (GF) in an effective field theory approach, where the Higgs
boson couples to gluons through loops of infinitely heavy quarks.
The LO rules have been obtained automatically by coding the effective
lagrangian in {\sc FeynRules}, while the UV and $R_2$ terms have
been coded by hand in the UFO format. This file is read by ALOHA~\cite{deAquino:2011ub}, 
which generates a library of helicity amplitudes and currents for a given process as requested by the user 
in {\sc MadGraph5\_aMC@NLO}.

In this note, we use the following conventions:
outgoing momenta for external particles;
the antisymmetric tensor $\epsilon^{0123}=+1$; and 
the metric tensor $g^{\mu\nu}=\text{diag}(1,-1,-1,-1)$. 

The relevant Higgs--gluon interaction lagrangian consists of
the first two operators in eq.~\eqref{L_loop}.
Since it is linear in the scalar and pseudoscalar components of $X_0$, 
we only need to consider the two separate cases of a pure scalar $X_0=H$  
(i.e. $c_\alpha=1,\,\kappa_{\sss Hgg}\ne0$ in eq.~\eqref{eq:0vv}), or a
pure pseudoscalar $X_0=A$ (i.e. $s_\alpha=1,\,\kappa_{\sss Agg}\ne0$).  
Thus, we start from the two effective lagrangians
\begin{align}
 \mathcal{L}_{H} & =-\frac{1}{4}\, g_{\sss Hgg}\, G_{\mu\nu}^a G^{a,\mu\nu}
 H\,,
\label{eq:ggHLagrangian} \\
\mathcal{L}_{A} & =-\frac{1}{4}\, g_{\sss Agg}\, G_{\mu\nu}^a \widetilde G^{a,\mu\nu} A \,,
\label{eq:ggALagrangian}
\end{align}
from which we obtain the interaction vertices listed in 
tables~\ref{tab:ggHvertices} and \ref{tab:ggAvertices}.

\begin{table*}
\begin{center}
\line(1,0){520}
\end{center}
\mbox{}\\[-13mm]
\begin{align}
 \begin{minipage}[l]{0.3\linewidth}
 \includegraphics[scale=0.28]{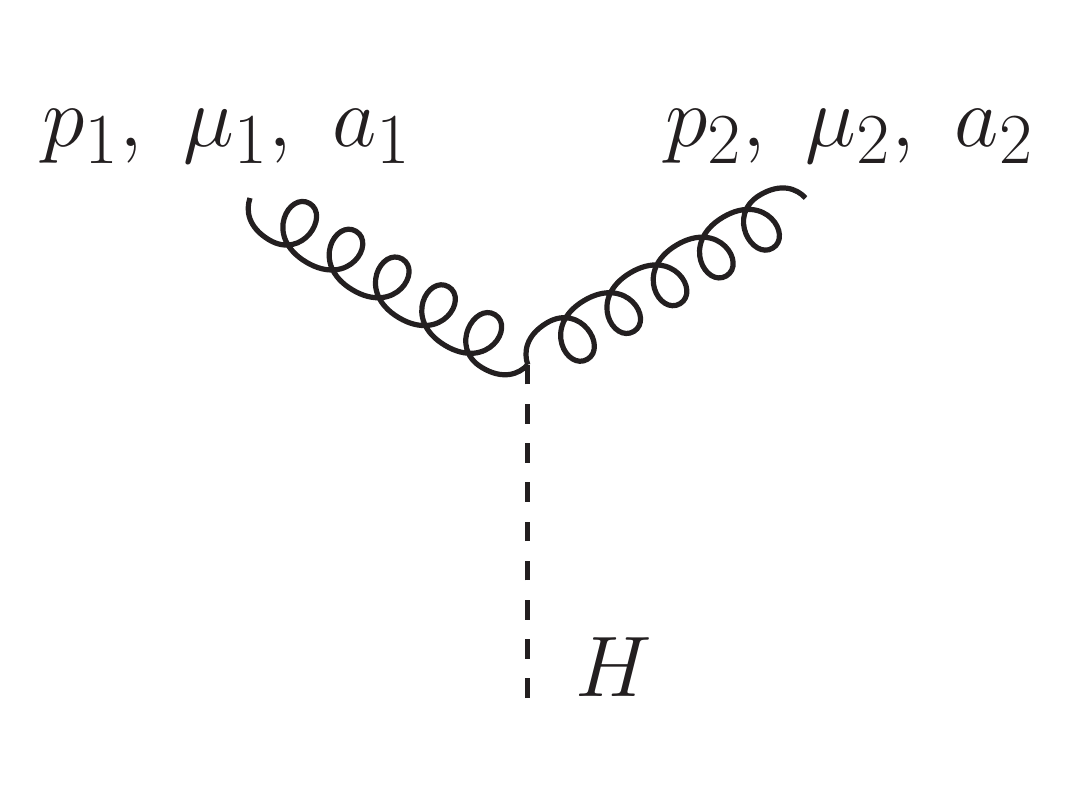}
 \end{minipage} 
 & \hspace*{-6em} =~ -i \, g_{\sss Hgg}  \, \delta^{a_1a_2} \, \big(\, p_2^{\mu_1} p_1^{\mu_2} - g^{\mu_1 \mu_2}\,p_1\!\cdot\! p_2 \,\big)
   \\[-3mm]
 \begin{minipage}[l]{0.3\linewidth}
 \includegraphics[scale=0.28]{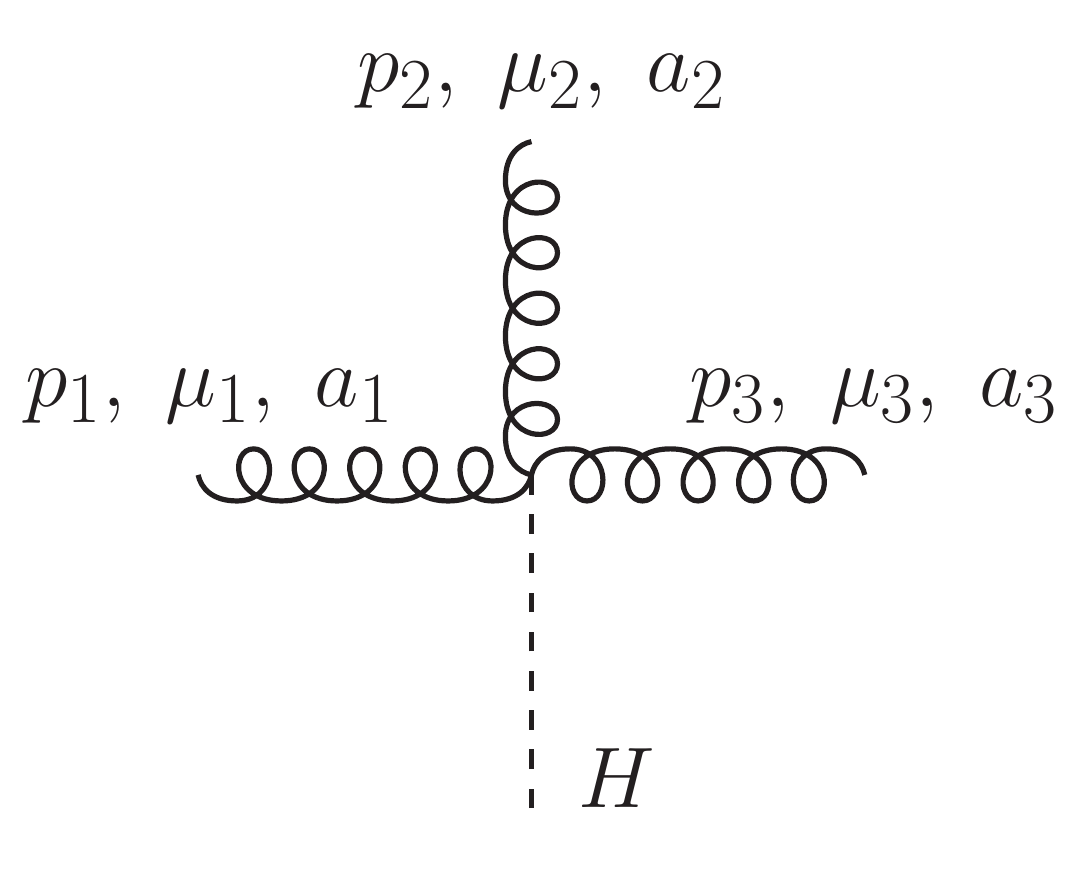}
 \end{minipage} 
 & \mbox{}\notag\\[-17mm]
 & \hspace*{-6em} =~ -g_{\sss Hgg}  \, g_s \, f^{a_1a_2a_3} \,\Big[ \hspace*{0.22em} 
    g^{\mu_1 \mu_2} \left(p_1-p_2\right)^{\mu_3} 
  + g^{\mu_2 \mu_3} \left(p_2-p_3\right)^{\mu_1} 
  + g^{\mu_3 \mu_1} \left(p_3-p_1\right)^{\mu_2} \,\Big] \\[6.mm]
 \begin{minipage}[l]{0.3\linewidth}
 \includegraphics[scale=0.28]{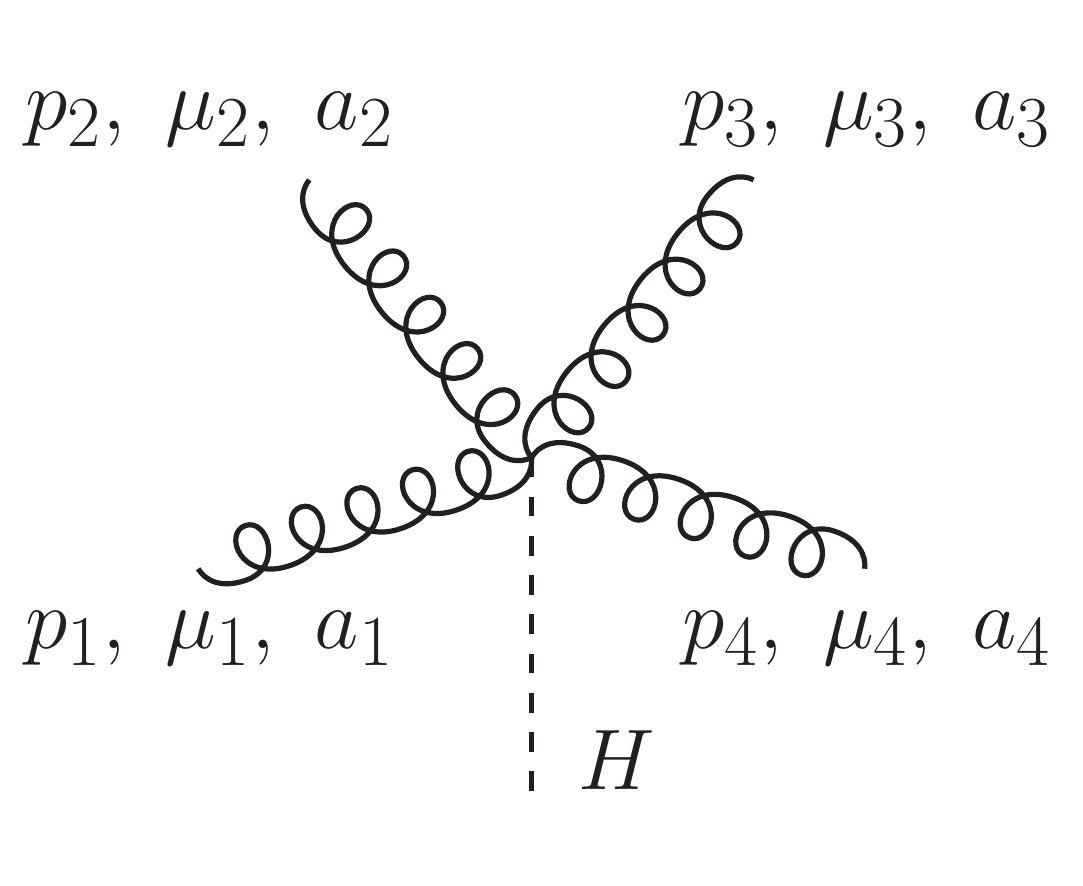}
 \end{minipage} 
 & \mbox{}\notag\\[-23.mm]
 & \hspace*{-6em} =~ -i \, g_{\sss Hgg} \, g_s^2  \, \Big[ \hspace*{1.1em}
    f^{a_1a_2b} f^{a_3a_4b} \left( g^{\mu_1 \mu_3} g^{\mu_2 \mu_4} - g^{\mu_1 \mu_4} g^{\mu_2 \mu_3} \right) \notag\\
 & \hspace*{1.em} + 
    f^{a_1a_3b} f^{a_2a_4b} \left( g^{\mu_1 \mu_2} g^{\mu_3 \mu_4} - g^{\mu_1 \mu_4} g^{\mu_2 \mu_3} \right) \notag\\
 & \hspace*{1.em}  + 
    f^{a_1a_4b} f^{a_2a_3b} \left( g^{\mu_1 \mu_2} g^{\mu_3 \mu_4} - g^{\mu_1 \mu_3} g^{\mu_2 \mu_4} \right)
    \, \Big]
\end{align}
\mbox{}\\[-14.5mm]
\begin{center}
\line(1,0){520}
\end{center}
\caption{Feynman rules derived from the lagrangian~\eqref{eq:ggHLagrangian}.} 
\label{tab:ggHvertices}
\end{table*}

\begin{table*}
\begin{center}
\line(1,0){520}
\end{center}
\mbox{}\\[-13mm]
\begin{align}
 \begin{minipage}[l]{0.3\linewidth}
 \includegraphics[scale=0.28]{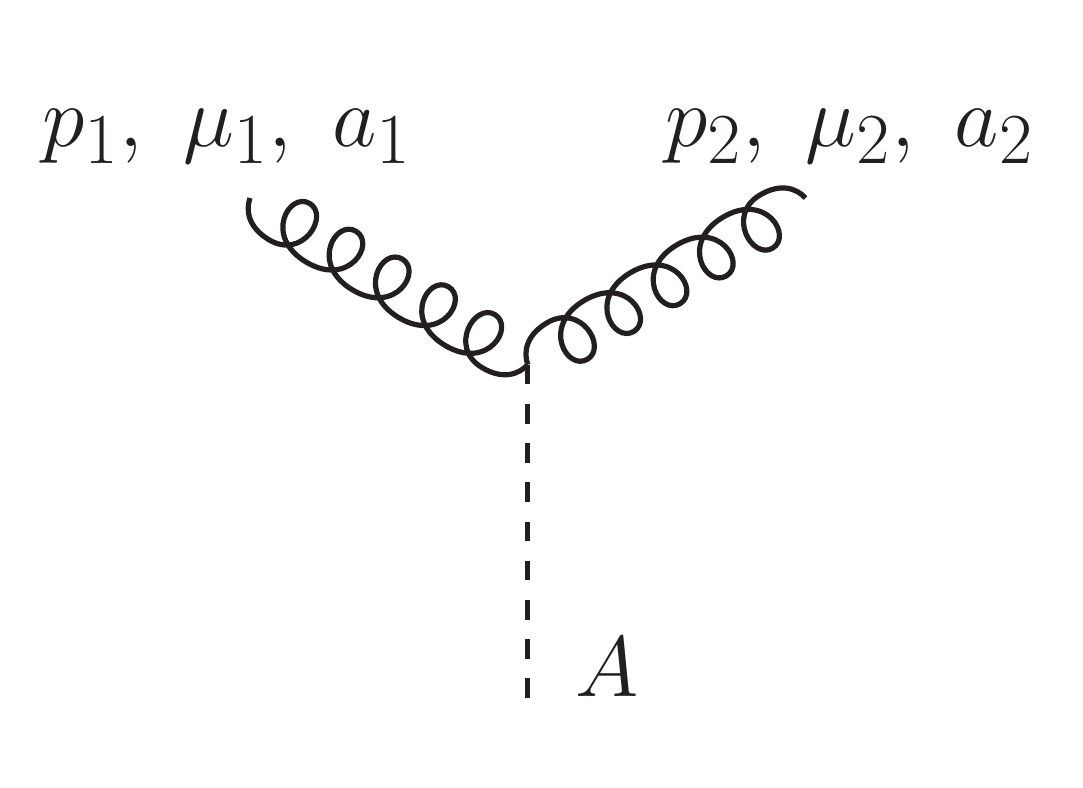}
 \end{minipage} 
 & \hspace*{-6em} =~ -i \, g_{\sss Agg} \, \delta^{a_1a_2} \, \epsilon^{\mu_1 \mu_2
 \rho \sigma} \, p_{1 \rho} \, p_{2 \sigma} \\[-3mm]
 \begin{minipage}[l]{0.3\linewidth}
 \includegraphics[scale=0.28]{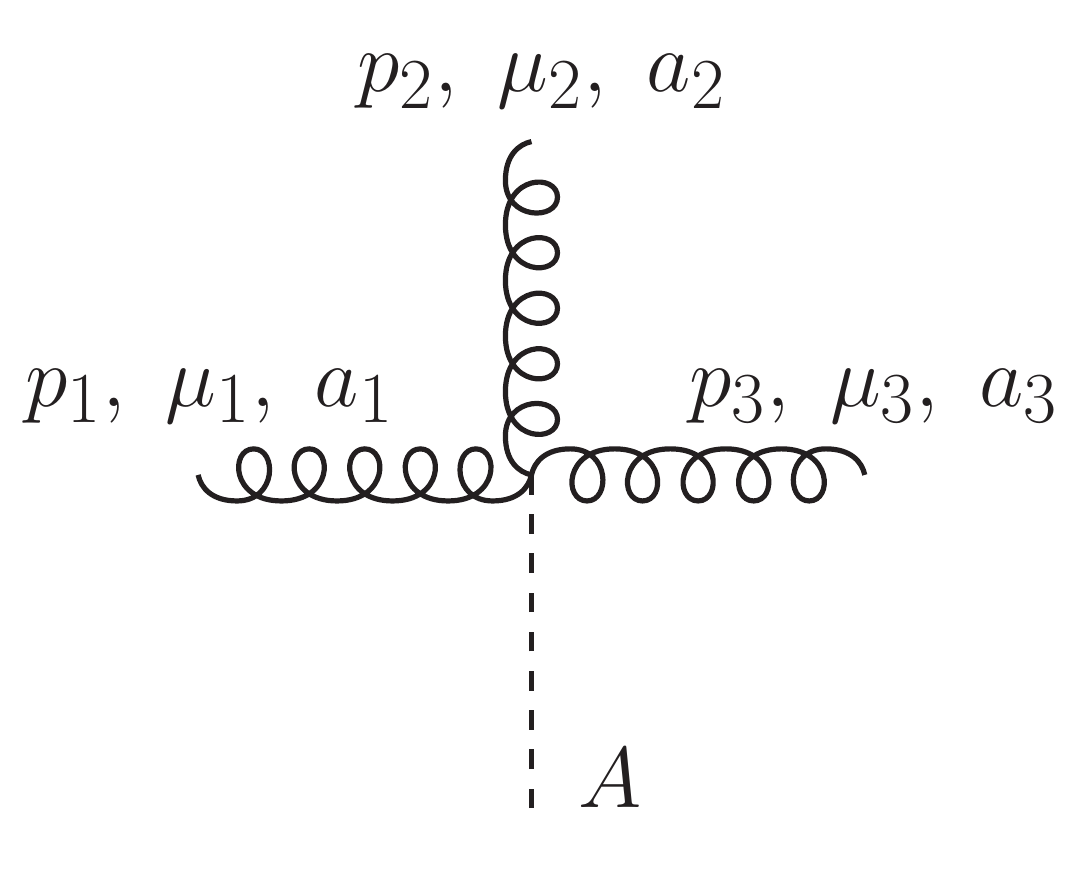}
 \end{minipage} 
 & \hspace*{-6em} =~ - g_{\sss Agg} \, g_s \, f^{a_1a_2a_3} \, \epsilon^{\mu_1 \mu_2 \mu_3 \rho} \left( p_1 + p_2 + p_3 \right)_\rho
\end{align}
\mbox{}\\[-15.mm]
\begin{center}
\line(1,0){520}
\end{center}
\caption{Feynman rules derived from the lagrangian~\eqref{eq:ggALagrangian}. 
 Note that all the pseudoscalar amplitudes vanish when $p_A^\mu \to (0,\mathbf{0})$.} 
\label{tab:ggAvertices}
\end{table*}

We match these effective vertices to the corresponding amplitudes
induced by a quark loop, which couples to the $H$ and $A$ components of
$X_0$ accordingly to eq.~\eqref{eq:0ff} ($\kappa_{\sss Htt,Att}=1$), in
the limit where this quark is infinitely massive. 
As a consequence, the effective couplings are fixed to the values
\begin{align}
 g_{\sss Hgg} = -\frac{\alpha_s}{3 \pi v} 
 \quad \mathrm{and} \quad 
 g_{\sss Agg} = \frac{\alpha_s}{2 \pi v} \,.
\label{eq:effcouplHAgg}
\end{align}

Our effective theory is invariant under $SU(3)_C$, so we can
consistently add higher order QCD corrections.
Going to NLO, we match again the result from the effective theory to the
corresponding case where the amplitude is induced by a heavy-quark
loop. 
In the latter case, virtual corrections consist of two-loop diagrams;
some of them appear explicitly in the effective theory as one-loop
diagrams, while the other ones simply result in a correction to the
value of the effective coupling.
This correction can be computed by means of a low-energy 
theorem~\cite{Spira:1995rr};
for the scalar we have
\begin{align}
 g_{\sss Hgg} & =  -\frac{\alpha_s}{3 \pi v} \Big( 1 +\frac{11}{4} \, \frac{\alpha_s}{\pi} \,
                + \mathcal{O}\big(\alpha_s^2 \big)   \Big) \,,
\intertext{while in the pseudoscalar case the effective coupling}
 g_{\sss Agg} & = \frac{\alpha_s}{2 \pi v} \,
\end{align}
is exact to all orders in $\alpha_s$~\cite{Adler:1969er}.
Together with this finite contribution to the UV renormalisation of the
effective couplings, we also need the UV polar terms that appear in
$D=4-2 \epsilon$ dimensional regularisation. 
Such counterterms are simply obtained by plugging into
eq.~\eqref{eq:effcouplHAgg} the known $\overline{\text{MS}}$
renormalisation of the strong coupling 
\begin{align}
 \alpha_s \to  \alpha_s \Big(1 - 
 \frac{1}{\epsilon}\, \frac{\alpha_s}{2 \pi}\, b_0 
 + \mathcal{O}\big(\alpha_s^2 \big)  \Big) \,,
\end{align}
where $b_0$ is the first coefficient of the QCD beta function
\begin{align}
  b_0 = \frac{11}{6}\, C_A - \frac{2}{3}\, T_F \, n_f \,.
\end{align}
Therefore, the UV counterterms have structures analogous to the 
tree-level Feynman rules in tables~\ref{tab:ggHvertices}
and \ref{tab:ggAvertices}.

To complete our set of rules, in tables~\ref{tab:ggHr2} and
\ref{tab:ggAr2} we report the $R_2$ counterterms~\cite{Ossola:2008xq,
Pittau:2011qp} of our effective theory, needed for the automatic
computation of one-loop amplitudes with the OPP
method~\cite{Ossola:2006us}. 
The $R_2$ vertices for $GGH$ have already been published
in~\cite{Page:2013xla} (with slightly different conventions),
while the $R_2$ vertices for the $G \widetilde G A$ operator are presented
here for the first time.

\begin{table*}
\begin{center}
\line(1,0){520}
\end{center}
\mbox{}\\[-13mm]
\begin{align}
 \begin{minipage}[l]{0.3\linewidth}
 \includegraphics[scale=0.28]{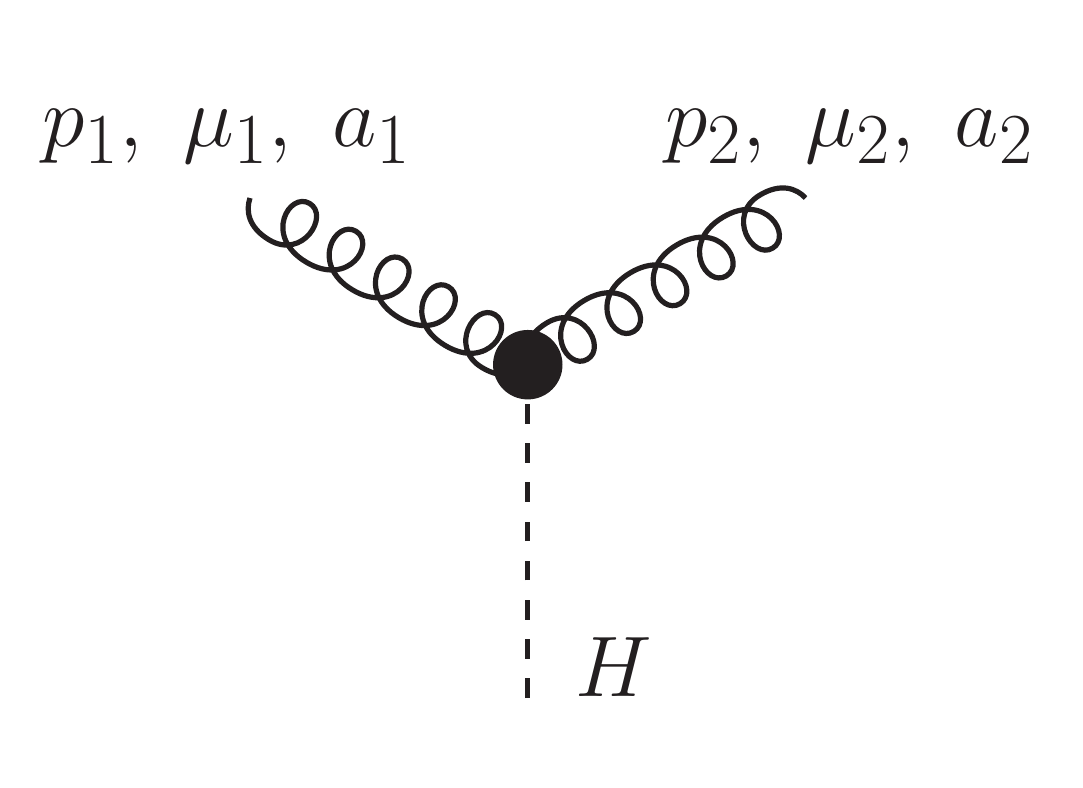}
 \end{minipage} 
 & \mbox{}\notag\\[-19mm]
 & \hspace*{-6em} =~ i \, g_{\sss Hgg}  \, \frac{g_s^2 N_c}{384 \pi^2} \, \delta^{a_1a_2} \, 
   \Big[
   -\big(\, 17 p_1^2 + 17 p_2^2 + 93 p_1\!\cdot\! p_2 \,\big) \, g^{\mu_1 \mu_2}
   \notag\\
 & \hspace*{4.2em}
   + p_1^{\mu_1} p_2^{\mu_2} \, + \, 89 p_2^{\mu_1} p_1^{\mu_2}
   + 14 \big(\, p_1^{\mu_1} p_1^{\mu_2} + p_2^{\mu_1} p_2^{\mu_2} \,\big)\,
   \Big] \\[0.5mm]
 \begin{minipage}[l]{0.3\linewidth}
 \includegraphics[scale=0.28]{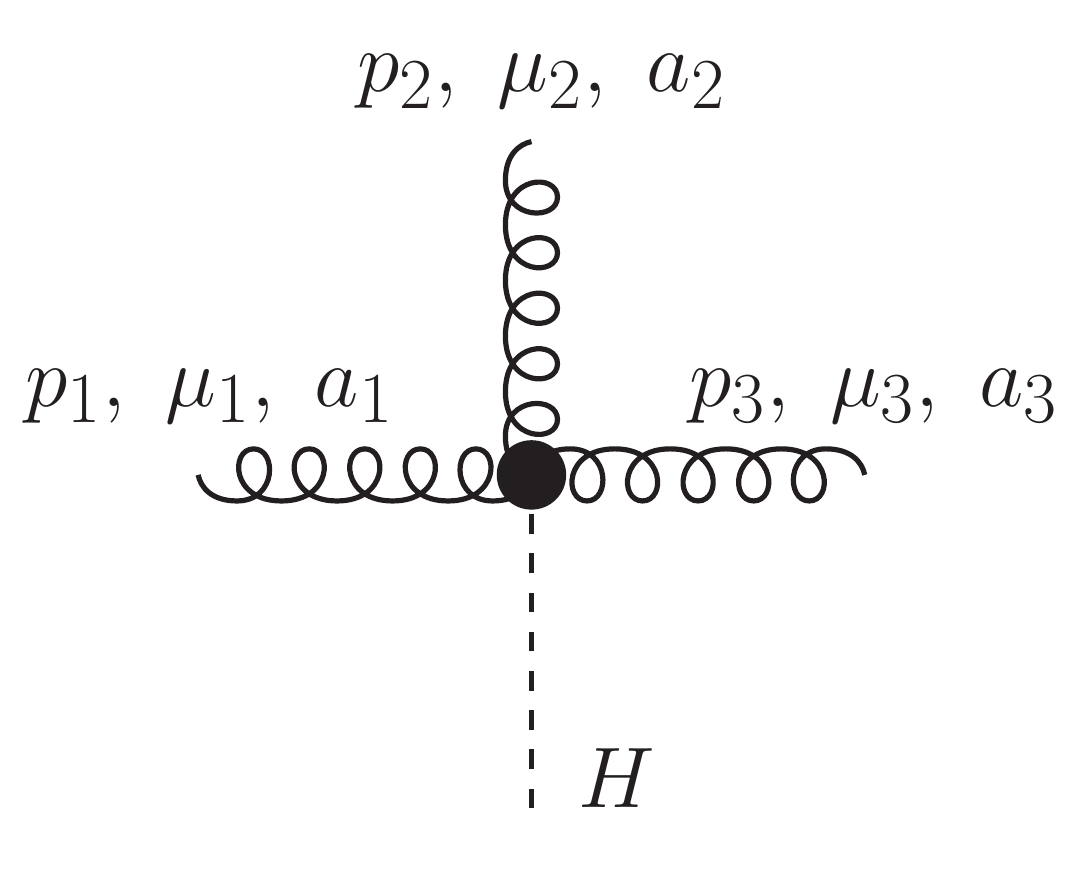}
 \end{minipage} 
 & \mbox{}\notag\\[-18.mm]
 & \hspace*{-6em} =~  g_{\sss Hgg}  \, \frac{15 g_s^3 N_c}{128 \pi^2} \,f^{a_1a_2a_3} \,\Big[  
    \hspace*{0.21em} g^{\mu_1 \mu_2} \left(p_1-p_2\right)^{\mu_3} 
   + g^{\mu_2 \mu_3} \left(p_2-p_3\right)^{\mu_1} 
   + g^{\mu_3 \mu_1} \left(p_3-p_1\right)^{\mu_2} 
   \,\Big] \\[6.mm]
 \begin{minipage}[l]{0.3\linewidth}
 \includegraphics[scale=0.28]{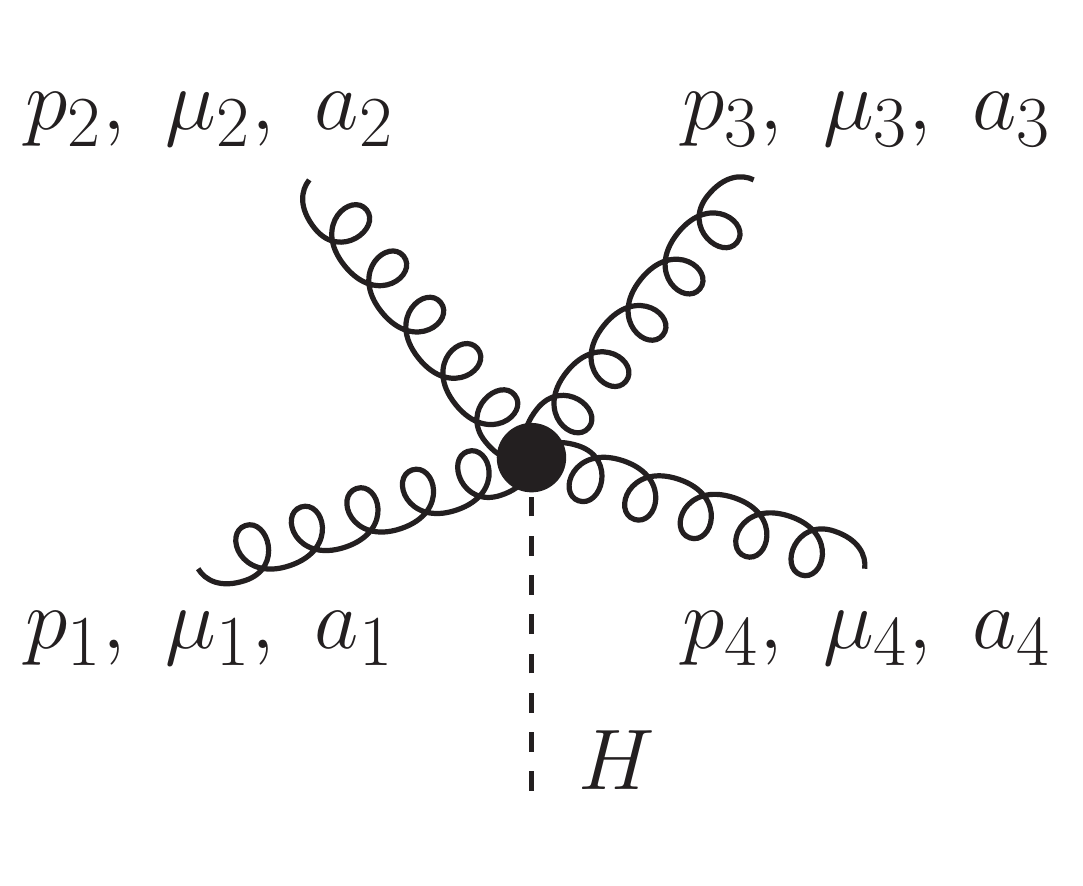}
 \end{minipage} 
 & \mbox{}\notag\\[-24.mm]
 & \hspace*{-6em} =~ i \, g_{\sss Hgg}  \, \frac{g_s^4}{128 \pi} \, \Big[ 
   \hspace*{1.43em} f^{a_1 b c} f^{a_2 c d} f^{a_3 d e} f^{a_4 e b} \,
  \big( \hspace*{1.22em} 21 g^{\mu_1 \mu_2} g^{\mu_3 \mu_4} 
  \,- 41 g^{\mu_1 \mu_3} g^{\mu_2 \mu_4} 
  \,+ 21 g^{\mu_1 \mu_4} g^{\mu_2 \mu_3}  \big) \, \notag\\
 & \hspace*{1.5em} + f^{a_1 b c} f^{a_2 c d} f^{a_4 d e} f^{a_3 e b} \,
  \big( \hspace*{1.22em} 21 g^{\mu_1 \mu_2} g^{\mu_3 \mu_4} 
  \,+ 21 g^{\mu_1 \mu_3} g^{\mu_2 \mu_4} 
  \,- 41 g^{\mu_1 \mu_4} g^{\mu_2 \mu_3}  \big) \, \notag\\
 & \hspace*{1.5em} + f^{a_1 b c} f^{a_3 c d} f^{a_2 d e} f^{a_4 e b} \, 
  \big(-41 g^{\mu_1 \mu_2} g^{\mu_3 \mu_4} 
  \,+ 21 g^{\mu_1 \mu_3} g^{\mu_2 \mu_4} 
  \,+ 21 g^{\mu_1 \mu_4} g^{\mu_2 \mu_3}  \big) \, \Big]
\\[-1.5mm]
 \begin{minipage}[l]{0.3\linewidth}
 \includegraphics[scale=0.28]{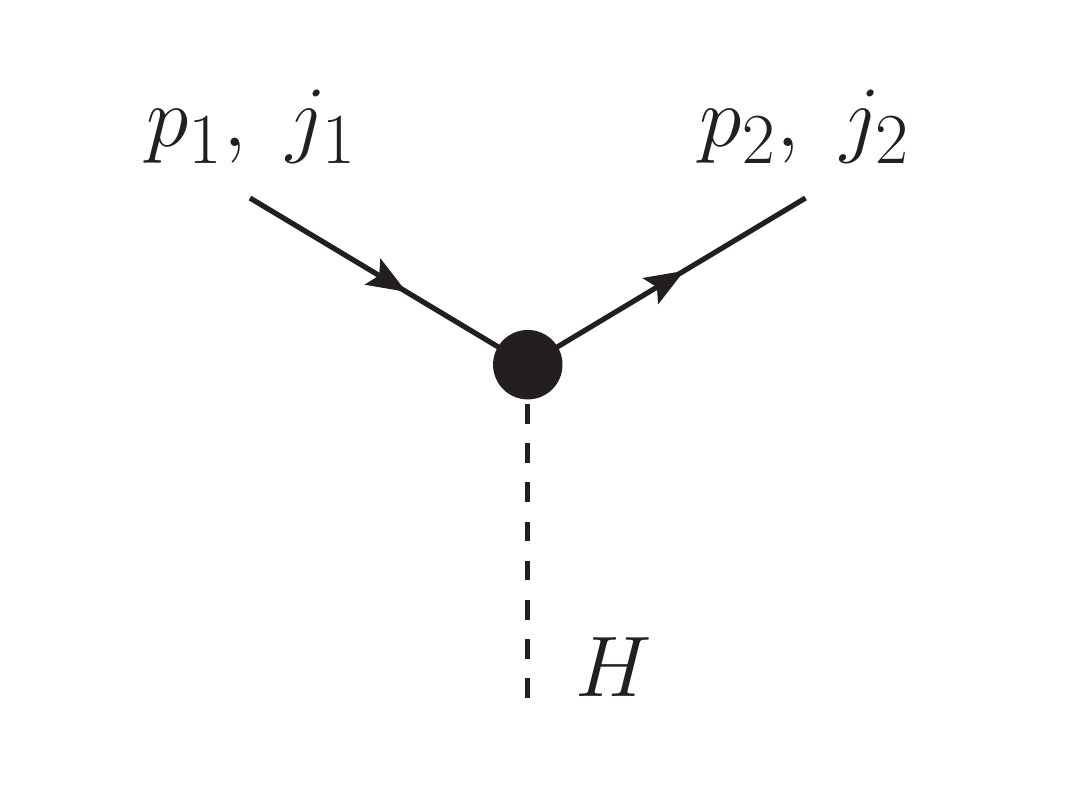}
 \end{minipage} 
 & \hspace*{-6em} =~ i \, g_{\sss Hgg}  \, \frac{g_s^2}{32 \pi^2} \, \frac{N_c^2-1}{2 N_c} \, 
 \lambda_{HV} \,\delta_{j_1j_2} \, \big(\, \slashed{p}_{^{\scriptstyle{2}}} - \slashed{p}_{^{\scriptstyle{1}}} \big)
   \\[-2.5mm]
 \begin{minipage}[l]{0.3\linewidth}
 \includegraphics[scale=0.28]{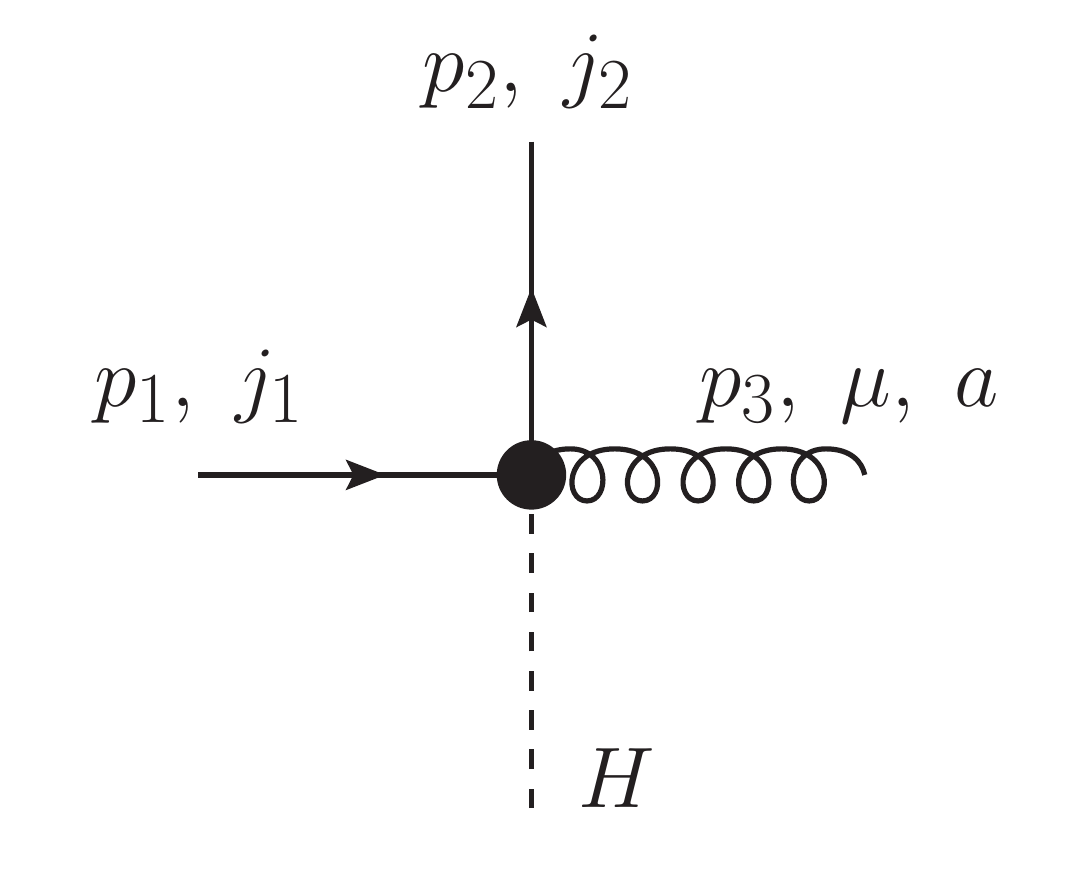}
 \end{minipage} 
 & \hspace*{-6em} =~ -i \, g_{\sss Hgg} \, \frac{g_s^3}{64 \pi^2}  \, t^a_{j_2 j_1} \gamma^\mu
   \left[ \frac{2 \lambda_{HV} + 1}{N_c} \, - \,
   \big( 2 \lambda_{HV} + 3 \big) N_c \right]
   \\[-5mm] \notag
\end{align}
%
%
\mbox{}\\[-17.5mm]
\begin{center}
\line(1,0){520}
\end{center}
\caption{$R_2$ counterterms for the lagrangian~\eqref{eq:ggHLagrangian}.
 $\lambda_{HV}=1$ is for dimensional regularisation, while
 $\lambda_{HV}=0$ for dimensional reduction.} 
\label{tab:ggHr2}
\end{table*}

\begin{table*}
\begin{center}
\line(1,0){520}
\end{center}
\mbox{}\\[-13mm]
\begin{align}
 \begin{minipage}[l]{0.3\linewidth}
 \includegraphics[scale=0.28]{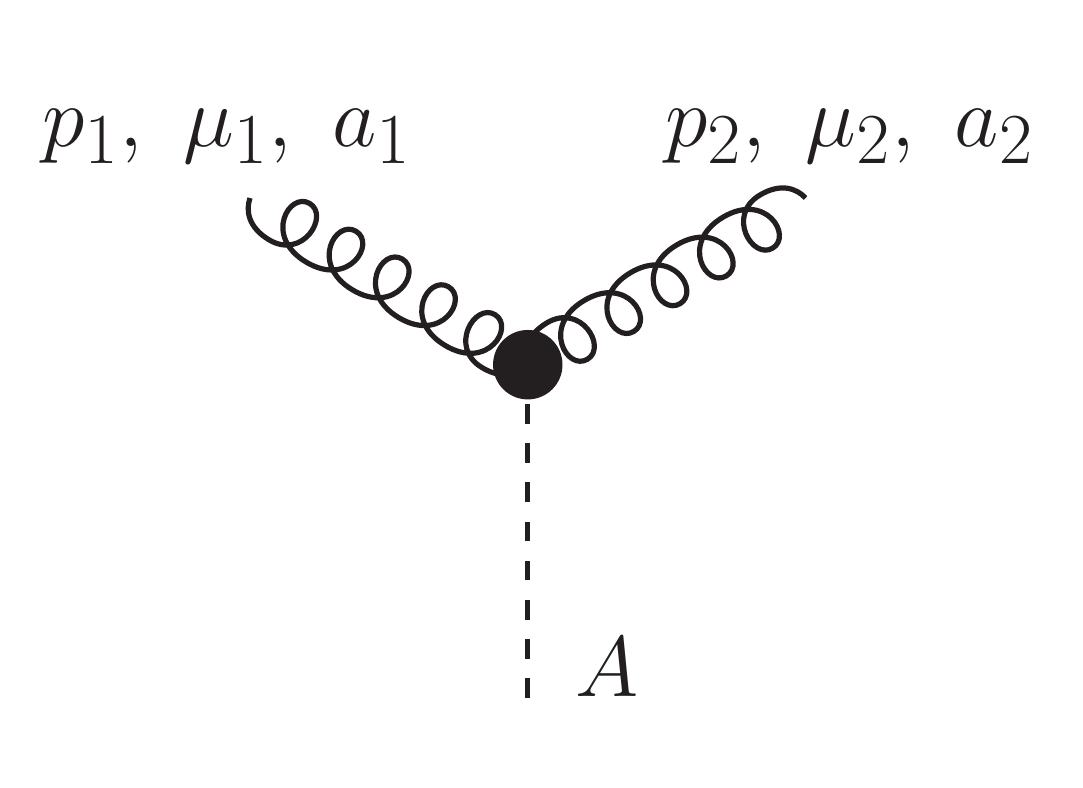}
 \end{minipage} 
 & \hspace*{-6em} =~ i \, g_{\sss Agg} \, \frac{g_s^2 N_c}{96 \pi^2} \, \delta^{a_1a_2}
 \, \epsilon^{\mu_1 \mu_2 \rho \sigma} \, p_{1 \rho} \, p_{2 \sigma} \\[-3mm]
 \begin{minipage}[l]{0.3\linewidth}
 \includegraphics[scale=0.28]{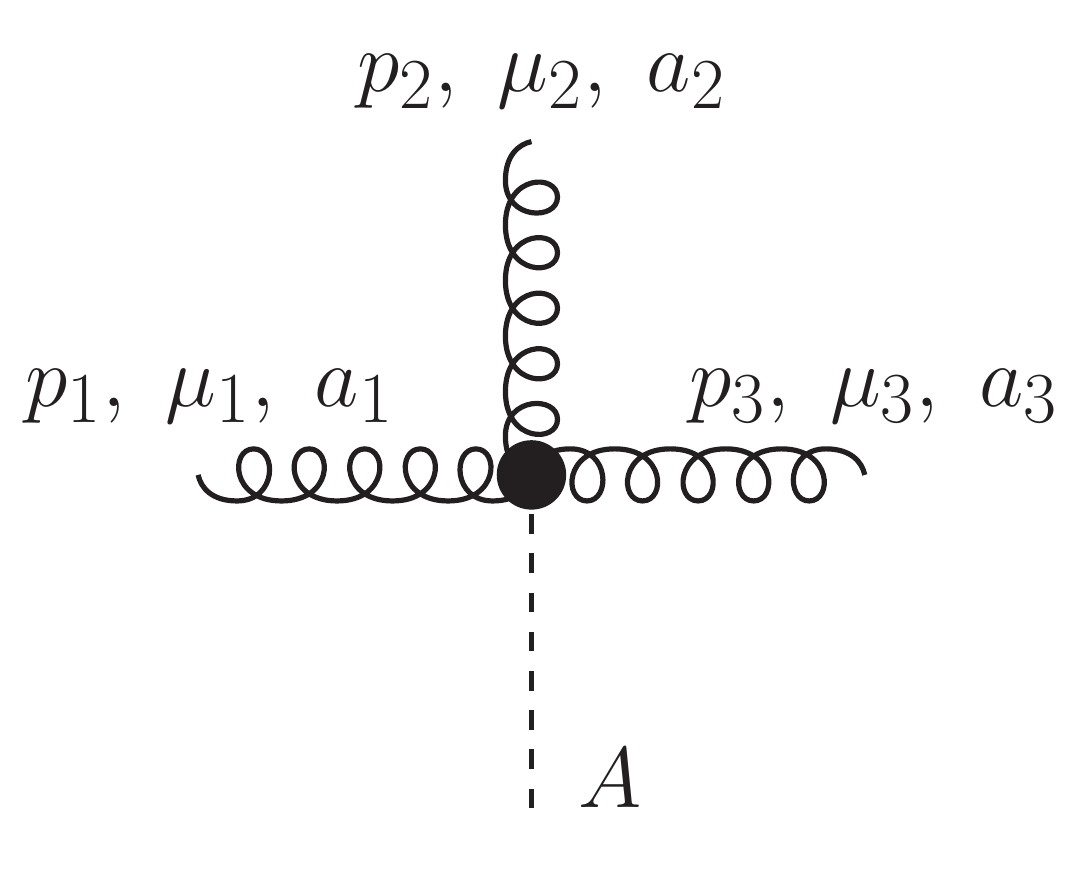}
 \end{minipage} 
 & \hspace*{-6em} =~ g_{\sss Agg} \, \frac{g_s^3 N_c}{64 \pi^2} \, f^{a_1a_2a_3} \, \epsilon^{\mu_1 \mu_2 \mu_3 \rho} \left( p_1 + p_2 + p_3 \right)_\rho
\end{align}
%
\mbox{}\\[-15.mm]
\begin{center}
\line(1,0){520}
\end{center}
\caption{$R_2$ counterterms for the lagrangian~\eqref{eq:ggALagrangian}.}
\label{tab:ggAr2}
\end{table*}

\bibliography{library}
\bibliographystyle{JHEP}

\end{document}